\documentclass[prd,nofootinbib,superscriptaddress,preprint]{revtex4-1}
\usepackage{amsmath,amssymb}
\usepackage{epsfig}
\usepackage{graphicx}
\usepackage[usenames,dvipsnames]{color}
\usepackage{slashed}
\usepackage[colorlinks,citecolor=blue]{hyperref}
\usepackage{color}

\usepackage{mathtools}
\usepackage{longtable}
\usepackage{booktabs}
\usepackage{tabularx}
\usepackage[section]{placeins}
\usepackage{amsmath}
\usepackage{graphicx}
\usepackage{dcolumn}
\usepackage{bm}
\usepackage{array,multirow}

\begin{document}


\title{Compatibility of $A_{4}$ Flavour Symmetric Minimal Extended Seesaw with $(3+1)$ Neutrino Data}

\author{Neelakshi Sarma}
\email{nsarma25@gmail.com}
\affiliation{Department of Physics, Gauhati University, Guwahati, Assam 781014, India.}


\author{Kalpana Bora}
\email{kalpana@gauhati.ac.in}
\affiliation{Department of Physics, Gauhati University, Guwahati, Assam 781014, India.}%

\author{Debasish Borah}%
 \email{dborah@iitg.ac.in}
\affiliation{Department of Physics, Indian Institute of Technology Guwahati, Assam 781039, India.}
%




\begin{abstract}
Motivated by the recent resurrection of the evidence for an eV scale sterile neutrino from the MiniBooNE experiment, we revisit one of the most minimal seesaw model known as the minimal extended seesaw that gives rise to a $3+1$ light neutrino mass matrix. We consider the presence of $A_4$ flavour symmetry which plays a non-trivial role in generating the structure of the neutrino mass matrix. Considering a diagonal charged lepton mass matrix and generic vacuum alignments of $A_4$ triplet flavons, we classify the resulting mass matrices based on their textures. Keeping aside the disallowed texture zeros based on earlier studies of $3+1$ neutrino textures, we categorise the remaining ones based on texture zeros, $\mu-\tau$ symmetry in the $3\times3$ block and hybrid textures. After pointing out the origin of such $3+1$ neutrino textures to $A_4$ vacuum alignments, we use the latest $3+1$ neutrino oscillation data and numerically analyse the texture zeros and $\mu-\tau$ symmetric cases. We find that a few of them are allowed from each category predicting interesting correlations between neutrino parameters. We also find that all of these allowed cases prefer normal hierarchical pattern of light neutrino masses over inverted hierarchy.
\end{abstract}

\maketitle


\section{Introduction}
\label{sec:level1}
Non-zero neutrino masses and large leptonic mixing have now become a well established fact, thanks to a series of results from several experiments \cite{PDG,PDG1,PDG2,PDG3,PDG4,kamland} over the last twenty years. While the solar and atmospheric mixing angles plus mass squared difference measurements have become more precise with time, the evidence for a non-zero reactor mixing angle emerged with the relatively recent experiments like MINOS~\cite{minos}, T2K~\cite{T2K}, NO$\nu$A~\cite{nova}, Double ChooZ~\cite{chooz}, Daya-Bay~\cite{daya} and RENO~\cite{reno}. Apart from the currently unknown parameters in the neutrino sector, like mass hierarchy, Dirac CP violating phase as the global fit data suggest \cite{schwetz17}, another interesting question in the neutrino sector is the possibility of additional neutrino species with eV scale mass. In fact, this has turned out to be not just a speculation, but has gathered considerable attention in the last two decades following some anomalies reported by a few experiments. The first such anomaly was reported by the Liquid Scintillator Neutrino Detector 
(LSND) experiment in their anti-neutrino flux measurements~\cite{Athanassopoulos:1996jb,Aguilar:2001ty}. 
The LSND experiment searched for $\bar{\nu}_{\mu} \rightarrow \bar{\nu}_e$ 
oscillations in the appearance mode and reported an excess of $\bar{\nu}_e$ 
interactions that could be explained 
by incorporating at least one additional light neutrino with mass in the eV 
range. This result was supported by the subsequent measurements at the MiniBooNE 
experiment~\cite{Aguilar-Arevalo:2013pmq}. 
Similar anomalies have also been observed at reactor neutrino
experiments~\cite{Mention:2011rk} as well as gallium solar neutrino 
experiments~\cite{Acero:2007su,Giunti:2010zu}. These anomalies received renewed attention recently after the MiniBooNE collaboration reported their new analysis incorporating twice the size data sample than before \cite{miniboone18}, confirming the anomaly at $4.8\sigma$ significance level which becomes $> 6\sigma$ effect if combined with LSND. Although an eV scale neutrino can explain this anomaly, such a neutrino can not have gauge interactions in the standard model (SM) from the requirement of being in agreement with precision measurement of $Z$ boson decay width at LEP experiment \cite{Agashe:2014kda}. Hence such a neutrinos is often referred to as a sterile neutrino while the usual light neutrinos are known as active neutrinos. Status of this framework with three active and one sterile or $3+1$ framework with respect to such short baseline neutrino anomalies can be found in several global fit studies \cite{Kopp:2013vaa,Giunti:2013aea,Gariazzo:2015rra, global3+1-18}. It is worth mentioning that the latest cosmology results from the Planck collaboration \cite{Aghanim:2018eyx} constrains the effective number of relativistic degrees of freedom $N_{\text{eff}} = 2.99 \pm 0.17$ at $68\%$ confidence level (CL), which is consistent with the SM prediction $N_{\text{eff}} = 3.046$ for three light neutrinos. Similarly, the constraint on the sum of absolute neutrino masses $\sum_i \lvert m_i \rvert < 0.12$ eV \cite{Aghanim:2018eyx} (at $95\%$ CL) does not leave any room for an additional light neutrino with mass in eV order. Although this latest bound from the Planck experiment can not accommodate one additional light sterile neutrino at eV scale within the standard $\Lambda$CDM model of cosmology, one can evade these tight bounds by considering the presence of some new physics beyond the standard model (BSM). For example, additional gauge interactions in order to suppress the production of sterile neutrinos through flavour oscillations were studied recently by the authors of~\cite{Dasgupta:2013zpn}.

Such experimental indications of an eV scale sterile neutrino having non-trivial mixing with active neutrinos have led to several BSM proposals that can account for the same. While the usual seesaw mechanisms like type I~\cite{ti,ti1,ti2,ti3,ti4}, type II~\cite{tii,tii1,tii2,tii3,tii4,tii5,tii6} and type III~\cite{tiii} explaining the lightness of active neutrinos were studied in details for a long time, their extensions to the $3+1$ case was not very straightforward primarily due to the gauge singlet nature of the sterile neutrino. Yet, there have been several proposals to generate a $4\times 4$ light neutrino mass matrix within different seesaw frameworks in recent times~\cite{sterileearlier1, sterileearlier2, sterileearlier3, 
sterileearlier4, sterileearlier5, sterileearlier6, 
sterileearlier7, sterileearlier8, sterileearlier9, sterileearlier10, 
Borah:2016xkc, Borah:2016lrl, Borah:2017azf, sterileearlier12, Borah:2016fqj}. Here we adopt a minimal framework known as the minimal extended seesaw proposed in the $3+1$ neutrino context by \cite{sterileearlier2, sterileearlier3} and study different possible realisations within the framework of non-abelian discrete flavour symmetry $A_4$. Flavour symmetry is needed to explain the observed flavour structure of different particles of the standard model. In the original proposal \cite{sterileearlier3} also, the $A_4$ flavour symmetry was utilised but within the limited discussion the issue of non-zero reactor mixing angle as well as different $A_4$ vacuum alignments were not addressed. In another recent work based on the same model with $A_4$ flavour symmetry \cite{Das:2018qyt}, some details of the associated neutrino phenomenology was discussed by sticking to the effective $3\times 3$ active neutrino mass matrix which can be generated by integrating out the sterile neutrino. In our present work, we consider the full $4\times 4$ mass matrix and do not integrate out the sterile neutrino as its mass may not lie far above the active ones always, as hinted by experiments mentioned above. We also classify different possible textures of the $4\times 4$ neutrino mass matrix based on generic $A_4$ vacuum alignments for triplet flavons. Similar but not texture specific work in three neutrino cases to constrain different $A_4$ vacuum alignments from three neutrino data was done by the authors of \cite{Chen:2012st} which was further constrained from successful leptogenesis in \cite{Kalita:2015jaa}. Here we extend such studies to the $3+1$ neutrino cases. Texture zeros in $3+1$ neutrino scenarios were discussed in different contexts earlier using flavour symmetries like $Z_N, U(1)$ etc. \cite{Borah:2016xkc, Borah:2017azf, sterileearlier12} but here we show that some of these textures can be realised (upto a few more constraints) just from the vacuum alignment of $A_4$ triplet flavons. We first make the classifications for allowed and disallowed textures based on already known texture results in $3+1$ neutrino frameworks \cite{Ghosh:2012pw,Ghosh:2013nya,Zhang:2013mb,Nath:2015emg,Borah:2016xkc,Borah:2017azf} and then numerically analyse some of the textures which have not been studied before. To be more specific, we categorise our textures based on $\mu-\tau$ symmetric cases, texture zero cases, hybrid cases and disallowed ones. Out of them, we numerically analyse all the textures belonging to $\mu-\tau$ symmetric and texture zero cases leaving the discussion on hybrid textures to future works. It should be noted that, although the discovery of non-zero reactor mixing angle has ruled out $\mu-\tau$ symmetry in the three neutrino scenarios, it is possible to retain it in a $3+1$ scenario where the $3\times 3$ neutrino block retains this symmetry while the active-sterile sector breaks it. This interesting but much less explored idea to generate non-zero $\theta_{13}$ by allowing the mixing of three active neutrinos with a eV scale sterile neutrino was proposed earlier in \cite{sterilemutau0, sterilemutau, sterilemutau1, sterilemutau2} and was also studied in details recently in \cite{Borah:2016fqj}. We find that many of the textures belonging to these categories are already ruled out by neutrino data while the ones which are allowed give interesting correlations between neutrino parameters which can be tested at ongoing and future experiments. 

This article is organised as follows. In section \ref{sec:level2} we discuss the details of the model followed by the classification of different textures in section \ref{sec3}. In section \ref{sec:numeric} we discuss the numerical analysis adopted in our work followed by results and discussions in section \ref{sec:results}. We finally conclude in section \ref{sec:conclude}.

\section{\label{sec:level2}The Model}
As mentioned before, here we adopt the model first proposed in \cite{sterileearlier3} but discuss it from a more general perspective taking all the allowed terms in the Lagrangian and all possible generic vacuum alignments of $A_4$ triplets. Here we note that the discrete non-abelian group $A_4$ is the group of even permutations of four objects or the symmetry group of a tetrahedron. It has twelve elements and four irreducible representations with dimensions $n_i$ such that $\sum_i n_i^2=12$. These four representations are denoted by $\bf{1}, \bf{1'}, \bf{1''}$ and $\bf{3}$ respectively. The product rules for these representations are given in Appendix \ref{appen1}.

\begin{table}[htbp]
\begin{center}
\begin{tabular}{|c|ccccc|cccccc|cccc|} \hline
&  $l$ & $e_{R}$ & $\mu_{R}$ & $\tau_{R}$ & H & $\phi$ & $\phi^{'}$ & $\phi^{''}$  & $\xi$ & $\xi^{'}$ & $\chi$ & $\nu_{R1}$  & $\nu_{R2}$ & $\nu_{R3}$ & S \\ \hline
$SU(2)_L$ & $2$ & $1$ & $1$ & $1$ & $2$ & $1$ & $1$ & $1$ & $1$ & $1$ & $1$ & $1$ & $1$ & $1$ & $1$\\ \hline
$A_4$ & $3$ & $1$ & $1^{''}$ & $1^{'}$ & $1$ & $3$ & $3$ & $3$ & $1$ & $1^{'}$ & $1$ & $1$ & $1^{'}$ & $1$ & $1$\\ \hline
$Z_4$ & $1$ & $1$ & $1$ & $1$ & $1$ & $1$ & $i$ & $-1$ & $1$ & $-1$ & $-i$ & $1$ & $-i$ & $-1$ & $i$\\ \hline
\end{tabular}
\caption{Fields and their transformations under the chosen symmetries.}
\label{table1}
\end{center}
\end{table}

The particle content of the model along with their transformations under the symmetries of the model are shown in table \ref{table1}. Apart from the SM gauge symmetry and $A_4$ flavour symmetry, an additional discrete symmetry $Z_4$ is also chosen in order to forbid certain unwanted terms. For example, the chosen $Z_4$ charge of the singlet neutrino $S$ keeps a bare mass term away from the Lagrangian. This is important because a bare mass term will be typically large, at least of electroweak scale and hence will not help us generate a $4\times 4$ light neutrino mass matrix with all terms at or below the eV scale. To have a seesaw mechanism at place, three right handed neutrinos $\nu_{Ri}, i=1,2,3$ are included into the model. Apart from the usual Higgs field $H$ responsible for electroweak symmetry breaking, there are six flavon fields $\phi$, $\phi^{'}$, $\phi^{''}$, $\xi$, $\xi^{'}$, $\chi$ responsible for spontaneous breaking of the flavour symmetries and generating the desired leptonic mass matrices. The leading order Lagrangian for the leptons can be written as
 \begin{align}
\mathcal{L}_Y & \supset \frac{y_{e}}{\Lambda}(\bar{l} H \phi)_{\underline{1}}e_{R} + \frac{y_{\mu}}{\Lambda}(\bar{l} H \phi)_{\underline{1'}}\mu_{R} + \frac{y_{\tau}}{\Lambda}(\bar{l} H \phi)_{\underline{1''}}\tau_{R} + \frac{y_{1}}{\Lambda}(\bar{l} H \phi)_{\underline{1}}\nu_{R1} + \frac{y_{2}}{\Lambda}(\bar{l} H \phi')_{\underline{1''}}\nu_{R2} \nonumber \\
& + \frac{y_{3}}{\Lambda}(\bar{\underline{l}} H \phi'')_{\underline{1}}\nu_{R3} + \frac{1}{2}\lambda_{1}\xi \overline{\nu_{R1}^{c}}\nu_{R1}
 + \frac{1}{2}\lambda_{2}\xi' \overline{\nu_{R2}}^{c}\nu_{R2}  + \frac{1}{2}\lambda_{3}\xi \overline{\nu_{R3}}^{c}\nu_{R3} + \frac{1}{2}\rho\chi\overline{S^{c}}\nu_{R1} + y_{4}\xi\overline{S^{c}}\nu_{R2} \nonumber \\
 &+ y_{5}\chi^{\dagger}\overline{S^{c}}\nu_{R3} + \text{h.c.} 
 \end{align}
where $\Lambda$ is the cut-off scale of the theory, $y_{e}, y_{\mu}, y_{\tau}, y_{1}, y_{2}, y_{3}, y_{4}, y_{5}, \lambda_{1}, \lambda_{2}, \lambda_{3}, \rho $ are the dimensionless Yukawa couplings. It is worth noting that the last two terms were not included in the original model \cite{sterileearlier3} although they are allowed by the chosen symmetry of the model. We include them here as they contribute non-trivially to the neutrino mass matrix as well as the generation of correct neutrino mixing. 

We denote a generic vacuum alignment of the flavon fields as follows
\begin{equation}
\langle \phi \rangle = v(n_{1}, n_{2}, n_{3}), \quad 
\langle \phi' \rangle = v(n_{4}, n_{5}, n_{6}), \quad 
\langle \phi'' \rangle = v(n_{7}, n_{8}, n_{9}), \quad 
\langle \xi \rangle = \langle \xi' \rangle = v,  \quad 
\langle \chi \rangle = u
\end{equation}
where $n_i, i=1-9$ are dimensionless numbers which we choose to take values as $n_i \in (-1, 0, 1)$, which are natural choices for alignments in such flavour symmetric models. Here $v$ or $u$ denotes the vacuum expectation value (VEV) of the flavon fields which typically characterises the scale of flavour symmetry breaking. Similar but more restricted alignments are chosen in the original proposal \cite{sterileearlier3}. Using such VEV alignments and the $A_4$ product rules given in Appendix \ref{appen1}, the charged lepton mass matrix can be written as
\begin{equation}
 m_{l} =\frac{\langle H \rangle v}{\Lambda}\begin{pmatrix}
n_{1}y_{e} & n_{2}y_{\mu} & n_{3}y_{\tau} \\
n_{3}y_{e} & n_{1}y_{\mu} & n_{2}y_{\tau} \\
n_{2}y_{e} & n_{3}y_{\mu} & n_{1}y_{\tau}  
\end{pmatrix}.  
\end{equation}
The neutral fermion mass matrix in the basis $(\nu_L, \nu_R, S)$ can be written 
as
\begin{equation}
\mathcal{M}= \left( \begin{array}{ccc}
              0 & M_{D} &  0 \\
              M^T_{D} & M_{R} & M^T_S \\
              0 & M_S & 0
                      \end{array} \right)
\label{eqn:numatrix}       
\end{equation}
where $M_D$, the Dirac neutrino mass matrix is 
\begin{equation}
M_{D} =\frac{\langle H \rangle v}{\Lambda}\begin{pmatrix}
y_{1} n_{1} & y_{2} n_{5} & y_{3} n_{7} \\
y_{1} n_{3} & y_{2} n_{4} & y_{3} n_{9} \\
y_{1} n_{2} & y_{2} n_{6} & y_{3} n_{8}
\end{pmatrix} =\sqrt{A}\begin{pmatrix}
y_{1} n_{1} & y_{2} n_{5} & y_{3} n_{7} \\
y_{1} n_{3} & y_{2} n_{4} & y_{3} n_{9} \\
y_{1} n_{2} & y_{2} n_{6} & y_{3} n_{8}
\end{pmatrix} 
\end{equation}
with $ A = \frac{\langle H \rangle^{2}v^{2}}{\Lambda^{2}} $. The right-handed neutrino mass matrix takes the diagonal form
\begin{equation}
M_{R} = \begin{pmatrix}
\lambda_{1}v & 0 & 0 \\
0 & \lambda_{2}v & 0 \\
0 & 0 & \lambda_{3}v
\end{pmatrix}, 
\end{equation}
and $M_S$ in the basis $(S, \nu_R)$ is given by 
\begin{equation}
M_{S} = (\rho u, y_{4}v, y_{5}u).
\end{equation}
In the case where $M_R \gg M_S > M_D$, the effective $4\times4$ light neutrino 
mass matrix in the basis $(\nu_L, \nu_s)$ can be written as~\cite{sterileearlier3}
\begin{equation}
M_{\nu}= -\left( \begin{array}{cc}
              M_{D} M^{-1}_R M^T_D &  M_D M^{-1}_R M^T_S\\
              M_S (M^{-1}_R)^TM^T_{D} & M_SM^{-1}_{R}M^T_S
                      \end{array} \right)
\label{eqn:numatrix2}       
\end{equation}
Using the expressions for $M_D, M_R, M_S$ mentioned above, the $4 \times 4$ active-sterile mass matrix can be written as
\begin{equation}
  m_{\nu}^{4\times4} = \begin{pmatrix}
 -Aa_{7} & -Aa_{8} & -Aa_{9} & -\sqrt{A}a_{1} \\
 -Aa_{10} & -Aa_{11} & -Aa_{12} & -\sqrt{A}a_{2} \\
 -Aa_{13} &  -Aa_{14} & -Aa_{15} & -\sqrt{A}a_{3} \\
 -\sqrt{A}a_{4} & -\sqrt{A}a_{5} & -\sqrt{A}a_{6} & -a_{0}
\end{pmatrix} 
\end{equation}
where
\begin{equation}
a_{0} =(\frac{\rho^{2}u^{2}}{\lambda_{1}v}+\frac{y_{4}^{2}v}{\lambda_{2}}+\frac{y_{5}^{2}u^{2}}{v\lambda_{3}}),
\end{equation}
\begin{equation}
a_{1}=a_4 = (\frac{\rho u y_{1}n_{1}}{v\lambda_{1}}+\frac{y_{4}y_{2}n_{5}}{\lambda_{2}}+\frac{uy_{5}y_{3}n_{7}}{v\lambda_{3}}), 
\end{equation}
\begin{equation}
 a_{2} =a_5=(\frac{\rho u y_{1}n_{3}}{v\lambda_{1}}+\frac{y_{4}y_{2}n_{4}}{\lambda_{2}}+\frac{uy_{5}y_{3}n_{9}}{v\lambda_{3}}), 
\end{equation}
\begin{equation}
 a_{3}=a_6 = (\frac{\rho u y_{1}n_{2}}{v\lambda_{1}}+\frac{y_{4}y_{2}n_{6}}{\lambda_{2}}+\frac{uy_{5}y_{3}n_{8}}{v\lambda_{3}}), 
\end{equation}
\begin{equation}
 a_{7} = (\frac{y_{1}^{2}n_{1}^{2}}{v\lambda_{1}}+\frac{y_{2}^{2}n_{5}^{2}}{v\lambda_{2}}+\frac{y_{3}^{2}n_{7}^{2}}{v\lambda_{3}}), 
\end{equation}
\begin{equation}
a_{8} =a_{10}= (\frac{y_{1}^{2}n_{3}n_{1}}{v\lambda_{1}}+\frac{y_{2}^{2}n_{4}n_{5}}{v\lambda_{2}}+\frac{y_{3}^{2}n_{7}n_{9}}{v\lambda_{3}}), 
\end{equation}
\begin{equation}
 a_{9} =a_{13}= (\frac{y_{1}^{2}n_{1}n_{2}}{v\lambda_{1}}+\frac{y_{2}^{2}n_{5}n_{6}}{v\lambda_{2}}+\frac{y_{3}^{2}n_{7}n_{8}}{v\lambda_{3}}), 
\end{equation}
\begin{equation}
 a_{11} = (\frac{y_{1}^{2}n_{3}^{2}}{v\lambda_{1}}+\frac{y_{2}^{2}n_{4}^{2}}{v\lambda_{2}}+\frac{y_{3}^{2}n_{9}^{2}}{v\lambda_{3}}), 
\end{equation}
\begin{equation}
 a_{12}=a_{14} = (\frac{y_{1}^{2}n_{2}n_{3}}{v\lambda_{1}}+\frac{y_{2}^{2}n_{4}n_{6}}{v\lambda_{2}}+\frac{y_{3}^{2}n_{8}n_{9}}{v\lambda_{3}}), 
\end{equation}
\begin{equation}
 a_{15} = (\frac{y_{1}^{2}n_{2}^{2}}{v\lambda_{1}}+\frac{y_{2}^{2}n_{6}^{2}}{v\lambda_{2}}+\frac{y_{3}^{2}n_{8}^{2}}{v\lambda_{3}}). 
\end{equation}\\
This is a $4\times4$ complex symmetric mass matrix, in general having ten independent elements. However, depending upon the vacuum alignments or the specific values of $n_i \in (-1, 0, 1)$, the mass matrix can have interesting textures which we discuss in details in the next section. 

The choice of vacuum alignment of the flavon fields required to achieve the desired structures of lepton mass matrices can be realised only when additional driving fields are incorporated as discussed in \cite{A4TBM1} for usual three neutrino scenarios. For similar discussion in a $3+1$ neutrino scenario, please refer to \cite{Borah:2016fqj}. Since these driving fields do not affect the general structure of the mass matrices, we have not incorporated them in the discussion above. The non-trivial vacuum alignment of the $\phi$, $\phi^{'}$, $\phi^{''}$ fields required to produce the specific structure of the charged lepton mass matrix and the $4 \times 4$ block of the light neutrino mass matrix is realised by introducing three additional driving fields. As shown in such works discussing the vacuum alignment of $A_4$ flavons, $n_i \in (-1, 0, 1)$ corresponds to generic alignments which can be naturally realised from the minimisation of the scalar potential (superpotential in supersymmetric scenarios).
 
\section{Classification of Textures}
\label{sec3}
We choose to work in the basis where the charged lepton mass matrix is diagonal. This allows the leptonic mixing matrix to be directly related to the diagonalising matrix of the light neutrino mass matrix. As discussed in the previous section, this corresponds to the VEV of the flavon field $\phi$ to be
$\langle \phi \rangle = v(n_{1}, n_{2}, n_{3})$, 
with $n_{1}=1$, $n_{2} = n_{3} = 0$.  In the most general case of the vacuum alignments of the flavon fields $\phi'$ and $\phi''$, each of $n_{4}, n_{5}, n_{6}, n_{7}, n_{8}, n_{9}$ can take 3 values, i.e. $0,1,-1$. Therefore we have $3^{6} = 729$ possible cases of different vacuum alignments, which will generate $729$ different $4\times4$ neutrino mass matrices. These 729 vacuum alignments are presented in Appendix \ref{appen2} and Appendix \ref{appen3}. We first single out the disallowed textures based on the known results from previous analysis  \cite{Ghosh:2012pw,Ghosh:2013nya,Zhang:2013mb,Nath:2015emg,Borah:2016xkc,Borah:2016lrl, Borah:2017azf}. They are given as follows.

\textbf{Disallowed cases:} 
\begin{enumerate}
\item Texture zero in entire second row and column. Total number of such textures is 71.
\item Texture zero in entire third row and column.  Total number of such textures is 73.
\item Texture zero in entire second and third rows and columns. Total number of such textures is 9.
\item $\mu-\tau$ symmetry in the entire $4\times4$ block. Total number of such textures is 72. 
\end{enumerate}
Total no of such disallowed mass matrices is 225. While the first three categories are inconsistent with the $3+1$ global fit neutrino data (see for example, \cite{2}), the last category is ruled out as it gives rise to vanishing reactor mixing angle. 

The textures which are not disallowed from the results of previous analysis can be categorised spontaneous breaking of flavour symmetries as follows.

\textbf{Allowed cases:} 
\begin{enumerate}
\item $\mu-\tau$ symmetry in $3\times3$ active neutrino block. Total number of such textures is 40.
\item One zero texture mass matrix. Total number of such textures is 96.
\item Two zero texture mass matrix. Total number of such textures is 64.
\item Three zero texture mass matrix. Total number of such textures is 8.
\item Hybrid texture mass matrix with no zeros but some constraints relating different elements. Total number of such textures is 296.
\end{enumerate}
Total number of such allowed mass matrices is 504. We further classify each of these allowed categories into different sub-categories based on the the constraints relating different elements of the light neutrino mass matrix. 

\subsection{Classification of Allowed Textures}

\subsubsection{$(\mu-\tau)$ symmetric textures}
The 40 $\mu-\tau$ symmetric textures can be classified into $4$ sub-categories depending upon the constraints that they satisfy. For representative purpose, we also mention one such VEV alignment and the corresponding mass matrix.

(i) 16 matrices with 5 complex constraints: 
\begin{center}
$M_{e\mu} = 0,\quad M_{e\tau}=0, \quad M_{\mu\mu}=M_{\tau\tau}, \quad M_{\mu\tau}=-M_{\tau\tau}, \quad M_{\mu s}=-M_{\tau s}$ 
\end{center}
$\bullet$ $n_1=1; n_{2 }=0;n_3=0; n_{4 }=1; n_5=0; n_6=-1; n_{7 }=0; n_8=1; n_9=-1; $\\

\noindent\(\left(
\begin{array}{cccc}
 -\frac{H^2 v y_1^2}{\Lambda ^2 \lambda _1} & 0 & 0 & -\frac{u \sqrt{\frac{H^2 v^2}{\Lambda ^2}} \rho  y_1}{v \lambda _1} \\
 0 & -\frac{H^2 v^2 \left(\frac{y_2^2}{v \lambda _2}+\frac{y_3^2}{v \lambda _3}\right)}{\Lambda ^2} & -\frac{H^2 v^2 \left(-\frac{y_2^2}{v \lambda
_2}-\frac{y_3^2}{v \lambda _3}\right)}{\Lambda ^2} & -\sqrt{\frac{H^2 v^2}{\Lambda ^2}} \left(\frac{y_2 y_4}{\lambda _2}-\frac{u y_3 y_5}{v \lambda
_3}\right) \\
 0 & -\frac{H^2 v^2 \left(-\frac{y_2^2}{v \lambda _2}-\frac{y_3^2}{v \lambda _3}\right)}{\Lambda ^2} & -\frac{H^2 v^2 \left(\frac{y_2^2}{v \lambda
_2}+\frac{y_3^2}{v \lambda _3}\right)}{\Lambda ^2} & -\sqrt{\frac{H^2 v^2}{\Lambda ^2}} \left(-\frac{y_2 y_4}{\lambda _2}+\frac{u y_3 y_5}{v \lambda
_3}\right) \\
 -\frac{u \sqrt{\frac{H^2 v^2}{\Lambda ^2}} \rho  y_1}{v \lambda _1} & -\sqrt{\frac{H^2 v^2}{\Lambda ^2}} \left(\frac{y_2 y_4}{\lambda _2}-\frac{u
y_3 y_5}{v \lambda _3}\right) & -\sqrt{\frac{H^2 v^2}{\Lambda ^2}} \left(-\frac{y_2 y_4}{\lambda _2}+\frac{u y_3 y_5}{v \lambda _3}\right) & -\frac{u^2
\rho ^2}{v \lambda _1}-\frac{v y_4^2}{\lambda _2}-\frac{u^2 y_5^2}{v \lambda _3} \\
\end{array}
\right)\) \\

\vspace{0.5cm}

(ii)  8 matrices with 3 complex constraints: 
\begin{center}
$M_{e\mu} = 0,\quad  M_{e\tau}=0, \quad M_{\mu\mu}=M_{\tau\tau}$ 
\end{center}

$\bullet$ $n_1=1; n_2=0;\text{  }n_3=0; n_4=1; n_5=0;\text{  }n_6=1; n_7=0; n_8=1;\text{  }n_9=-1;\text{  } $ \\

\noindent\(\left(
\begin{array}{cccc}
 -\frac{H^2 v y_1^2}{\Lambda ^2 \lambda _1} & 0 & 0 & -\frac{u \sqrt{\frac{H^2 v^2}{\Lambda ^2}} \rho  y_1}{v \lambda _1} \\
 0 & -\frac{H^2 v^2 \left(\frac{y_2^2}{v \lambda _2}+\frac{y_3^2}{v \lambda _3}\right)}{\Lambda ^2} & -\frac{H^2 v^2 \left(\frac{y_2^2}{v \lambda
_2}-\frac{y_3^2}{v \lambda _3}\right)}{\Lambda ^2} & -\sqrt{\frac{H^2 v^2}{\Lambda ^2}} \left(\frac{y_2 y_4}{\lambda _2}-\frac{u y_3 y_5}{v \lambda
_3}\right) \\
 0 & -\frac{H^2 v^2 \left(\frac{y_2^2}{v \lambda _2}-\frac{y_3^2}{v \lambda _3}\right)}{\Lambda ^2} & -\frac{H^2 v^2 \left(\frac{y_2^2}{v \lambda
_2}+\frac{y_3^2}{v \lambda _3}\right)}{\Lambda ^2} & -\sqrt{\frac{H^2 v^2}{\Lambda ^2}} \left(\frac{y_2 y_4}{\lambda _2}+\frac{u y_3 y_5}{v \lambda
_3}\right) \\
 -\frac{u \sqrt{\frac{H^2 v^2}{\Lambda ^2}} \rho  y_1}{v \lambda _1} & -\sqrt{\frac{H^2 v^2}{\Lambda ^2}} \left(\frac{y_2 y_4}{\lambda _2}-\frac{u
y_3 y_5}{v \lambda _3}\right) & -\sqrt{\frac{H^2 v^2}{\Lambda ^2}} \left(\frac{y_2 y_4}{\lambda _2}+\frac{u y_3 y_5}{v \lambda _3}\right) & -\frac{u^2
\rho ^2}{v \lambda _1}-\frac{v y_4^2}{\lambda _2}-\frac{u^2 y_5^2}{v \lambda _3} \\
\end{array}
\right)\) \\

\vspace{0.5cm}

(iii) 8 matrices with 3 complex constraints: 
\begin{center}
$M_{e\mu}=M_{e\tau},\quad  M_{\mu\mu}=M_{\tau\tau}, \quad M_{\tau\tau}+M_{\mu\tau} = 2 M_{e\tau}$ 
\end{center}

$\bullet$ $n_1=1; n_2=0;\text{  }n_3=0; n_4=1; n_5=0;\text{  }n_6=-1; n_7=1; n_8=1;\text{  }n_9=1;\text{  }$\\

\noindent\(\left(
\begin{array}{cccc}
 -\frac{H^2 v^2 \left(\frac{y_1^2}{v \lambda _1}+\frac{y_3^2}{v \lambda _3}\right)}{\Lambda ^2} & -\frac{H^2 v y_3^2}{\Lambda ^2 \lambda _3} & -\frac{H^2
v y_3^2}{\Lambda ^2 \lambda _3} & -\sqrt{\frac{H^2 v^2}{\Lambda ^2}} \left(\frac{u \rho  y_1}{v \lambda _1}+\frac{u y_3 y_5}{v \lambda _3}\right)
\\
 -\frac{H^2 v y_3^2}{\Lambda ^2 \lambda _3} & -\frac{H^2 v^2 \left(\frac{y_2^2}{v \lambda _2}+\frac{y_3^2}{v \lambda _3}\right)}{\Lambda ^2} & -\frac{H^2
v^2 \left(-\frac{y_2^2}{v \lambda _2}+\frac{y_3^2}{v \lambda _3}\right)}{\Lambda ^2} & -\sqrt{\frac{H^2 v^2}{\Lambda ^2}} \left(\frac{y_2 y_4}{\lambda
_2}+\frac{u y_3 y_5}{v \lambda _3}\right) \\
 -\frac{H^2 v y_3^2}{\Lambda ^2 \lambda _3} & -\frac{H^2 v^2 \left(-\frac{y_2^2}{v \lambda _2}+\frac{y_3^2}{v \lambda _3}\right)}{\Lambda ^2} & -\frac{H^2
v^2 \left(\frac{y_2^2}{v \lambda _2}+\frac{y_3^2}{v \lambda _3}\right)}{\Lambda ^2} & -\sqrt{\frac{H^2 v^2}{\Lambda ^2}} \left(-\frac{y_2 y_4}{\lambda
_2}+\frac{u y_3 y_5}{v \lambda _3}\right) \\
 -\sqrt{\frac{H^2 v^2}{\Lambda ^2}} \left(\frac{u \rho  y_1}{v \lambda _1}+\frac{u y_3 y_5}{v \lambda _3}\right) & -\sqrt{\frac{H^2 v^2}{\Lambda
^2}} \left(\frac{y_2 y_4}{\lambda _2}+\frac{u y_3 y_5}{v \lambda _3}\right) & -\sqrt{\frac{H^2 v^2}{\Lambda ^2}} \left(-\frac{y_2 y_4}{\lambda _2}+\frac{u
y_3 y_5}{v \lambda _3}\right) & -\frac{u^2 \rho ^2}{v \lambda _1}-\frac{v y_4^2}{\lambda _2}-\frac{u^2 y_5^2}{v \lambda _3} \\
\end{array}
\right)\) \\

\vspace{0.5cm}

(iv) 8 matrices with 3 complex constraints:
\begin{center}
$M_{e\mu}=M_{e\tau},\quad  M_{\mu\mu}=M_{\tau\tau}, \quad M_{\tau\tau}+M_{\mu\tau} = -2 M_{e\tau}$
\end{center}

$\bullet$ $n_1=1; n_2=0;\text{  }n_3=0; n_4=1; n_5=0;\text{  }n_6=-1; n_7=1; n_8=-1;\text{  }n_9=-1;\text{  }$\\

\noindent\(\left(
\begin{array}{cccc}
 -\frac{H^2 v^2 \left(\frac{y_1^2}{v \lambda _1}+\frac{y_3^2}{v \lambda _3}\right)}{\Lambda ^2} & \frac{H^2 v y_3^2}{\Lambda ^2 \lambda _3} & \frac{H^2
v y_3^2}{\Lambda ^2 \lambda _3} & -\sqrt{\frac{H^2 v^2}{\Lambda ^2}} \left(\frac{u \rho  y_1}{v \lambda _1}+\frac{u y_3 y_5}{v \lambda _3}\right)
\\
 \frac{H^2 v y_3^2}{\Lambda ^2 \lambda _3} & -\frac{H^2 v^2 \left(\frac{y_2^2}{v \lambda _2}+\frac{y_3^2}{v \lambda _3}\right)}{\Lambda ^2} & -\frac{H^2
v^2 \left(-\frac{y_2^2}{v \lambda _2}+\frac{y_3^2}{v \lambda _3}\right)}{\Lambda ^2} & -\sqrt{\frac{H^2 v^2}{\Lambda ^2}} \left(\frac{y_2 y_4}{\lambda
_2}-\frac{u y_3 y_5}{v \lambda _3}\right) \\
 \frac{H^2 v y_3^2}{\Lambda ^2 \lambda _3} & -\frac{H^2 v^2 \left(-\frac{y_2^2}{v \lambda _2}+\frac{y_3^2}{v \lambda _3}\right)}{\Lambda ^2} & -\frac{H^2
v^2 \left(\frac{y_2^2}{v \lambda _2}+\frac{y_3^2}{v \lambda _3}\right)}{\Lambda ^2} & -\sqrt{\frac{H^2 v^2}{\Lambda ^2}} \left(-\frac{y_2 y_4}{\lambda
_2}-\frac{u y_3 y_5}{v \lambda _3}\right) \\
 -\sqrt{\frac{H^2 v^2}{\Lambda ^2}} \left(\frac{u \rho  y_1}{v \lambda _1}+\frac{u y_3 y_5}{v \lambda _3}\right) & -\sqrt{\frac{H^2 v^2}{\Lambda
^2}} \left(\frac{y_2 y_4}{\lambda _2}-\frac{u y_3 y_5}{v \lambda _3}\right) & -\sqrt{\frac{H^2 v^2}{\Lambda ^2}} \left(-\frac{y_2 y_4}{\lambda _2}-\frac{u
y_3 y_5}{v \lambda _3}\right) & -\frac{u^2 \rho ^2}{v \lambda _1}-\frac{v y_4^2}{\lambda _2}-\frac{u^2 y_5^2}{v \lambda _3} \\
\end{array}
\right)\)

\subsubsection{Texture 1 zero case}

All 96  texture 1 zero cases can be classified into 12 categories depending upon constraints satisfied by them. For representative purpose, we also mention one such VEV alignment and the corresponding mass matrix.

(i)  8 matrices with 3 complex constraints:
\begin{center}
$M_{e\mu}= 0,\quad  M_{\mu\mu}= -M_{\mu\tau},\quad M_{e\tau}+M_{\mu\tau}=-M_{\tau\tau}$
\end{center}
$\bullet$ $ n_1=1; n_{2 }=0;n_3=0; n_{4 }=0; n_5=1; n_6=-1; n_{7 }=0; n_8=1; n_9=-1; $ \\

\vspace{0.1cm}

\noindent\(\left(
\begin{array}{cccc}
 -\frac{H^2 v^2 \left(\frac{y_1^2}{v \lambda _1}+\frac{y_2^2}{v \lambda _2}\right)}{\Lambda ^2} & 0 & \frac{H^2 v y_2^2}{\Lambda ^2 \lambda _2} &
-\sqrt{\frac{H^2 v^2}{\Lambda ^2}} \left(\frac{u \rho  y_1}{v \lambda _1}+\frac{y_2 y_4}{\lambda _2}\right) \\
 0 & -\frac{H^2 v y_3^2}{\Lambda ^2 \lambda _3} & \frac{H^2 v y_3^2}{\Lambda ^2 \lambda _3} & \frac{u \sqrt{\frac{H^2 v^2}{\Lambda ^2}} y_3 y_5}{v
\lambda _3} \\
 \frac{H^2 v y_2^2}{\Lambda ^2 \lambda _2} & \frac{H^2 v y_3^2}{\Lambda ^2 \lambda _3} & -\frac{H^2 v^2 \left(\frac{y_2^2}{v \lambda _2}+\frac{y_3^2}{v
\lambda _3}\right)}{\Lambda ^2} & -\sqrt{\frac{H^2 v^2}{\Lambda ^2}} \left(-\frac{y_2 y_4}{\lambda _2}+\frac{u y_3 y_5}{v \lambda _3}\right) \\
 -\sqrt{\frac{H^2 v^2}{\Lambda ^2}} \left(\frac{u \rho  y_1}{v \lambda _1}+\frac{y_2 y_4}{\lambda _2}\right) & \frac{u \sqrt{\frac{H^2 v^2}{\Lambda
^2}} y_3 y_5}{v \lambda _3} & -\sqrt{\frac{H^2 v^2}{\Lambda ^2}} \left(-\frac{y_2 y_4}{\lambda _2}+\frac{u y_3 y_5}{v \lambda _3}\right) & -\frac{u^2
\rho ^2}{v \lambda _1}-\frac{v y_4^2}{\lambda _2}-\frac{u^2 y_5^2}{v \lambda _3} \\
\end{array}
\right)\)
\vspace{0.2cm}

(ii) 8 matrices with 3 complex constraints: 
 \begin{center}
$M_{e\tau}= 0,\quad  M_{\tau\tau}= -M_{\mu\tau},\quad M_{e\mu}+M_{\mu\tau}=-M_{\mu\mu}$
\end{center}

$\bullet$ $ n_1=1; n_{2 }=0;_3=0; n_{4 }=1; n_5=0; n_6=-1; n_{7 }=1; n_8=0; n_9=-1;$ \\

\vspace{0.2cm}

\noindent\(\left(
\begin{array}{cccc}
 -\frac{H^2 v^2 \left(\frac{y_1^2}{v \lambda _1}+\frac{y_3^2}{v \lambda _3}\right)}{\Lambda ^2} & \frac{H^2 v y_3^2}{\Lambda ^2 \lambda _3} & 0 &
-\sqrt{\frac{H^2 v^2}{\Lambda ^2}} \left(\frac{u \rho  y_1}{v \lambda _1}+\frac{u y_3 y_5}{v \lambda _3}\right) \\
 \frac{H^2 v y_3^2}{\Lambda ^2 \lambda _3} & -\frac{H^2 v^2 \left(\frac{y_2^2}{v \lambda _2}+\frac{y_3^2}{v \lambda _3}\right)}{\Lambda ^2} & \frac{H^2
v y_2^2}{\Lambda ^2 \lambda _2} & -\sqrt{\frac{H^2 v^2}{\Lambda ^2}} \left(\frac{y_2 y_4}{\lambda _2}-\frac{u y_3 y_5}{v \lambda _3}\right) \\
 0 & \frac{H^2 v y_2^2}{\Lambda ^2 \lambda _2} & -\frac{H^2 v y_2^2}{\Lambda ^2 \lambda _2} & \frac{\sqrt{\frac{H^2 v^2}{\Lambda ^2}} y_2 y_4}{\lambda
_2} \\
 -\sqrt{\frac{H^2 v^2}{\Lambda ^2}} \left(\frac{u \rho  y_1}{v \lambda _1}+\frac{u y_3 y_5}{v \lambda _3}\right) & -\sqrt{\frac{H^2 v^2}{\Lambda
^2}} \left(\frac{y_2 y_4}{\lambda _2}-\frac{u y_3 y_5}{v \lambda _3}\right) & \frac{\sqrt{\frac{H^2 v^2}{\Lambda ^2}} y_2 y_4}{\lambda _2} & -\frac{u^2
\rho ^2}{v \lambda _1}-\frac{v y_4^2}{\lambda _2}-\frac{u^2 y_5^2}{v \lambda _3} \\
\end{array}
\right)\)\\

\vspace{0.2cm}

(iii)  8 matrices with 3 complex constraints: 
\begin{center}
$M_{e\mu}= 0,\quad  M_{\mu\mu}= M_{\mu\tau},\quad M_{e\tau}+M_{\mu\tau}=M_{\tau\tau}$
\end{center}

$\bullet$ $ n_1=1; n_2=0;\text{  }n_3=0; n_4=0; n_5=1;\text{  }n_6=1; n_7=0; n_8=1;\text{  }n_9=1; $\\

\noindent\(\left(
\begin{array}{cccc}
 -\frac{H^2 v^2 \left(\frac{y_1^2}{v \lambda _1}+\frac{y_2^2}{v \lambda _2}\right)}{\Lambda ^2} & 0 & -\frac{H^2 v y_2^2}{\Lambda ^2 \lambda _2}
& -\sqrt{\frac{H^2 v^2}{\Lambda ^2}} \left(\frac{u \rho  y_1}{v \lambda _1}+\frac{y_2 y_4}{\lambda _2}\right) \\
 0 & -\frac{H^2 v y_3^2}{\Lambda ^2 \lambda _3} & -\frac{H^2 v y_3^2}{\Lambda ^2 \lambda _3} & -\frac{u \sqrt{\frac{H^2 v^2}{\Lambda ^2}} y_3 y_5}{v
\lambda _3} \\
 -\frac{H^2 v y_2^2}{\Lambda ^2 \lambda _2} & -\frac{H^2 v y_3^2}{\Lambda ^2 \lambda _3} & -\frac{H^2 v^2 \left(\frac{y_2^2}{v \lambda _2}+\frac{y_3^2}{v
\lambda _3}\right)}{\Lambda ^2} & -\sqrt{\frac{H^2 v^2}{\Lambda ^2}} \left(\frac{y_2 y_4}{\lambda _2}+\frac{u y_3 y_5}{v \lambda _3}\right) \\
 -\sqrt{\frac{H^2 v^2}{\Lambda ^2}} \left(\frac{u \rho  y_1}{v \lambda _1}+\frac{y_2 y_4}{\lambda _2}\right) & -\frac{u \sqrt{\frac{H^2 v^2}{\Lambda
^2}} y_3 y_5}{v \lambda _3} & -\sqrt{\frac{H^2 v^2}{\Lambda ^2}} \left(\frac{y_2 y_4}{\lambda _2}+\frac{u y_3 y_5}{v \lambda _3}\right) & -\frac{u^2
\rho ^2}{v \lambda _1}-\frac{v y_4^2}{\lambda _2}-\frac{u^2 y_5^2}{v \lambda _3} \\
\end{array}
\right)\)\\

\vspace{0.2cm}

(iv)  8 matrices with 3 complex constraints: 
\begin{center}
$M_{e\mu}= 0,\quad  M_{\mu\mu}=- M_{\mu\tau},\quad M_{e\tau}-M_{\mu\tau}=M_{\tau\tau}$
\end{center}

$\bullet$ $ n_1=1; n_2=0;\text{  }n_3=0; n_4=0; n_5=1;\text{  }n_6=1; n_7=0; n_8=1;\text{  }n_9=-1;$ \\

\vspace{0.2cm}

\noindent\(\left(
\begin{array}{cccc}
 -\frac{H^2 v^2 \left(\frac{y_1^2}{v \lambda _1}+\frac{y_2^2}{v \lambda _2}\right)}{\Lambda ^2} & 0 & -\frac{H^2 v y_2^2}{\Lambda ^2 \lambda _2}
& -\sqrt{\frac{H^2 v^2}{\Lambda ^2}} \left(\frac{u \rho  y_1}{v \lambda _1}+\frac{y_2 y_4}{\lambda _2}\right) \\
 0 & -\frac{H^2 v y_3^2}{\Lambda ^2 \lambda _3} & \frac{H^2 v y_3^2}{\Lambda ^2 \lambda _3} & \frac{u \sqrt{\frac{H^2 v^2}{\Lambda ^2}} y_3 y_5}{v
\lambda _3} \\
 -\frac{H^2 v y_2^2}{\Lambda ^2 \lambda _2} & \frac{H^2 v y_3^2}{\Lambda ^2 \lambda _3} & -\frac{H^2 v^2 \left(\frac{y_2^2}{v \lambda _2}+\frac{y_3^2}{v
\lambda _3}\right)}{\Lambda ^2} & -\sqrt{\frac{H^2 v^2}{\Lambda ^2}} \left(\frac{y_2 y_4}{\lambda _2}+\frac{u y_3 y_5}{v \lambda _3}\right) \\
 -\sqrt{\frac{H^2 v^2}{\Lambda ^2}} \left(\frac{u \rho  y_1}{v \lambda _1}+\frac{y_2 y_4}{\lambda _2}\right) & \frac{u \sqrt{\frac{H^2 v^2}{\Lambda
^2}} y_3 y_5}{v \lambda _3} & -\sqrt{\frac{H^2 v^2}{\Lambda ^2}} \left(\frac{y_2 y_4}{\lambda _2}+\frac{u y_3 y_5}{v \lambda _3}\right) & -\frac{u^2
\rho ^2}{v \lambda _1}-\frac{v y_4^2}{\lambda _2}-\frac{u^2 y_5^2}{v \lambda _3} \\
\end{array}
\right)\)\\

\vspace{0.2cm}

(v) 8 matrices with 3 complex constraints: 
\begin{center}
$M_{e\mu}= 0,\quad  M_{\mu\mu}= M_{\mu\tau},\quad M_{e\tau}-M_{\mu\tau}=-M_{\tau\tau}$
\end{center}

$\bullet$ $ n_1=1; n_2=0;\text{  }n_3=0; n_4=0; n_5=1;\text{  }n_6=-1; n_7=0; n_8=1;\text{  }n_9=1;$ \\

\vspace{0.2cm}

\noindent\(\left(
\begin{array}{cccc}
 -\frac{H^2 v^2 \left(\frac{y_1^2}{v \lambda _1}+\frac{y_2^2}{v \lambda _2}\right)}{\Lambda ^2} & 0 & \frac{H^2 v y_2^2}{\Lambda ^2 \lambda _2} &
-\sqrt{\frac{H^2 v^2}{\Lambda ^2}} \left(\frac{u \rho  y_1}{v \lambda _1}+\frac{y_2 y_4}{\lambda _2}\right) \\
 0 & -\frac{H^2 v y_3^2}{\Lambda ^2 \lambda _3} & -\frac{H^2 v y_3^2}{\Lambda ^2 \lambda _3} & -\frac{u \sqrt{\frac{H^2 v^2}{\Lambda ^2}} y_3 y_5}{v
\lambda _3} \\
 \frac{H^2 v y_2^2}{\Lambda ^2 \lambda _2} & -\frac{H^2 v y_3^2}{\Lambda ^2 \lambda _3} & -\frac{H^2 v^2 \left(\frac{y_2^2}{v \lambda _2}+\frac{y_3^2}{v
\lambda _3}\right)}{\Lambda ^2} & -\sqrt{\frac{H^2 v^2}{\Lambda ^2}} \left(-\frac{y_2 y_4}{\lambda _2}+\frac{u y_3 y_5}{v \lambda _3}\right) \\
 -\sqrt{\frac{H^2 v^2}{\Lambda ^2}} \left(\frac{u \rho  y_1}{v \lambda _1}+\frac{y_2 y_4}{\lambda _2}\right) & -\frac{u \sqrt{\frac{H^2 v^2}{\Lambda
^2}} y_3 y_5}{v \lambda _3} & -\sqrt{\frac{H^2 v^2}{\Lambda ^2}} \left(-\frac{y_2 y_4}{\lambda _2}+\frac{u y_3 y_5}{v \lambda _3}\right) & -\frac{u^2
\rho ^2}{v \lambda _1}-\frac{v y_4^2}{\lambda _2}-\frac{u^2 y_5^2}{v \lambda _3} \\
\end{array}
\right)\)\\

\vspace{0.2cm}

(vi)  8 matrices with 3 complex constraints: 
\begin{center}
$M_{e\tau}= 0,\quad  M_{\tau\tau}= M_{\mu\tau},\quad M_{e\mu}+ M_{\mu\tau}=M_{\mu\mu}$
\end{center}

\vspace{0.2cm}

$\bullet$ $ n_1=1; n_2=0;\text{  }n_3=0; n_4=1; n_5=0;\text{  }n_6=1; n_7=1; n_8=0;\text{  }n_9=1;\text{  }$ \\

\noindent\(\left(
\begin{array}{cccc}
 -\frac{H^2 v^2 \left(\frac{y_1^2}{v \lambda _1}+\frac{y_3^2}{v \lambda _3}\right)}{\Lambda ^2} & -\frac{H^2 v y_3^2}{\Lambda ^2 \lambda _3} & 0
& -\sqrt{\frac{H^2 v^2}{\Lambda ^2}} \left(\frac{u \rho  y_1}{v \lambda _1}+\frac{u y_3 y_5}{v \lambda _3}\right) \\
 -\frac{H^2 v y_3^2}{\Lambda ^2 \lambda _3} & -\frac{H^2 v^2 \left(\frac{y_2^2}{v \lambda _2}+\frac{y_3^2}{v \lambda _3}\right)}{\Lambda ^2} & -\frac{H^2
v y_2^2}{\Lambda ^2 \lambda _2} & -\sqrt{\frac{H^2 v^2}{\Lambda ^2}} \left(\frac{y_2 y_4}{\lambda _2}+\frac{u y_3 y_5}{v \lambda _3}\right) \\
 0 & -\frac{H^2 v y_2^2}{\Lambda ^2 \lambda _2} & -\frac{H^2 v y_2^2}{\Lambda ^2 \lambda _2} & -\frac{\sqrt{\frac{H^2 v^2}{\Lambda ^2}} y_2 y_4}{\lambda
_2} \\
 -\sqrt{\frac{H^2 v^2}{\Lambda ^2}} \left(\frac{u \rho  y_1}{v \lambda _1}+\frac{u y_3 y_5}{v \lambda _3}\right) & -\sqrt{\frac{H^2 v^2}{\Lambda
^2}} \left(\frac{y_2 y_4}{\lambda _2}+\frac{u y_3 y_5}{v \lambda _3}\right) & -\frac{\sqrt{\frac{H^2 v^2}{\Lambda ^2}} y_2 y_4}{\lambda _2} & -\frac{u^2
\rho ^2}{v \lambda _1}-\frac{v y_4^2}{\lambda _2}-\frac{u^2 y_5^2}{v \lambda _3} \\
\end{array}
\right)\)\\

\vspace{0.2cm}

(vii) 7 matrices with 3 complex constraints:
\begin{center}
$M_{e\tau}= 0,\quad  M_{\tau\tau}= M_{\mu\tau},\quad M_{e\mu}- M_{\mu\tau}=-M_{\mu\mu}$
\end{center}

$\bullet$ $ n_1=1; n_2=0;\text{  }n_3=0; n_4=1; n_5=0;\text{  }n_6=1; n_7=1; n_8=0;\text{  }n_9=-1;\text{  }$ \\

\noindent\(\left(
\begin{array}{cccc}
 -\frac{H^2 v^2 \left(\frac{y_1^2}{v \lambda _1}+\frac{y_3^2}{v \lambda _3}\right)}{\Lambda ^2} & \frac{H^2 v y_3^2}{\Lambda ^2 \lambda _3} & 0 &
-\sqrt{\frac{H^2 v^2}{\Lambda ^2}} \left(\frac{u \rho  y_1}{v \lambda _1}+\frac{u y_3 y_5}{v \lambda _3}\right) \\
 \frac{H^2 v y_3^2}{\Lambda ^2 \lambda _3} & -\frac{H^2 v^2 \left(\frac{y_2^2}{v \lambda _2}+\frac{y_3^2}{v \lambda _3}\right)}{\Lambda ^2} & -\frac{H^2
v y_2^2}{\Lambda ^2 \lambda _2} & -\sqrt{\frac{H^2 v^2}{\Lambda ^2}} \left(\frac{y_2 y_4}{\lambda _2}-\frac{u y_3 y_5}{v \lambda _3}\right) \\
 0 & -\frac{H^2 v y_2^2}{\Lambda ^2 \lambda _2} & -\frac{H^2 v y_2^2}{\Lambda ^2 \lambda _2} & -\frac{\sqrt{\frac{H^2 v^2}{\Lambda ^2}} y_2 y_4}{\lambda
_2} \\
 -\sqrt{\frac{H^2 v^2}{\Lambda ^2}} \left(\frac{u \rho  y_1}{v \lambda _1}+\frac{u y_3 y_5}{v \lambda _3}\right) & -\sqrt{\frac{H^2 v^2}{\Lambda
^2}} \left(\frac{y_2 y_4}{\lambda _2}-\frac{u y_3 y_5}{v \lambda _3}\right) & -\frac{\sqrt{\frac{H^2 v^2}{\Lambda ^2}} y_2 y_4}{\lambda _2} & -\frac{u^2
\rho ^2}{v \lambda _1}-\frac{v y_4^2}{\lambda _2}-\frac{u^2 y_5^2}{v \lambda _3} \\
\end{array}
\right)\)\\

\vspace{0.2cm}

(viii) 8 matrices with 3 complex constraints: 
\begin{center}
$M_{e\tau}= 0,\quad  M_{\tau\tau}=- M_{\mu\tau},\quad M_{e\mu}- M_{\mu\tau}=M_{\mu\mu}$
\end{center}

\vspace{0.2cm}

$\bullet$ $ n_1=1; n_2=0;\text{  }n_3=0; n_4=1; n_5=0;\text{  }n_6=-1; n_7=1; n_8=0;\text{  }n_9=1;\text{  }$ \\

\noindent\(\left(
\begin{array}{cccc}
 -\frac{H^2 v^2 \left(\frac{y_1^2}{v \lambda _1}+\frac{y_3^2}{v \lambda _3}\right)}{\Lambda ^2} & -\frac{H^2 v y_3^2}{\Lambda ^2 \lambda _3} & 0
& -\sqrt{\frac{H^2 v^2}{\Lambda ^2}} \left(\frac{u \rho  y_1}{v \lambda _1}+\frac{u y_3 y_5}{v \lambda _3}\right) \\
 -\frac{H^2 v y_3^2}{\Lambda ^2 \lambda _3} & -\frac{H^2 v^2 \left(\frac{y_2^2}{v \lambda _2}+\frac{y_3^2}{v \lambda _3}\right)}{\Lambda ^2} & \frac{H^2
v y_2^2}{\Lambda ^2 \lambda _2} & -\sqrt{\frac{H^2 v^2}{\Lambda ^2}} \left(\frac{y_2 y_4}{\lambda _2}+\frac{u y_3 y_5}{v \lambda _3}\right) \\
 0 & \frac{H^2 v y_2^2}{\Lambda ^2 \lambda _2} & -\frac{H^2 v y_2^2}{\Lambda ^2 \lambda _2} & \frac{\sqrt{\frac{H^2 v^2}{\Lambda ^2}} y_2 y_4}{\lambda
_2} \\
 -\sqrt{\frac{H^2 v^2}{\Lambda ^2}} \left(\frac{u \rho  y_1}{v \lambda _1}+\frac{u y_3 y_5}{v \lambda _3}\right) & -\sqrt{\frac{H^2 v^2}{\Lambda
^2}} \left(\frac{y_2 y_4}{\lambda _2}+\frac{u y_3 y_5}{v \lambda _3}\right) & \frac{\sqrt{\frac{H^2 v^2}{\Lambda ^2}} y_2 y_4}{\lambda _2} & -\frac{u^2
\rho ^2}{v \lambda _1}-\frac{v y_4^2}{\lambda _2}-\frac{u^2 y_5^2}{v \lambda _3} \\
\end{array}
\right)\)\\

\vspace{0.2cm}

(ix)  8 matrices with 3 complex constraints: 
\begin{center}
$M_{\mu\tau}= 0,\quad  M_{e\mu}= -M_{\mu\mu},\quad M_{e\tau}=-M_{\tau\tau}$
\end{center}

\vspace{0.2cm}

$\bullet$ $  n_1=1; n_{2 }=0;n_3=0; n_{4 }=1; n_5=-1; n_6=0; n_{7 }=1; n_8=-1; n_9=0;$ \\

\noindent\(\left(
\begin{array}{cccc}
 -\frac{H^2 v^2 \left(\frac{y_1^2}{v \lambda _1}+\frac{y_2^2}{v \lambda _2}+\frac{y_3^2}{v \lambda _3}\right)}{\Lambda ^2} & \frac{H^2 v y_2^2}{\Lambda
^2 \lambda _2} & \frac{H^2 v y_3^2}{\Lambda ^2 \lambda _3} & -\sqrt{\frac{H^2 v^2}{\Lambda ^2}} \left(\frac{u \rho  y_1}{v \lambda _1}-\frac{y_2
y_4}{\lambda _2}+\frac{u y_3 y_5}{v \lambda _3}\right) \\
 \frac{H^2 v y_2^2}{\Lambda ^2 \lambda _2} & -\frac{H^2 v y_2^2}{\Lambda ^2 \lambda _2} & 0 & -\frac{\sqrt{\frac{H^2 v^2}{\Lambda ^2}} y_2 y_4}{\lambda
_2} \\
 \frac{H^2 v y_3^2}{\Lambda ^2 \lambda _3} & 0 & -\frac{H^2 v y_3^2}{\Lambda ^2 \lambda _3} & \frac{u \sqrt{\frac{H^2 v^2}{\Lambda ^2}} y_3 y_5}{v
\lambda _3} \\
 -\sqrt{\frac{H^2 v^2}{\Lambda ^2}} \left(\frac{u \rho  y_1}{v \lambda _1}-\frac{y_2 y_4}{\lambda _2}+\frac{u y_3 y_5}{v \lambda _3}\right) & -\frac{\sqrt{\frac{H^2
v^2}{\Lambda ^2}} y_2 y_4}{\lambda _2} & \frac{u \sqrt{\frac{H^2 v^2}{\Lambda ^2}} y_3 y_5}{v \lambda _3} & -\frac{u^2 \rho ^2}{v \lambda _1}-\frac{v
y_4^2}{\lambda _2}-\frac{u^2 y_5^2}{v \lambda _3} \\
\end{array}
\right)\)\\

\vspace{0.2cm}

(x)  8 matrices with 3 complex constraints: 
\begin{center}
$M_{\mu\tau}= 0,\quad  M_{e\mu}= M_{\mu\mu},\quad M_{e\tau}=M_{\tau\tau}$
\end{center}

\vspace{0.2cm}

$\bullet$ $ n_1=1; n_2=0;\text{  }n_3=0; n_4=0; n_5=1;\text{  }n_6=1; n_7=1; n_8=0;\text{  }n_9=1;$ \\

\noindent\(\left(
\begin{array}{cccc}
 -\frac{H^2 v^2 \left(\frac{y_1^2}{v \lambda _1}+\frac{y_2^2}{v \lambda _2}+\frac{y_3^2}{v \lambda _3}\right)}{\Lambda ^2} & -\frac{H^2 v y_3^2}{\Lambda
^2 \lambda _3} & -\frac{H^2 v y_2^2}{\Lambda ^2 \lambda _2} & -\sqrt{\frac{H^2 v^2}{\Lambda ^2}} \left(\frac{u \rho  y_1}{v \lambda _1}+\frac{y_2
y_4}{\lambda _2}+\frac{u y_3 y_5}{v \lambda _3}\right) \\
 -\frac{H^2 v y_3^2}{\Lambda ^2 \lambda _3} & -\frac{H^2 v y_3^2}{\Lambda ^2 \lambda _3} & 0 & -\frac{u \sqrt{\frac{H^2 v^2}{\Lambda ^2}} y_3 y_5}{v
\lambda _3} \\
 -\frac{H^2 v y_2^2}{\Lambda ^2 \lambda _2} & 0 & -\frac{H^2 v y_2^2}{\Lambda ^2 \lambda _2} & -\frac{\sqrt{\frac{H^2 v^2}{\Lambda ^2}} y_2 y_4}{\lambda
_2} \\
 -\sqrt{\frac{H^2 v^2}{\Lambda ^2}} \left(\frac{u \rho  y_1}{v \lambda _1}+\frac{y_2 y_4}{\lambda _2}+\frac{u y_3 y_5}{v \lambda _3}\right) & -\frac{u
\sqrt{\frac{H^2 v^2}{\Lambda ^2}} y_3 y_5}{v \lambda _3} & -\frac{\sqrt{\frac{H^2 v^2}{\Lambda ^2}} y_2 y_4}{\lambda _2} & -\frac{u^2 \rho ^2}{v
\lambda _1}-\frac{v y_4^2}{\lambda _2}-\frac{u^2 y_5^2}{v \lambda _3} \\
\end{array}
\right)\)\\

\vspace{0.2cm}

(xi)  9 matrices with 3 complex constraints:
\begin{center}
$M_{\mu\tau}= 0,\quad  M_{e\mu}= -M_{\mu\mu},\quad M_{e\tau}=M_{\tau\tau}$
\end{center}

\vspace{0.2cm}

$\bullet$ $ n_1=1; n_2=0;\text{  }n_3=0; n_4=0; n_5=1;\text{  }n_6=1; n_7=1; n_8=0;\text{  }n_9=-1;$ \\

\noindent\(\left(
\begin{array}{cccc}
 -\frac{H^2 v^2 \left(\frac{y_1^2}{v \lambda _1}+\frac{y_2^2}{v \lambda _2}+\frac{y_3^2}{v \lambda _3}\right)}{\Lambda ^2} & \frac{H^2 v y_3^2}{\Lambda
^2 \lambda _3} & -\frac{H^2 v y_2^2}{\Lambda ^2 \lambda _2} & -\sqrt{\frac{H^2 v^2}{\Lambda ^2}} \left(\frac{u \rho  y_1}{v \lambda _1}+\frac{y_2
y_4}{\lambda _2}+\frac{u y_3 y_5}{v \lambda _3}\right) \\
 \frac{H^2 v y_3^2}{\Lambda ^2 \lambda _3} & -\frac{H^2 v y_3^2}{\Lambda ^2 \lambda _3} & 0 & \frac{u \sqrt{\frac{H^2 v^2}{\Lambda ^2}} y_3 y_5}{v
\lambda _3} \\
 -\frac{H^2 v y_2^2}{\Lambda ^2 \lambda _2} & 0 & -\frac{H^2 v y_2^2}{\Lambda ^2 \lambda _2} & -\frac{\sqrt{\frac{H^2 v^2}{\Lambda ^2}} y_2 y_4}{\lambda
_2} \\
 -\sqrt{\frac{H^2 v^2}{\Lambda ^2}} \left(\frac{u \rho  y_1}{v \lambda _1}+\frac{y_2 y_4}{\lambda _2}+\frac{u y_3 y_5}{v \lambda _3}\right) & \frac{u
\sqrt{\frac{H^2 v^2}{\Lambda ^2}} y_3 y_5}{v \lambda _3} & -\frac{\sqrt{\frac{H^2 v^2}{\Lambda ^2}} y_2 y_4}{\lambda _2} & -\frac{u^2 \rho ^2}{v
\lambda _1}-\frac{v y_4^2}{\lambda _2}-\frac{u^2 y_5^2}{v \lambda _3} \\
\end{array}
\right)\)\\

\vspace{0.2cm}

(xii)  8 matrices with 3 complex constraints: 
\begin{center}
$M_{\mu\tau}= 0,\quad  M_{e\mu}= M_{\mu\mu},\quad M_{e\tau}=-M_{\tau\tau}$
\end{center}

\vspace{0.2cm}

$\bullet$ $ n_1=1; n_2=0;\text{  }n_3=0; n_4=0; n_5=1;\text{  }n_6=-1; n_7=1; n_8=0;\text{  }n_9=1; $\\

\noindent\(\left(
\begin{array}{cccc}
 -\frac{H^2 v^2 \left(\frac{y_1^2}{v \lambda _1}+\frac{y_2^2}{v \lambda _2}+\frac{y_3^2}{v \lambda _3}\right)}{\Lambda ^2} & -\frac{H^2 v y_3^2}{\Lambda
^2 \lambda _3} & \frac{H^2 v y_2^2}{\Lambda ^2 \lambda _2} & -\sqrt{\frac{H^2 v^2}{\Lambda ^2}} \left(\frac{u \rho  y_1}{v \lambda _1}+\frac{y_2
y_4}{\lambda _2}+\frac{u y_3 y_5}{v \lambda _3}\right) \\
 -\frac{H^2 v y_3^2}{\Lambda ^2 \lambda _3} & -\frac{H^2 v y_3^2}{\Lambda ^2 \lambda _3} & 0 & -\frac{u \sqrt{\frac{H^2 v^2}{\Lambda ^2}} y_3 y_5}{v
\lambda _3} \\
 \frac{H^2 v y_2^2}{\Lambda ^2 \lambda _2} & 0 & -\frac{H^2 v y_2^2}{\Lambda ^2 \lambda _2} & \frac{\sqrt{\frac{H^2 v^2}{\Lambda ^2}} y_2 y_4}{\lambda
_2} \\
 -\sqrt{\frac{H^2 v^2}{\Lambda ^2}} \left(\frac{u \rho  y_1}{v \lambda _1}+\frac{y_2 y_4}{\lambda _2}+\frac{u y_3 y_5}{v \lambda _3}\right) & -\frac{u
\sqrt{\frac{H^2 v^2}{\Lambda ^2}} y_3 y_5}{v \lambda _3} & \frac{\sqrt{\frac{H^2 v^2}{\Lambda ^2}} y_2 y_4}{\lambda _2} & -\frac{u^2 \rho ^2}{v \lambda
_1}-\frac{v y_4^2}{\lambda _2}-\frac{u^2 y_5^2}{v \lambda _3} \\
\end{array}
\right)\)

\subsubsection{Texture 2 zero case}
All 64 texture 2 zero cases can be classified into following categories. For representative purpose, we also mention one such VEV alignment and the corresponding mass matrix.

(i)  8 matrices with 3 complex constraints: 
\begin{center}
$M_{e\mu}= 0,\quad  M_{e\tau}= 0,\quad M_{\mu\mu}=M_{\mu\tau}$
\end{center}

$\bullet$ $ n_1=1; n_2=0;\text{  }n_3=0; n_4=0; n_5=0;\text{  }n_6=1; n_7=0; n_8=1;\text{  }n_9=1;$ \\

\noindent\(\left(
\begin{array}{cccc}
 -\frac{H^2 v y_1^2}{\Lambda ^2 \lambda _1} & 0 & 0 & -\frac{u \sqrt{\frac{H^2 v^2}{\Lambda ^2}} \rho  y_1}{v \lambda _1} \\
 0 & -\frac{H^2 v y_3^2}{\Lambda ^2 \lambda _3} & -\frac{H^2 v y_3^2}{\Lambda ^2 \lambda _3} & -\frac{u \sqrt{\frac{H^2 v^2}{\Lambda ^2}} y_3 y_5}{v
\lambda _3} \\
 0 & -\frac{H^2 v y_3^2}{\Lambda ^2 \lambda _3} & -\frac{H^2 v^2 \left(\frac{y_2^2}{v \lambda _2}+\frac{y_3^2}{v \lambda _3}\right)}{\Lambda ^2}
& -\sqrt{\frac{H^2 v^2}{\Lambda ^2}} \left(\frac{y_2 y_4}{\lambda _2}+\frac{u y_3 y_5}{v \lambda _3}\right) \\
 -\frac{u \sqrt{\frac{H^2 v^2}{\Lambda ^2}} \rho  y_1}{v \lambda _1} & -\frac{u \sqrt{\frac{H^2 v^2}{\Lambda ^2}} y_3 y_5}{v \lambda _3} & -\sqrt{\frac{H^2
v^2}{\Lambda ^2}} \left(\frac{y_2 y_4}{\lambda _2}+\frac{u y_3 y_5}{v \lambda _3}\right) & -\frac{u^2 \rho ^2}{v \lambda _1}-\frac{v y_4^2}{\lambda
_2}-\frac{u^2 y_5^2}{v \lambda _3} \\
\end{array}
\right)\)\\

\vspace{0.5cm}

(ii)  8 matrices with 3 complex constraints: 
\begin{center}
$M_{e\mu}= 0,\quad  M_{e\tau}= 0,\quad M_{\mu\mu}=-M_{\mu\tau}$
\end{center}

\vspace{0.2cm}

$\bullet$ $ n_1=1; n_2=0;\text{  }n_3=0; n_4=0; n_5=0;\text{  }n_6=1; n_7=0; n_8=1;\text{  }n_9=-1;$ \\

\noindent\(\left(
\begin{array}{cccc}
 -\frac{H^2 v y_1^2}{\Lambda ^2 \lambda _1} & 0 & 0 & -\frac{u \sqrt{\frac{H^2 v^2}{\Lambda ^2}} \rho  y_1}{v \lambda _1} \\
 0 & -\frac{H^2 v y_3^2}{\Lambda ^2 \lambda _3} & \frac{H^2 v y_3^2}{\Lambda ^2 \lambda _3} & \frac{u \sqrt{\frac{H^2 v^2}{\Lambda ^2}} y_3 y_5}{v
\lambda _3} \\
 0 & \frac{H^2 v y_3^2}{\Lambda ^2 \lambda _3} & -\frac{H^2 v^2 \left(\frac{y_2^2}{v \lambda _2}+\frac{y_3^2}{v \lambda _3}\right)}{\Lambda ^2} &
-\sqrt{\frac{H^2 v^2}{\Lambda ^2}} \left(\frac{y_2 y_4}{\lambda _2}+\frac{u y_3 y_5}{v \lambda _3}\right) \\
 -\frac{u \sqrt{\frac{H^2 v^2}{\Lambda ^2}} \rho  y_1}{v \lambda _1} & \frac{u \sqrt{\frac{H^2 v^2}{\Lambda ^2}} y_3 y_5}{v \lambda _3} & -\sqrt{\frac{H^2
v^2}{\Lambda ^2}} \left(\frac{y_2 y_4}{\lambda _2}+\frac{u y_3 y_5}{v \lambda _3}\right) & -\frac{u^2 \rho ^2}{v \lambda _1}-\frac{v y_4^2}{\lambda
_2}-\frac{u^2 y_5^2}{v \lambda _3} \\
\end{array}
\right)\)\\

\vspace{0.5cm}

(iii) 8 matrices with 3 complex constraints: 
\begin{center}
$M_{e\tau}= 0,\quad  M_{\mu\tau}= 0,\quad M_{e\mu}=-M_{\mu\mu}$
\end{center}

\vspace{0.2cm}

$\bullet$ $ n_1=1; n_2=0;\text{  }n_3=0; n_4=0; n_5=0;\text{  }n_6=1; n_7=1; n_8=0;\text{  }n_9=-1;$ \\

\noindent\(\left(
\begin{array}{cccc}
 -\frac{H^2 v^2 \left(\frac{y_1^2}{v \lambda _1}+\frac{y_3^2}{v \lambda _3}\right)}{\Lambda ^2} & \frac{H^2 v y_3^2}{\Lambda ^2 \lambda _3} & 0 &
-\sqrt{\frac{H^2 v^2}{\Lambda ^2}} \left(\frac{u \rho  y_1}{v \lambda _1}+\frac{u y_3 y_5}{v \lambda _3}\right) \\
 \frac{H^2 v y_3^2}{\Lambda ^2 \lambda _3} & -\frac{H^2 v y_3^2}{\Lambda ^2 \lambda _3} & 0 & \frac{u \sqrt{\frac{H^2 v^2}{\Lambda ^2}} y_3 y_5}{v
\lambda _3} \\
 0 & 0 & -\frac{H^2 v y_2^2}{\Lambda ^2 \lambda _2} & -\frac{\sqrt{\frac{H^2 v^2}{\Lambda ^2}} y_2 y_4}{\lambda _2} \\
 -\sqrt{\frac{H^2 v^2}{\Lambda ^2}} \left(\frac{u \rho  y_1}{v \lambda _1}+\frac{u y_3 y_5}{v \lambda _3}\right) & \frac{u \sqrt{\frac{H^2 v^2}{\Lambda
^2}} y_3 y_5}{v \lambda _3} & -\frac{\sqrt{\frac{H^2 v^2}{\Lambda ^2}} y_2 y_4}{\lambda _2} & -\frac{u^2 \rho ^2}{v \lambda _1}-\frac{v y_4^2}{\lambda
_2}-\frac{u^2 y_5^2}{v \lambda _3} \\
\end{array}
\right)\)\\

\vspace{0.5cm}

(iv) 8 matrices with 3 complex constraints: 
\begin{center}
$M_{e\tau}= 0,\quad  M_{\mu\tau}= 0,\quad M_{e\mu}=M_{\mu\mu}$
\end{center}

\vspace{0.2cm}

$\bullet$ $ n_1=1; n_2=0;\text{  }n_3=0; n_4=0; n_5=0;\text{  }n_6=1; n_7=1; n_8=0;\text{  }n_9=1;$ \\

\noindent\(\left(
\begin{array}{cccc}
 -\frac{H^2 v^2 \left(\frac{y_1^2}{v \lambda _1}+\frac{y_3^2}{v \lambda _3}\right)}{\Lambda ^2} & -\frac{H^2 v y_3^2}{\Lambda ^2 \lambda _3} & 0
& -\sqrt{\frac{H^2 v^2}{\Lambda ^2}} \left(\frac{u \rho  y_1}{v \lambda _1}+\frac{u y_3 y_5}{v \lambda _3}\right) \\
 -\frac{H^2 v y_3^2}{\Lambda ^2 \lambda _3} & -\frac{H^2 v y_3^2}{\Lambda ^2 \lambda _3} & 0 & -\frac{u \sqrt{\frac{H^2 v^2}{\Lambda ^2}} y_3 y_5}{v
\lambda _3} \\
 0 & 0 & -\frac{H^2 v y_2^2}{\Lambda ^2 \lambda _2} & -\frac{\sqrt{\frac{H^2 v^2}{\Lambda ^2}} y_2 y_4}{\lambda _2} \\
 -\sqrt{\frac{H^2 v^2}{\Lambda ^2}} \left(\frac{u \rho  y_1}{v \lambda _1}+\frac{u y_3 y_5}{v \lambda _3}\right) & -\frac{u \sqrt{\frac{H^2 v^2}{\Lambda
^2}} y_3 y_5}{v \lambda _3} & -\frac{\sqrt{\frac{H^2 v^2}{\Lambda ^2}} y_2 y_4}{\lambda _2} & -\frac{u^2 \rho ^2}{v \lambda _1}-\frac{v y_4^2}{\lambda
_2}-\frac{u^2 y_5^2}{v \lambda _3} \\
\end{array}
\right)\)\\

\vspace{0.5cm}

(v)  8 matrices with 3 complex constraints: 
\begin{center}
$M_{e\mu}= 0,\quad  M_{\mu\tau}= 0,\quad M_{e\tau}=M_{\tau\tau}$
\end{center}

\vspace{0.2cm}

$\bullet$ $ n_1=1; n_2=0;\text{  }n_3=0; n_4=0; n_5=1;\text{  }n_6=1; n_7=0; n_8=0;\text{  }n_9=-1;$ \\

\noindent\(\left(
\begin{array}{cccc}
 -\frac{H^2 v^2 \left(\frac{y_1^2}{v \lambda _1}+\frac{y_2^2}{v \lambda _2}\right)}{\Lambda ^2} & 0 & -\frac{H^2 v y_2^2}{\Lambda ^2 \lambda _2}
& -\sqrt{\frac{H^2 v^2}{\Lambda ^2}} \left(\frac{u \rho  y_1}{v \lambda _1}+\frac{y_2 y_4}{\lambda _2}\right) \\
 0 & -\frac{H^2 v y_3^2}{\Lambda ^2 \lambda _3} & 0 & \frac{u \sqrt{\frac{H^2 v^2}{\Lambda ^2}} y_3 y_5}{v \lambda _3} \\
 -\frac{H^2 v y_2^2}{\Lambda ^2 \lambda _2} & 0 & -\frac{H^2 v y_2^2}{\Lambda ^2 \lambda _2} & -\frac{\sqrt{\frac{H^2 v^2}{\Lambda ^2}} y_2 y_4}{\lambda
_2} \\
 -\sqrt{\frac{H^2 v^2}{\Lambda ^2}} \left(\frac{u \rho  y_1}{v \lambda _1}+\frac{y_2 y_4}{\lambda _2}\right) & \frac{u \sqrt{\frac{H^2 v^2}{\Lambda
^2}} y_3 y_5}{v \lambda _3} & -\frac{\sqrt{\frac{H^2 v^2}{\Lambda ^2}} y_2 y_4}{\lambda _2} & -\frac{u^2 \rho ^2}{v \lambda _1}-\frac{v y_4^2}{\lambda
_2}-\frac{u^2 y_5^2}{v \lambda _3} \\
\end{array}
\right)\)\\

\vspace{0.5cm}

(vi) 8 matrices with 3 complex constraints: 
\begin{center}
$M_{e\mu}= 0,\quad  M_{\mu\tau}= 0,\quad M_{e\tau}=-M_{\tau\tau}$
\end{center}

\vspace{0.2cm}

$\bullet$ $ n_1=1; n_2=0;\text{  }n_3=0; n_4=0; n_5=1;\text{  }n_6=-1; n_7=0; n_8=0;\text{  }n_9=1;$ \\

\noindent\(\left(
\begin{array}{cccc}
 -\frac{H^2 v^2 \left(\frac{y_1^2}{v \lambda _1}+\frac{y_2^2}{v \lambda _2}\right)}{\Lambda ^2} & 0 & \frac{H^2 v y_2^2}{\Lambda ^2 \lambda _2} &
-\sqrt{\frac{H^2 v^2}{\Lambda ^2}} \left(\frac{u \rho  y_1}{v \lambda _1}+\frac{y_2 y_4}{\lambda _2}\right) \\
 0 & -\frac{H^2 v y_3^2}{\Lambda ^2 \lambda _3} & 0 & -\frac{u \sqrt{\frac{H^2 v^2}{\Lambda ^2}} y_3 y_5}{v \lambda _3} \\
 \frac{H^2 v y_2^2}{\Lambda ^2 \lambda _2} & 0 & -\frac{H^2 v y_2^2}{\Lambda ^2 \lambda _2} & \frac{\sqrt{\frac{H^2 v^2}{\Lambda ^2}} y_2 y_4}{\lambda
_2} \\
 -\sqrt{\frac{H^2 v^2}{\Lambda ^2}} \left(\frac{u \rho  y_1}{v \lambda _1}+\frac{y_2 y_4}{\lambda _2}\right) & -\frac{u \sqrt{\frac{H^2 v^2}{\Lambda
^2}} y_3 y_5}{v \lambda _3} & \frac{\sqrt{\frac{H^2 v^2}{\Lambda ^2}} y_2 y_4}{\lambda _2} & -\frac{u^2 \rho ^2}{v \lambda _1}-\frac{v y_4^2}{\lambda
_2}-\frac{u^2 y_5^2}{v \lambda _3} \\
\end{array}
\right)\)\\

\vspace{0.5cm}

(vii) 9 matrices with 3 complex constraints: 
\begin{center}
$M_{e\mu}= 0,\quad  M_{e\tau}= 0,\quad M_{\mu\tau}=M_{\tau\tau}$
\end{center}

\vspace{0.5cm}

$\bullet$ $ n_1=1; n_2=0;\text{  }n_3=0; n_4=1; n_5=0;\text{  }n_6=0; n_7=0; n_8=1;\text{  }n_9=1;\text{  }$ \\

\noindent\(\left(
\begin{array}{cccc}
 -\frac{H^2 v y_1^2}{\Lambda ^2 \lambda _1} & 0 & 0 & -\frac{u \sqrt{\frac{H^2 v^2}{\Lambda ^2}} \rho  y_1}{v \lambda _1} \\
 0 & -\frac{H^2 v^2 \left(\frac{y_2^2}{v \lambda _2}+\frac{y_3^2}{v \lambda _3}\right)}{\Lambda ^2} & -\frac{H^2 v y_3^2}{\Lambda ^2 \lambda _3}
& -\sqrt{\frac{H^2 v^2}{\Lambda ^2}} \left(\frac{y_2 y_4}{\lambda _2}+\frac{u y_3 y_5}{v \lambda _3}\right) \\
 0 & -\frac{H^2 v y_3^2}{\Lambda ^2 \lambda _3} & -\frac{H^2 v y_3^2}{\Lambda ^2 \lambda _3} & -\frac{u \sqrt{\frac{H^2 v^2}{\Lambda ^2}} y_3 y_5}{v
\lambda _3} \\
 -\frac{u \sqrt{\frac{H^2 v^2}{\Lambda ^2}} \rho  y_1}{v \lambda _1} & -\sqrt{\frac{H^2 v^2}{\Lambda ^2}} \left(\frac{y_2 y_4}{\lambda _2}+\frac{u
y_3 y_5}{v \lambda _3}\right) & -\frac{u \sqrt{\frac{H^2 v^2}{\Lambda ^2}} y_3 y_5}{v \lambda _3} & -\frac{u^2 \rho ^2}{v \lambda _1}-\frac{v y_4^2}{\lambda
_2}-\frac{u^2 y_5^2}{v \lambda _3} \\
\end{array}
\right)\)\\

\vspace{0.5cm}

(viii)  7 matrices with 3 complex constraints: 
\begin{center}
$M_{e\mu}= 0,\quad  M_{e\tau}= 0,\quad M_{\mu\tau}=-M_{\tau\tau}$
\end{center}

\vspace{0.5cm}

$\bullet$ $ n_1=1; n_2=0;\text{  }n_3=0; n_4=1; n_5=0;\text{  }n_6=0; n_7=0; n_8=-1;\text{  }n_9=1;\text{  }$ \\

\noindent\(\left(
\begin{array}{cccc}
 -\frac{H^2 v y_1^2}{\Lambda ^2 \lambda _1} & 0 & 0 & -\frac{u \sqrt{\frac{H^2 v^2}{\Lambda ^2}} \rho  y_1}{v \lambda _1} \\
 0 & -\frac{H^2 v^2 \left(\frac{y_2^2}{v \lambda _2}+\frac{y_3^2}{v \lambda _3}\right)}{\Lambda ^2} & \frac{H^2 v y_3^2}{\Lambda ^2 \lambda _3} &
-\sqrt{\frac{H^2 v^2}{\Lambda ^2}} \left(\frac{y_2 y_4}{\lambda _2}+\frac{u y_3 y_5}{v \lambda _3}\right) \\
 0 & \frac{H^2 v y_3^2}{\Lambda ^2 \lambda _3} & -\frac{H^2 v y_3^2}{\Lambda ^2 \lambda _3} & \frac{u \sqrt{\frac{H^2 v^2}{\Lambda ^2}} y_3 y_5}{v
\lambda _3} \\
 -\frac{u \sqrt{\frac{H^2 v^2}{\Lambda ^2}} \rho  y_1}{v \lambda _1} & -\sqrt{\frac{H^2 v^2}{\Lambda ^2}} \left(\frac{y_2 y_4}{\lambda _2}+\frac{u
y_3 y_5}{v \lambda _3}\right) & \frac{u \sqrt{\frac{H^2 v^2}{\Lambda ^2}} y_3 y_5}{v \lambda _3} & -\frac{u^2 \rho ^2}{v \lambda _1}-\frac{v y_4^2}{\lambda
_2}-\frac{u^2 y_5^2}{v \lambda _3} \\
\end{array}
\right)\)

\subsubsection{Texture 3 zero case}
All 8 texture 3 zero cases can be classified into the following category. For representative purpose, we mention the VEV alignment and the corresponding mass matrix.

(i) 8 matrices with 3 complex constraints: 
\begin{center}
$M_{e\mu}= 0,\quad  M_{e\tau}= 0,\quad M_{\mu\tau}=0$
\end{center}

\vspace{0.2cm}

$\bullet$ $ n_1=1; n_2=0;\text{  }n_3=0; n_4=0; n_5=0;\text{  }n_6=1; n_7=0; n_8=0;\text{  }v_9=-1;$ \\

\noindent\(\left(
\begin{array}{cccc}
 -\frac{H^2 v y_1^2}{\Lambda ^2 \lambda _1} & 0 & 0 & -\frac{u \sqrt{\frac{H^2 v^2}{\Lambda ^2}} \rho  y_1}{v \lambda _1} \\
 0 & -\frac{H^2 v y_3^2}{\Lambda ^2 \lambda _3} & 0 & \frac{u \sqrt{\frac{H^2 v^2}{\Lambda ^2}} y_3 y_5}{v \lambda _3} \\
 0 & 0 & -\frac{H^2 v y_2^2}{\Lambda ^2 \lambda _2} & -\frac{\sqrt{\frac{H^2 v^2}{\Lambda ^2}} y_2 y_4}{\lambda _2} \\
 -\frac{u \sqrt{\frac{H^2 v^2}{\Lambda ^2}} \rho  y_1}{v \lambda _1} & \frac{u \sqrt{\frac{H^2 v^2}{\Lambda ^2}} y_3 y_5}{v \lambda _3} & -\frac{\sqrt{\frac{H^2
v^2}{\Lambda ^2}} y_2 y_4}{\lambda _2} & -\frac{u^2 \rho ^2}{v \lambda _1}-\frac{v y_4^2}{\lambda _2}-\frac{u^2 y_5^2}{v \lambda _3} \\
\end{array}
\right)\)

\subsubsection{Hybrid texture case}
All 296 hybrid textures can be classified into following categories depending upon constraints satisfied by them. 

(i) 12 matrices with 6 complex constraints:
\begin{center}
$ M_{e \mu} = -M_{e \tau}$, \quad
$ M_{\mu \mu} = M_{\tau \tau}$, \quad 
$ M_{e \tau} = M_{\tau \tau}$, \quad
$ M_{e \mu} = M_{\mu \tau} $,\quad
$ M_{e \mu} + M_{e \tau} = M_{\mu \mu} + M_{\mu \tau}$,
$ M_{\mu s} = - M_{\tau s} $ 
\end{center}
(ii) 6 matrices with 6 complex constraints:
\begin{center}
$ M_{e \mu} = -M_{e \tau} $, \quad  
$ M_{\mu \mu} = M_{\tau \tau} $ , \quad  
$ M_{e \tau} =  M_{\mu \tau}  $ , \quad  
$ M_{e \mu} =  M_{\tau \tau} $ , \quad  
$  M_{e \mu} + M_{e \tau} = M_{\mu \mu} + M_{\mu \tau} $, \quad  
$ M_{\mu s} = - M_{\tau s} $  
\end{center}
(iii) 6 matrices with 6 complex constraints: 
\begin{center}
$ M_{e \mu} = -M_{e \tau} $, \quad  
$ M_{\mu \mu} = M_{\tau \tau} $, \quad  
$ M_{e \tau} =  M_{\mu \tau} $, \quad  
$ M_{e \mu} = M_{\mu \mu}   $, \quad  
$  M_{e \mu} + M_{e \tau} = M_{\mu \mu} + M_{\mu \tau} $, \quad  
$ M_{\mu s} = - M_{\tau s} $  
\end{center}
(iv) 8 matrices with 3 complex constraints: 
\begin{center}
$ M_{e \mu} = M_{e \tau} $, \quad   
$ M_{\mu \mu} = M_{\mu \tau} $, \quad  
$ M_{e \mu} = M_{\mu \tau} $ 
\end{center}
(v) 8 matrices with 4 complex constraints:
\begin{center}
$ M_{e \mu} = M_{\mu \tau} $, \quad   
$ M_{e \tau} = M_{\mu \mu} $, \quad   
$ M_{e \mu} = -M_{e \tau} $, \quad    
$ M_{\mu \mu} = -M_{\tau \tau} $ 
\end{center}
(vi) 8 matrices with 3 complex constraints:
\begin{center}
$ M_{e \mu} = -M_{e \tau} $, \quad     
$ M_{e \mu} = M_{\mu \mu} $, \quad    
$ M_{e \tau} = M_{\mu \tau} $ 
\end{center}
(vii) 8 matrices with 4 complex constraints: 
\begin{center}
$ M_{e \mu} = M_{e \tau} $, \quad   
$ M_{\mu \mu} = M_{\mu \tau} $, \quad  
$ M_{e \mu} = -M_{\mu \tau} $, \quad  
$ M_{e \mu} = -M_{\mu \mu} $  
\end{center}
(viii) 8 matrices with 3 complex constraints: 
\begin{center}
$ M_{\mu \mu} = M_{\mu \tau} $, \quad     
$ M_{e \mu} = M_{\mu \tau} $, \quad    
$ M_{e \tau} = M_{\tau \tau} $ 
\end{center}  
(ix) 8 matrices with 3 complex constraints: 
\begin{center} 
$ M_{e \mu} = M_{\mu \tau} $, \quad     
$ M_{e \mu} = -M_{\mu \mu} $, \quad      
$ M_{e \tau} = M_{\tau \tau} $ 
\end{center}
(x) 8 matrices with 3 complex constraints: 
\begin{center} 
$ M_{e \mu} = M_{\mu \mu} $, \quad     
$ M_{e \mu} + M_{e \tau} =  M_{\mu \tau} + M_{\tau \tau}$, \quad    
$ M_{e \mu} = -M_{\mu \tau}  $ 
\end{center}
(xi) 8 matrices with 3 complex constraints:
\begin{center}
$ M_{\mu \mu} = M_{\mu \tau} $, \quad      
$ M_{e \mu} = -M_{\mu \mu} $ , \quad     
$ M_{e \tau} + M_{\mu \mu} = M_{e \mu} + M_{\tau \tau } $ 
\end{center}
(xii) 7 matrices with 2 complex constraints: 
\begin{center}
$ M_{e \mu} = M_{\mu \mu} $, \quad    
$ M_{e \mu} = M_{\mu \tau} $ 
\end{center}
(xiii) 8 matrices with 3 complex constraints: 
\begin{center}
$ M_{\mu \mu} - M_{e \tau}= M_{\mu \tau} + M_{\tau \tau } $ , \quad   
$  M_{\mu \mu} = - M_{\mu \tau} $ , \quad    
$  M_{e \mu} =  M_{\mu \tau} $ 
\end{center}
(xiv) 8 matrices with 3 complex constraints: 
\begin{center}
$ M_{e \mu} =  M_{\mu \mu}   $, \quad    
$ M_{e \mu} = - M_{\mu \tau} $, \quad    
$ M_{e \tau} = - M_{\tau \tau} $ 
\end{center}
(xv) 8 matrices with 3 complex constraints: 
\begin{center}
$ M_{\mu \mu} =  M_{\mu \tau}  $, \quad   
$ M_{e \mu} =  -M_{\mu \mu}   $, \quad    
$ M_{\tau \tau} = -M_{e \tau}  $
\end{center}
(xvi) 8 matrices with 3 complex constraints: 
\begin{center} 
$ M_{e \mu} =  M_{e \tau} $, \quad    
$ M_{e \tau} =  M_{\mu \tau} $, \quad    
$ M_{e \tau} = M_{\tau \tau} $ 
\end{center}
(xvii) 8 matrices with 3 complex constraints: 
\begin{center}
$ M_{e \mu} =  M_{\mu \tau} $, \quad   
$ M_{e \tau} =  M_{\tau \tau} $, \quad    
$ M_{e \mu} = -M_{e \tau} $ 
\end{center}
(xviii) 8 matrices with 4 complex constraints: 
\begin{center}
$ M_{e \mu} =  -M_{e \tau}  $, \quad    
$ M_{e \mu} =  -M_{\mu \tau}   $, \quad   
$ M_{\mu \tau} = -M_{\tau \tau}  $, \quad    
$ M_{e \mu} = M_{\tau \tau}  $ 
\end{center}
(xix) 8 matrices with 3 complex constraints: 
\begin{center}
$ M_{e \mu} =  M_{e \tau}  $, \quad    
$ M_{\mu \tau} =  M_{\tau \tau}   $, \quad   
$ M_{e \mu} = -M_{\mu \tau}  $
\end{center}
(xx) 16 matrices with 3 complex constraints: 
\begin{center}
$ M_{e \mu} = -M_{e \tau}  $, \quad    
$ M_{\mu \mu} = M_{\tau \tau}  $ , \quad   
$ 2 M_{e \mu} + M_{\mu \mu} - M_{\mu \tau} =  0  $ 
\end{center}
(xxi) 16 matrices with 4 complex constraints: 
\begin{center}
$ M_{e \mu} = -M_{e \tau}  $, \quad  
$ M_{\mu \mu} = M_{\tau \tau}  $, \quad  
$ M_{\mu s} = - M_{\tau s} $, \quad   
$ M_{\mu \tau} = -M_{\tau \tau}  $ 
\end{center}
(xxii) 8 matrices with 4 complex constraints: 
\begin{center}
$ M_{e \mu} = -M_{\mu \mu}  $, \quad  
$ M_{e \tau} = M_{\mu \tau}  $, \quad  
$ M_{e \tau} =  M_{\tau \tau} $ , \quad  
$ M_{e \mu} -M_{\mu \mu} + M_{e \tau} + M _{\mu \tau} - 2M_{\tau \tau} = 0 $ 
\end{center}
(xxiii) 8 matrices with 3 complex constraints:
\begin{center}
$ M_{e \tau} = M_{\tau \tau}  $, \quad  
$ M_{e \tau} = - M_{\mu \tau} $, \quad   
$ M_{e \mu} -M_{\mu \mu} - 2M_{\mu \tau} = 0 $
\end{center}
(xxiv) 8 matrices with 3 complex constraints:
\begin{center} 
$ M_{e \mu} = M_{\mu \mu} $ , \quad  
$ M_{e \tau} = -M_{\tau \tau} $ , \quad  
$ M_{e \tau} =  M_{\mu \tau} $  
\end{center}
(xxv) 8 matrices with 3 complex constraints:
 \begin{center}
$ M_{e \mu} = M_{\mu \mu} $, \quad   
$ M_{\mu \mu} = M_{\tau \tau} $, \quad   
$ M_{e \tau} =  M_{\mu \tau} $ 
\end{center}
(xxvi) 8 matrices with 3 complex constraints: 
\begin{center} 
$ M_{e \tau} = -M_{\tau \tau} $, \quad   
$ M_{e \tau} = - M_{\mu \tau} $  , \quad  
$ M_{e \mu} -M_{\mu \mu} - 2 M_{e \tau} = 0 $ 
\end{center}
(xxvii) 3 matrices with 3 complex constraints: 
\begin{center}
$ M_{e \tau} =  M_{\tau \tau} $ , \quad  
$ M_{e \tau} =  M_{\mu \tau} $, \quad    
$ M_{e \mu} -M_{\mu \mu} + M_{\mu \tau} + M _{\tau \tau}= 0 $ 
\end{center}
(xxviii) 1 matrix with 5 complex constraints:  
\begin{center}
$ M_{e \tau} = M_{\tau \tau}  $, \quad   
$ M_{\mu \mu} = M_{\tau \tau}  $, \quad  
$ M_{\mu s} = - M_{\tau s} $, \quad   
$ M_{e \mu} = M_{\mu \mu}  $, \quad  
$ M_{e \mu} + M_{\mu \mu} = M_{e \tau} + M _{\mu \tau} $ 
\end{center}
(xxix) 2 matrices with 3 complex constraints: 
\begin{center}
$ M_{e \tau} = M_{\mu \tau}  $, \quad   
$ M_{\mu \tau} = -M_{\tau \tau}  $, \quad  
$ M_{e \mu} + M_{\mu \mu} - 2 M _{e \tau} = 0 $
\end{center}
(xxx) 6 matrices with 3 complex constraints:
\begin{center} 
$ M_{e \tau} = M_{\mu \tau}  $, \quad   
$ M_{e \tau} = -M_{\tau \tau}  $, \quad   
$ M_{e \mu} + M_{\mu \mu} - 2 M _{\tau \tau} = 0 $ 
\end{center}
(xxxi) 5 matrices with 3 complex constraints: 
\begin{center}
$ M_{e \tau} = M_{\mu \tau}  $ , \quad  
$ M_{e \tau} = M_{\tau \tau}  $ , \quad  
$ M_{e \mu} + M_{\mu \mu} - 2 M _{e \tau} = 0 $ 
\end{center}
(xxxii) 8 matrices with 3 complex constraints: 
\begin{center}
$ M_{e \tau} = -M_{\mu \tau}  $, \quad   
$ M_{e \tau} = M_{\tau \tau}  $, \quad   
$ M_{e \mu} + M _{\mu \mu} = 0 $ 
\end{center}
(xxxiii) 7 matrices with 3 complex constraints: 
\begin{center}
$ M_{e \tau} = M_{\tau \tau}  $ , \quad  
$ M_{\mu \mu} = M_{\tau \tau}  $, \quad   
$ M_{e \mu} =  M_{\mu \tau}  $
\end{center}
(xxxiv) 4 matrices with 5 complex constraints:
\begin{center} 
$ M_{e \mu} = M_{\mu \mu}  $, \quad  
$ M_{e \tau} = M_{\mu \tau}  $, \quad   
$ M_{e \mu} =  M_{\tau \tau}  $, \quad  
$ M_{\mu s} + M _{\tau s} = 0  $, \quad  
$ M_{e \tau} + M_{\mu \mu} = 0 $ 
\end{center}
(xxxv) 8 matrices with 4 complex constraints:
\begin{center}
$ M_{\mu \mu} = M_{\tau \tau}  $ , \quad  
$ M_{e \mu} + M_{\mu \tau} = 0 $ , \quad  
$ M_{\mu \mu} +  M_{e \tau} = 0 $ , \quad  
$ M_{e \tau} + M _{\tau \tau} = 0  $ 
\end{center}
(xxxvi) 4 matrices with 4 complex constraints: 
\begin{center}
$ M_{\mu \mu} = M_{\tau \tau} $, \quad  
$ M_{e \mu} = M_{\mu \tau}  $, \quad  
$ M_{e \tau} =  M_{\mu \mu}  $, \quad  
$ M_{e \mu} + M_{e \tau} = 0 $ 
\end{center}
(xxxvii) 8 matrices with 4 complex constraints:
\begin{center}
$ M_{\mu \mu} = M_{\tau \tau}  $ , \quad  
$ M_{\mu s} + M _{\tau s} = 0  $ , \quad  
$ M_{\mu \mu} = - M_{\mu \tau}  $ , \quad  
$ M_{e \mu} + M_{\mu \mu} = M_{\tau \tau} - M_{e \tau} $
\end{center}
(xxxviii) 8 matrices with 3 complex constraints: 
\begin{center}
$ M_{\mu \mu} = M_{\tau \tau} $ , \quad  
$ M_{e \tau} = - M_{\mu \tau}  $ , \quad  
$ M_{e \mu} = - M_{\tau \tau}  $ \
\end{center}
(xxxix) 8 matrices with 3 complex constraints: 
\begin{center}
$ M_{e \tau} = - M_{\tau \tau} $, \quad   
$ M_{e \tau} = - M_{\mu \tau}  $ , \quad  
$ M_{e \mu} + M_{\mu \mu} = 0 $ 
\end{center}
(xxxx) 1 matrix with 3 complex constraints: 
\begin{center}
$ M_{\mu \mu} =  M_{\mu \tau} $, \quad   
$ M_{e \mu} =  M_{\mu \tau}  $ , \quad  
$ M_{e \tau} - M_{\mu \mu} = M_{\mu \tau}- M_{\tau \tau} $ 
\end{center}

\section{Numerical analysis}
 \label{sec:numeric}
In this section, we present the method adopted for numerical analysis for $(\mu-\tau)$ symmetric textures, texture 1, texture 2 and texture 3 zero cases, in order to check their consistency with $3+1$ neutrino data. It is well known that $4\times4$ unitary mixing matrix can be parametrised as \cite{sterilemutau}

\begin{equation}
U = R_{34} \tilde R_{24} \tilde R_{14}R_{23}\tilde R_{13}R_{12}P
\end{equation}
where 
\begin{equation}
R_{34}= \begin{pmatrix}
1 & 0 & 0 & 0  \\
0 & 1 & 0 & 0  \\
0 & 0 & c_{34} & s_{34}  \\
0 & 0 & -s_{34} & c_{34} \\ 
\end{pmatrix} 
\end{equation}

\begin{equation}
\tilde R_{14}= \begin{pmatrix}
c_{14} & 0 & 0 & s_{14}e^{-i\delta_{14}} \\
0 & 1 & 0 & 0  \\
0 & 0 & 1 & 0  \\
-s_{14}e^{i \delta_{14}} & 0 & 0 & c_{14} \\ 
\end{pmatrix} 
\end{equation}\\
with $c_{ij} = \cos{\theta_{ij}}$, $s_{ij} = \sin{\theta_{ij}}$ , $\delta_{ij}$ being the Dirac CP phases, and 
$$ P = \text{diag}(1, e^{-i\frac{\alpha}{2}}, e^{-i(\frac{\beta}{2}-\delta_{13})}, e^{-i(\frac{\gamma}{2}-\delta_{14})})$$
is the diagonal phase matrix containing the three Majorana phases $\alpha, \beta, \gamma$. In this parametrisation, the six CP phases vary from $-\pi$ to $\pi$. Using the above form of mixing matrix, the $4\times4$ complex symmetric Majorana light neutrino mass matrix can be written as
\begin{eqnarray}
M_{\nu} &=& U M^{\text{diag}}_{\nu} U^T \\
        &=&\begin{pmatrix}
m_{ee} & m_{e\mu} & m_{e\tau} & m_{es}\\
m_{\mu e} & m_{\mu\mu} & m_{\mu\tau} & m_{\mu s} \\
m_{\tau e} & m_{\tau\mu} & m_{\tau\tau} & m_{\tau s} \\
m_{se} & m_{s\mu} & m_{s\tau} & m_{ss}
\end{pmatrix},
\label{mnu}
\end{eqnarray}
 where $M_{\nu}^{diag} = \text{diag}(m_{1}, m_{2}, m_{3}, m_{4})$ 
 is the diagonal light neutrino mass matrix. For normal hierarchy (NH) of active neutrinos i.e., $m_4 > m_3 > m_2 > m_1$, the neutrino mass eigenvalues can be written in terms of the lightest neutrino mass $m_1$ as
  $$m_{2}=\sqrt{m_{1}^{2}+\Delta m_{21}^{2}},\quad  m_{3}=\sqrt{m_{1}^{2}+\Delta m_{31}^{2}},\quad  m_{4}=\sqrt{m_{1}^{2}+\Delta m_{41}^{2}}. $$
Similarly for inverted hierarchy (IH) of active neutrinos i.e., $m_4 > m_2 > m_1 > m_3$, the neutrino mass eigenvalues can be written in terms of the lightest neutrino mass $m_3$ as
 $$m_{1}=\sqrt{m_{3}^{2}-\Delta m_{32}^{2}-\Delta m_{21}^{2}},\quad  m_{2}=\sqrt{m_{3}^{2}-\Delta m_{32}^{2}},\quad  m_{4}=\sqrt{m_{3}^{2}+\Delta m_{43}^{2}}.$$
 Using these, one can analytically write down the $4\times4$ light neutrino mass matrix in terms of three mass squared differences, lightest neutrino mass $m_1 (m_3)$, six mixing angles i.e., $
\theta_{13}$, $\theta_{12}$, $\theta_{23}$, $\theta_{14}$, $\theta_{24}$,
$\theta_{34}$, three Dirac type CP phases i.e., $\delta_{13}$, $\delta_{14}$, 
$\delta_{24}$ and three Majorana type CP phases i.e., $\alpha$, $\beta$, 
$\gamma$.
The analytical expressions of the $4\times4$ light neutrino mass matrix elements are given in Appendix \ref{appen4}.
 
For each class of neutrino mass matrix with textures that we analyse, there exists several constraints relating the mass matrix elements or equating some of them to zero. Since the mass matrix is complex symmetric, each such constraint gives rise to two real equations that can be solved for two unknown parameters. Depending upon the number of constraints, we choose the set of input parameters and solve for the remaining ones. We have varied our input parameters for the usual three neutrino part in the $3 \sigma$ allowed range as given in the global analysis of the world neutrino data~\cite{schwetz17, Forero:2014bxa, Capozzi:2013csa} and varied $\Delta m^2_{\rm LSND}$ from 0.7 eV$^2$ to 2.5 eV$^2$. If the output of $\theta_{14}$, $\theta_{24}$ and $\theta_{34}$ falls between $0^\circ$ to $20^\circ$, $0^\circ$ to $11.5^\circ$ and $0^\circ$ to $30^\circ$ respectively~\cite{Kopp:2013vaa,An:2014bik,Adamson:2011ku} with the condition $m_1 (m_3) =0$ (as the model predicts vanishing lightest neutrino mass), then we say this texture is allowed in NH (IH). Note that according to the global analysis of the short-baseline data \cite{Kopp:2013vaa} we have $6^\circ < \theta_{14} < 20^\circ$ and $3^\circ < \theta_{24} < 11.5^\circ$ at $3 \sigma$. However the Refs. \cite{An:2014bik,Adamson:2011ku}, give only an upper limit on $\theta_{14}$ and $\theta_{24}$ as they analyse stand-alone data. Thus for a conservative approach, in our analysis we have taken the upper limits of $\theta_{14}$ and $\theta_{24}$ from the global analysis and allowed them to have lower limits as zero.
 
 \section{Results and Discussion}
 \label{sec:results}
Adopting the classifications of different textures in $4\times4$ mass matrix not ruled out from previous studies and the method of numerical analysis discussed in the previous section we have analysed all possible mass matrices either with texture zeros or with $\mu-\tau$ symmetry in the $3\times 3$ block. Here we show some of the numerical results we have obtained for those textures which are found to be allowed in our analysis after taking into account the $3+1$ neutrino data.

\begin{figure}
  \centering

    \includegraphics[width=0.45\textwidth]{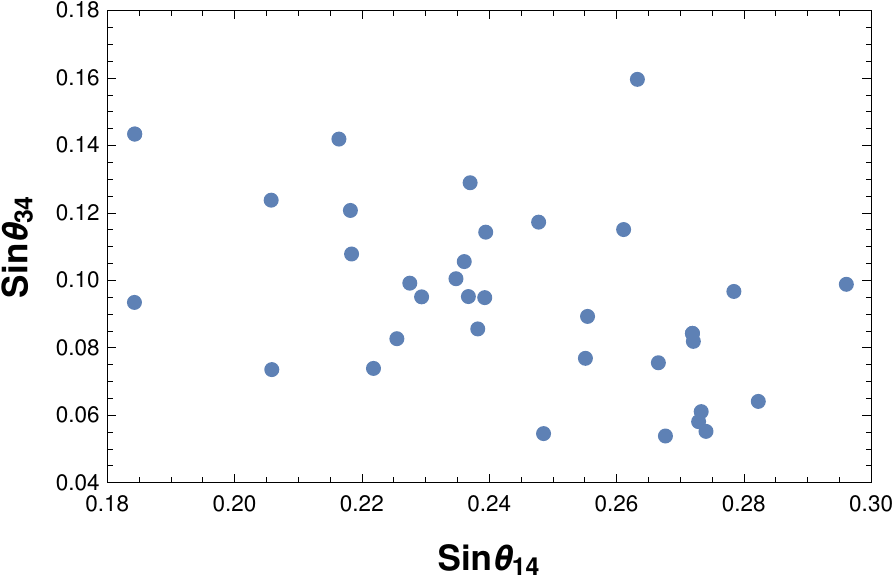}
    \includegraphics[width=0.45\textwidth]{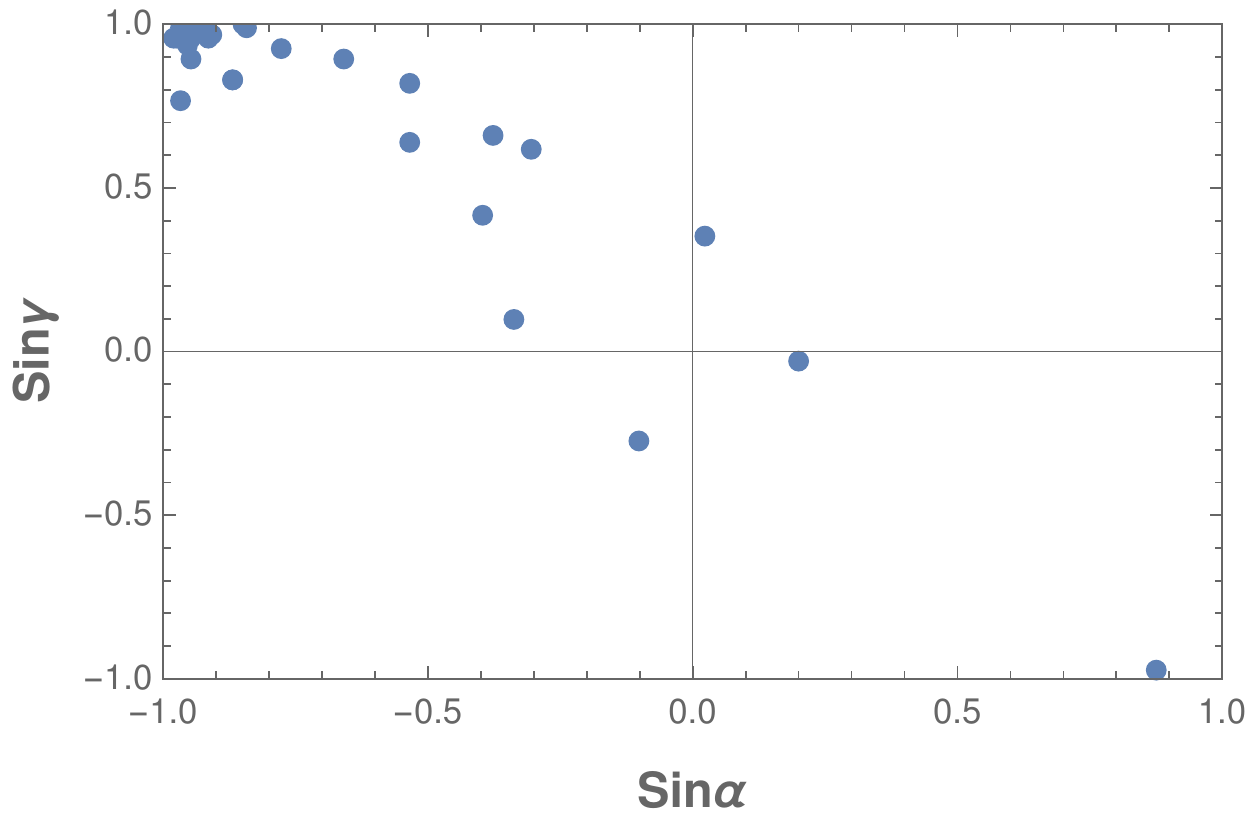}\\

      \caption{Neutrino oscillation parameters in active-sterile sector for case (ii) from $\mu-\tau$ symmetric category for NH.}
      \label{fig1}
  \end{figure}

\begin{figure}
  \centering

    \includegraphics[width=0.45\textwidth]{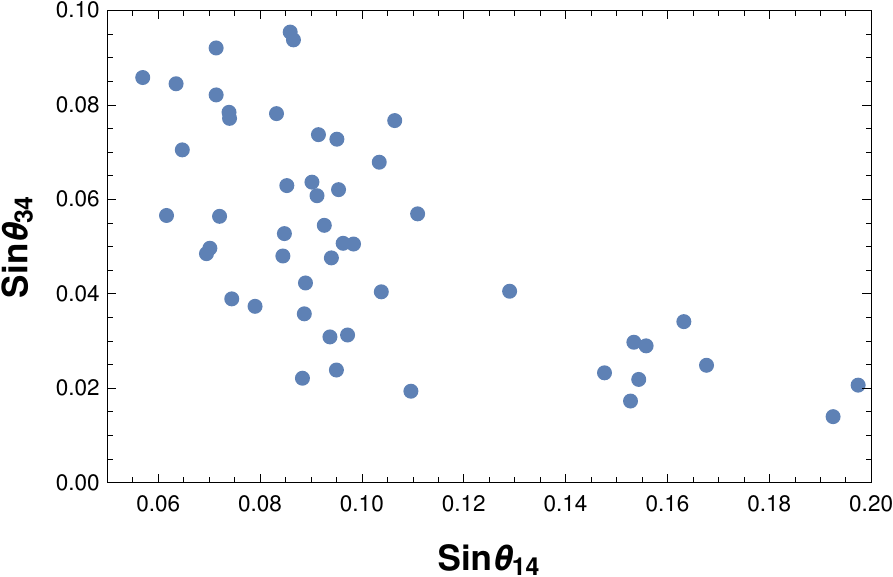}
    \includegraphics[width=0.45\textwidth]{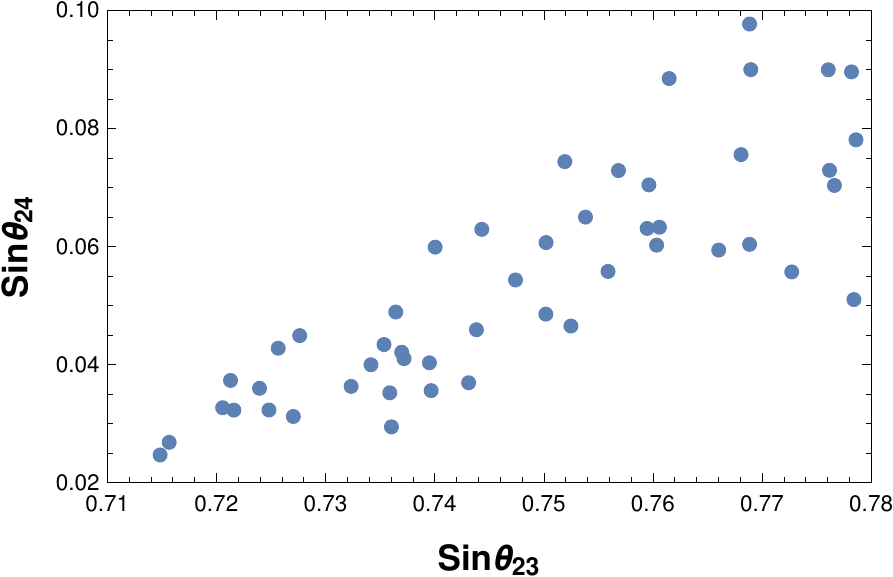} \\
    \includegraphics[width=0.45\textwidth]{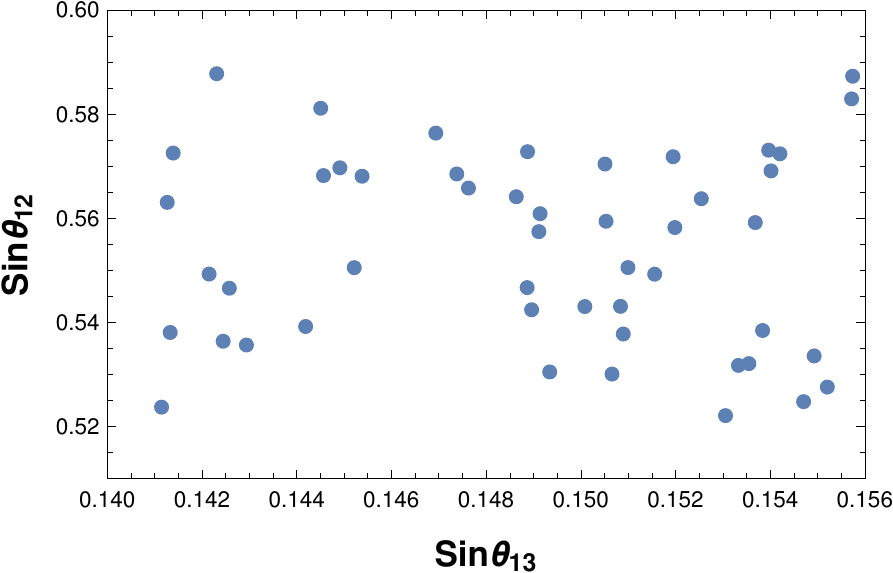}
    \includegraphics[width=0.45\textwidth]{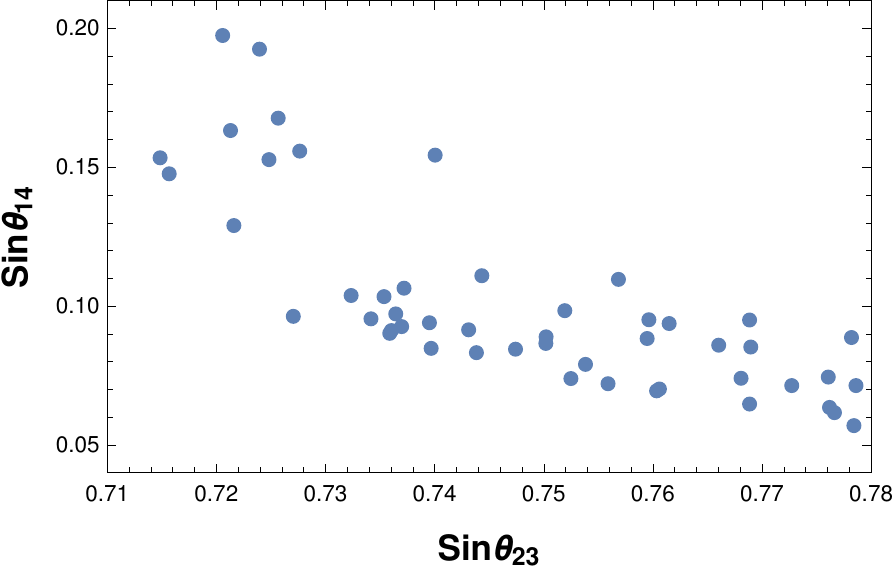} \\
    \includegraphics[width=0.45\textwidth]{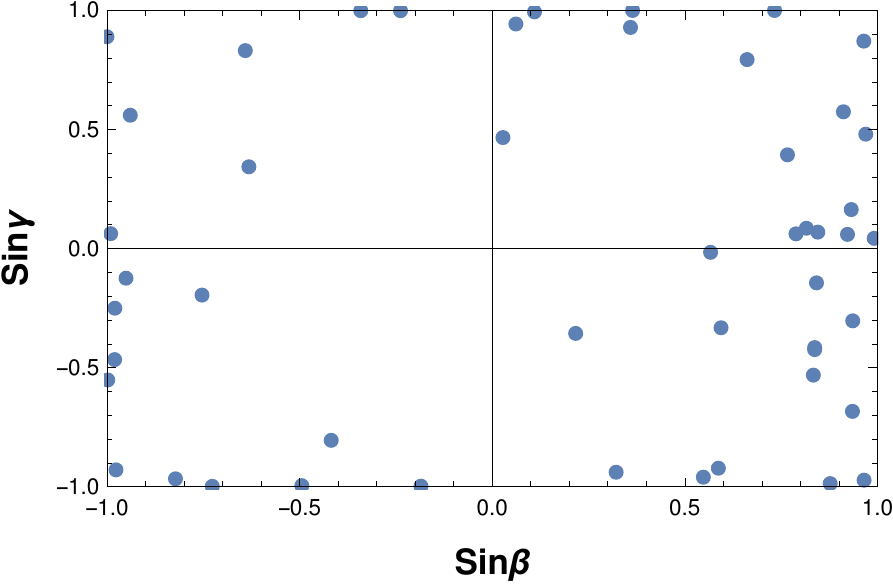}
    \includegraphics[width=0.45\textwidth]{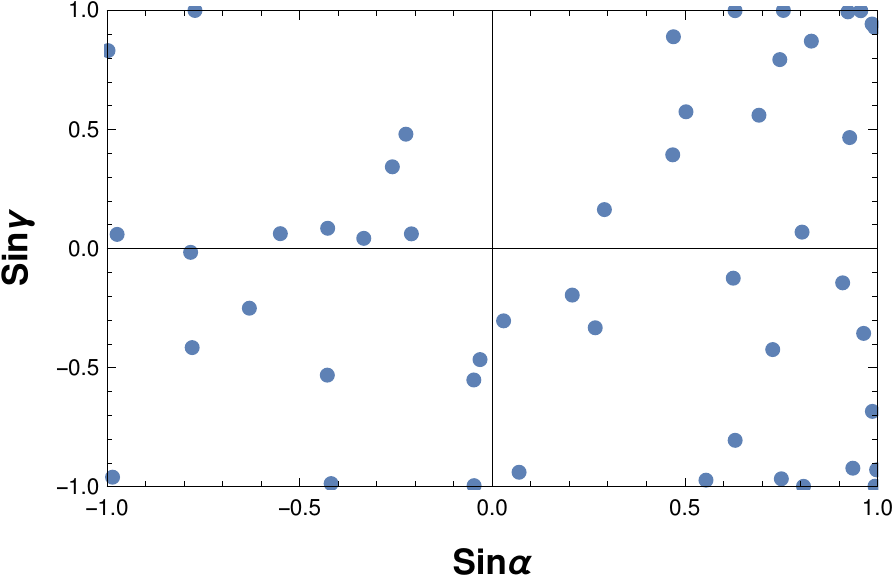} \\
    \includegraphics[width=0.45\textwidth]{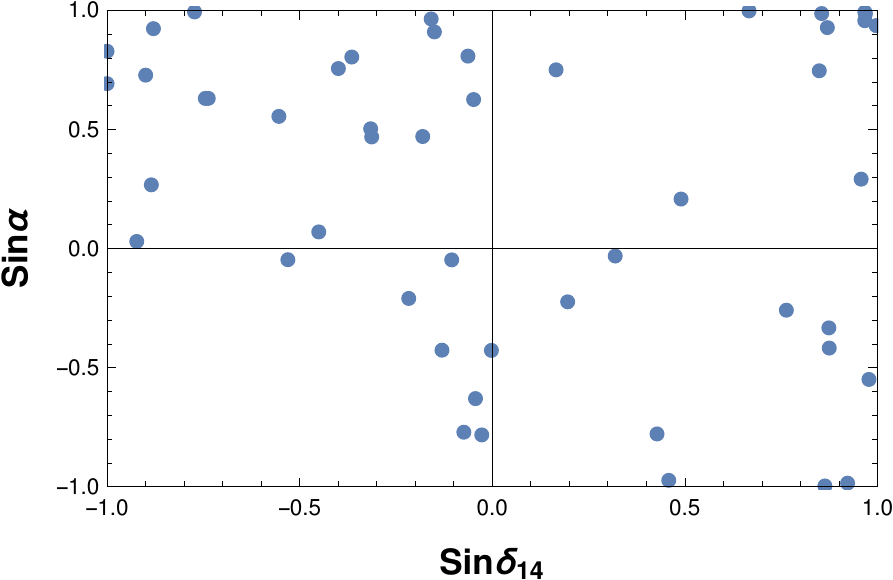}
    \includegraphics[width=0.45\textwidth]{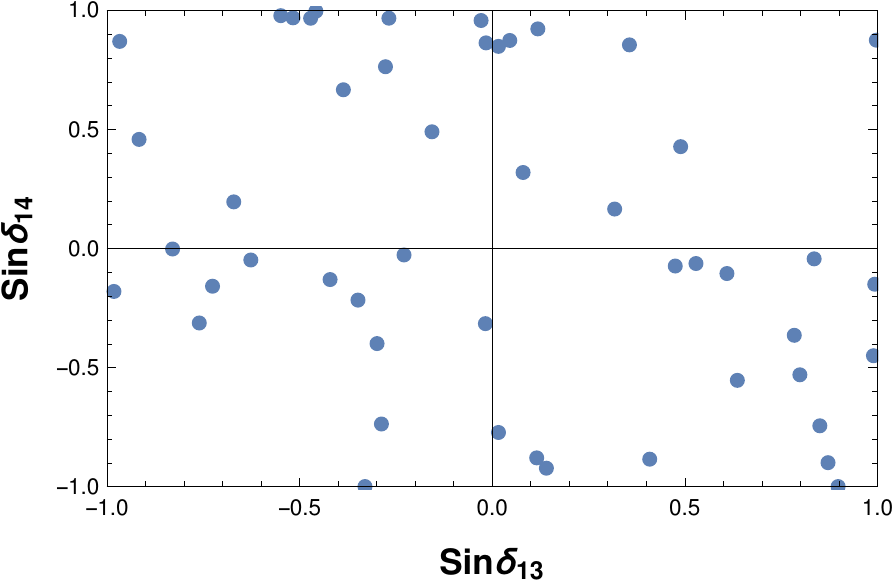} \\
    
  \caption{Neutrino oscillation parameters in active-sterile sector for case (iii) from $\mu-\tau$ symmetric category for NH.}
   \label{fig2}
 \end{figure}

\begin{figure}
  \centering

    \includegraphics[width=0.45\textwidth]{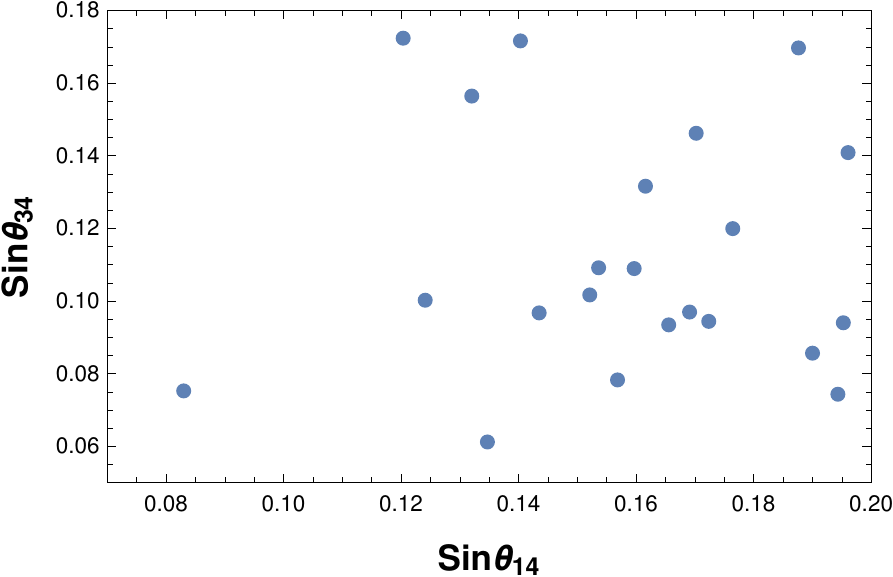}
    \includegraphics[width=0.45\textwidth]{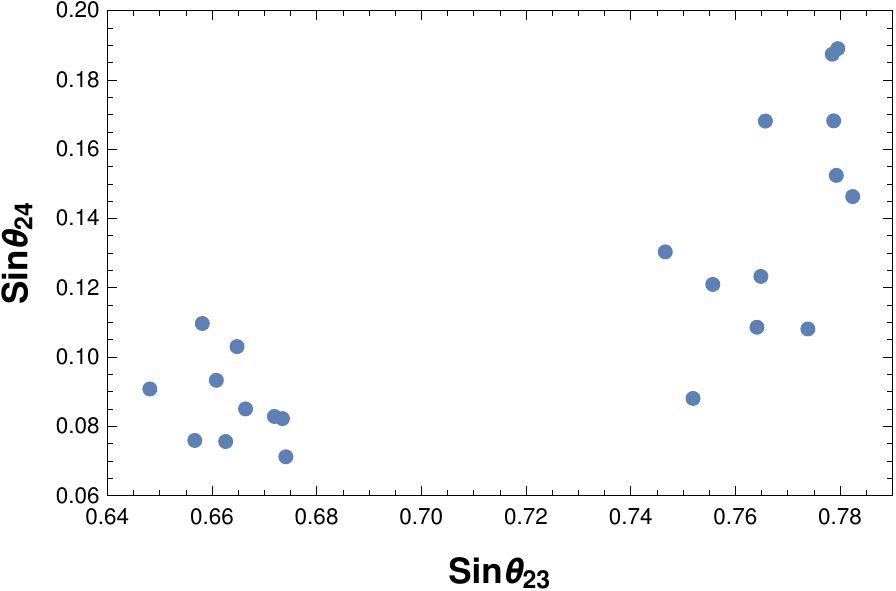} \\
    \includegraphics[width=0.45\textwidth]{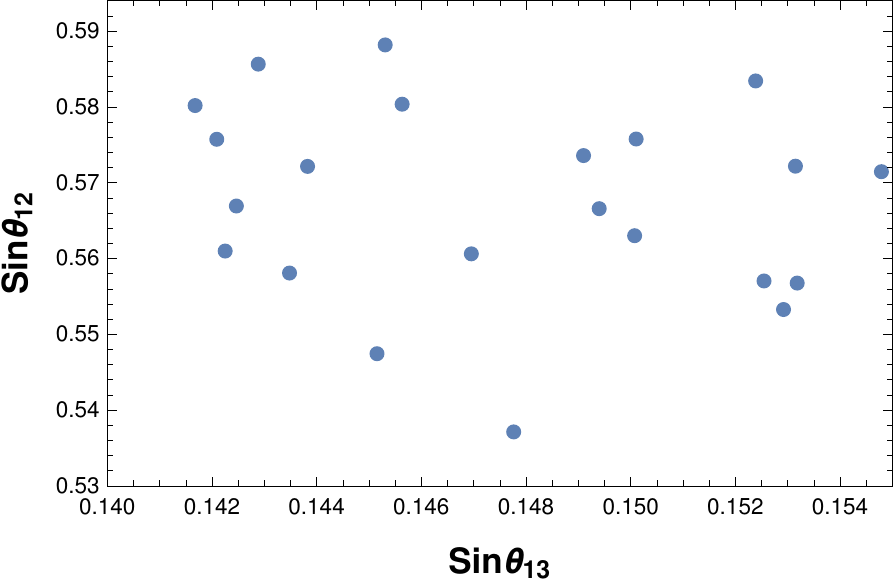}
    \includegraphics[width=0.45\textwidth]{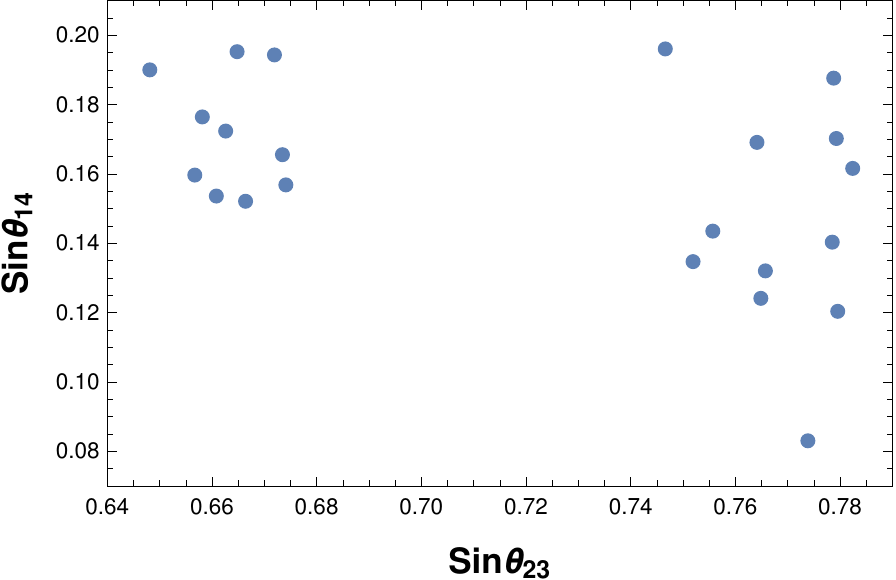}\\

  \caption{Neutrino oscillation parameters in active-sterile sector for case (iv) from $\mu-\tau$ symmetric category for NH}
   \label{fig3}
 \end{figure}

We first show the results for $\mu-\tau$ symmetric textures. Out of four different classes belonging to this type of texture we found three of them to be allowed for NH of active neutrino masses. They are namely, the subclasses (ii), (iii), (iv) of $\mu-\tau$ symmetric textures discussed earlier. It is not surprising that the texture subclass (i) is not allowed due to too many constraints (5 complex and hence 10 real constraint equations) it has, which become difficult to be satisfied simultaneously while keeping all neutrino parameters in allowed range. For IH of active neutrino masses, none of these textures are allowed. We show some correlation plots between neutrino parameters for $\mu-\tau$ symmetric subclasses (ii), (iii), (iv) with NH in figure \ref{fig1}, \ref{fig2}, \ref{fig3} respectively. Figure \ref{fig1} shows the correlations between active-sterile mixing angles $\theta_{34}-\theta_{14}$ and between Majorana CP phases $\alpha-\gamma$ after the constraint equations corresponding to $\mu-\tau$ symmetric subclass (ii) were solved for 1 million random points. The resulting acceptable number of solutions is only a handful, as can be seen from the plots. The same trend is repeated for the other two subclasses (iii), (iv) as well, the correlations for which are shown in figure \ref{fig2}, \ref{fig3} respectively.

\begin{figure}
  \centering

    \includegraphics[width=0.45\textwidth]{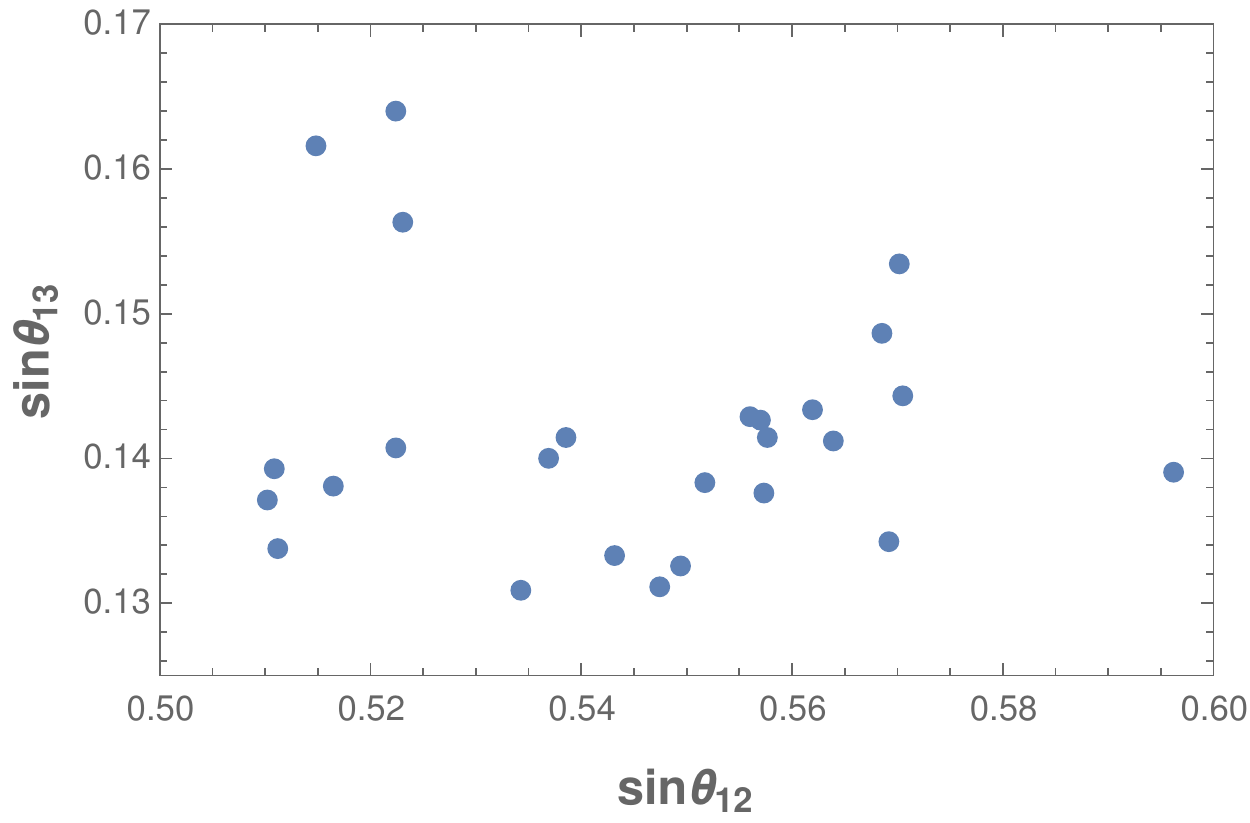}
    \includegraphics[width=0.45\textwidth]{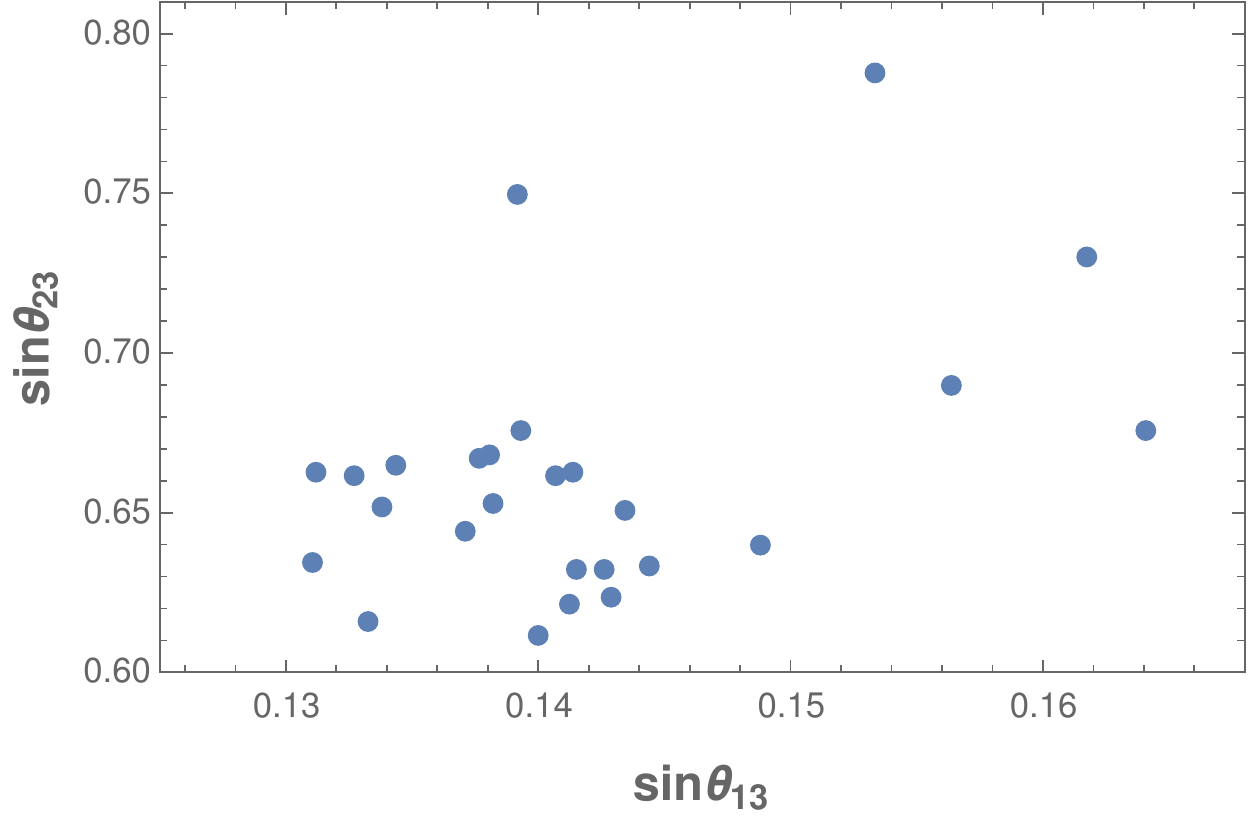} \\
    \includegraphics[width=0.45\textwidth]{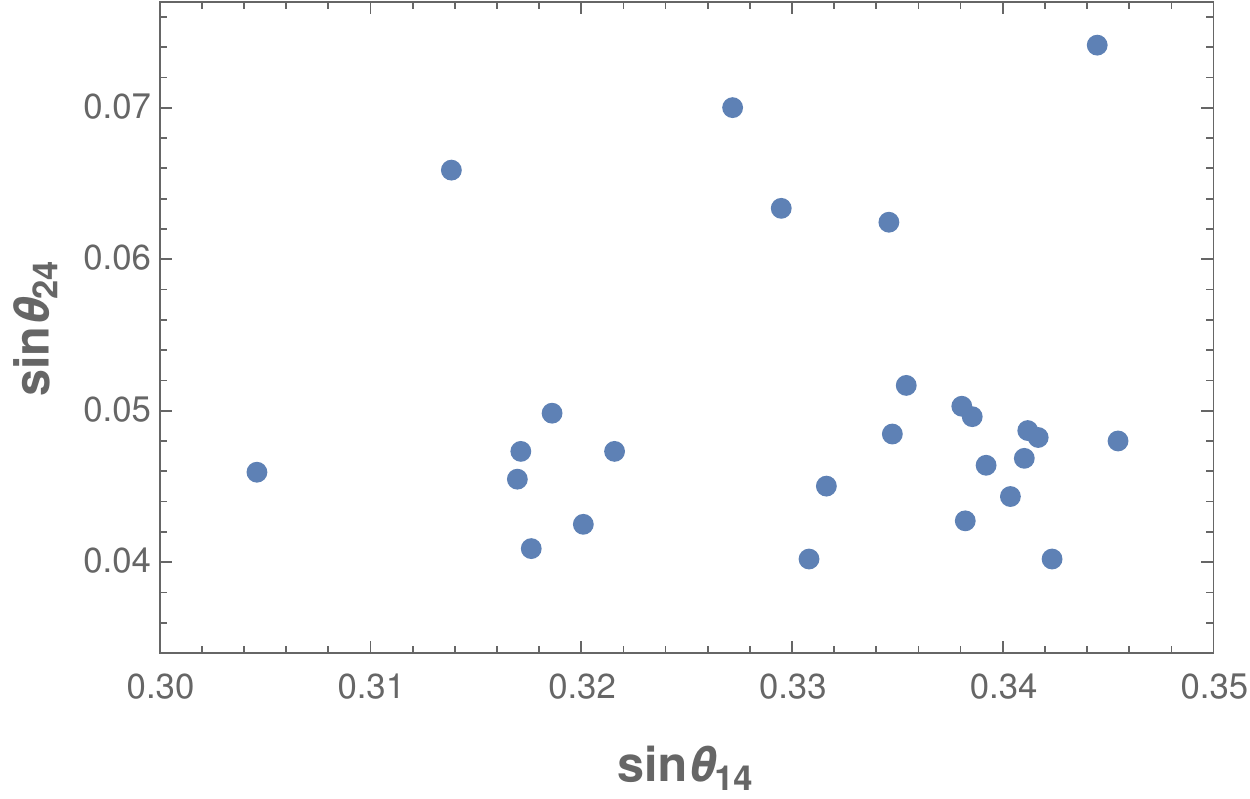}
     \includegraphics[width=0.45\textwidth]{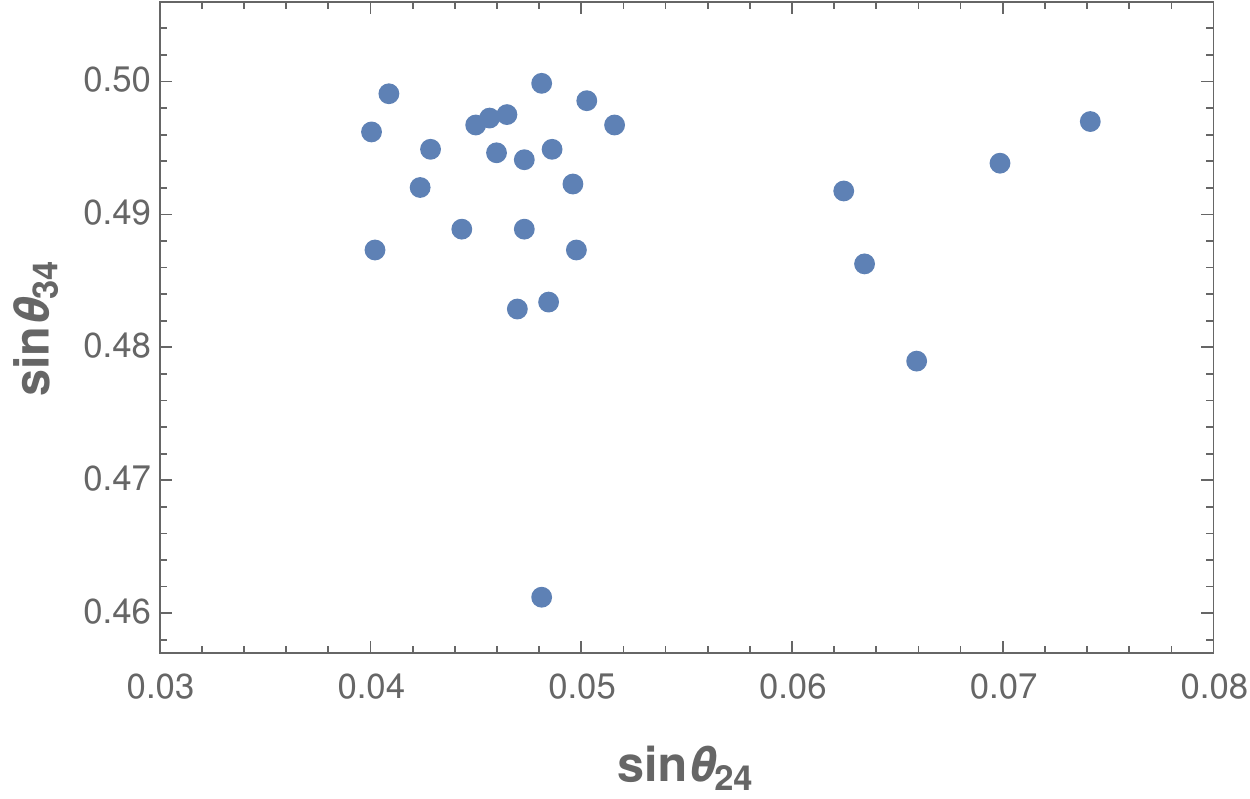} \\
    \includegraphics[width=0.45\textwidth]{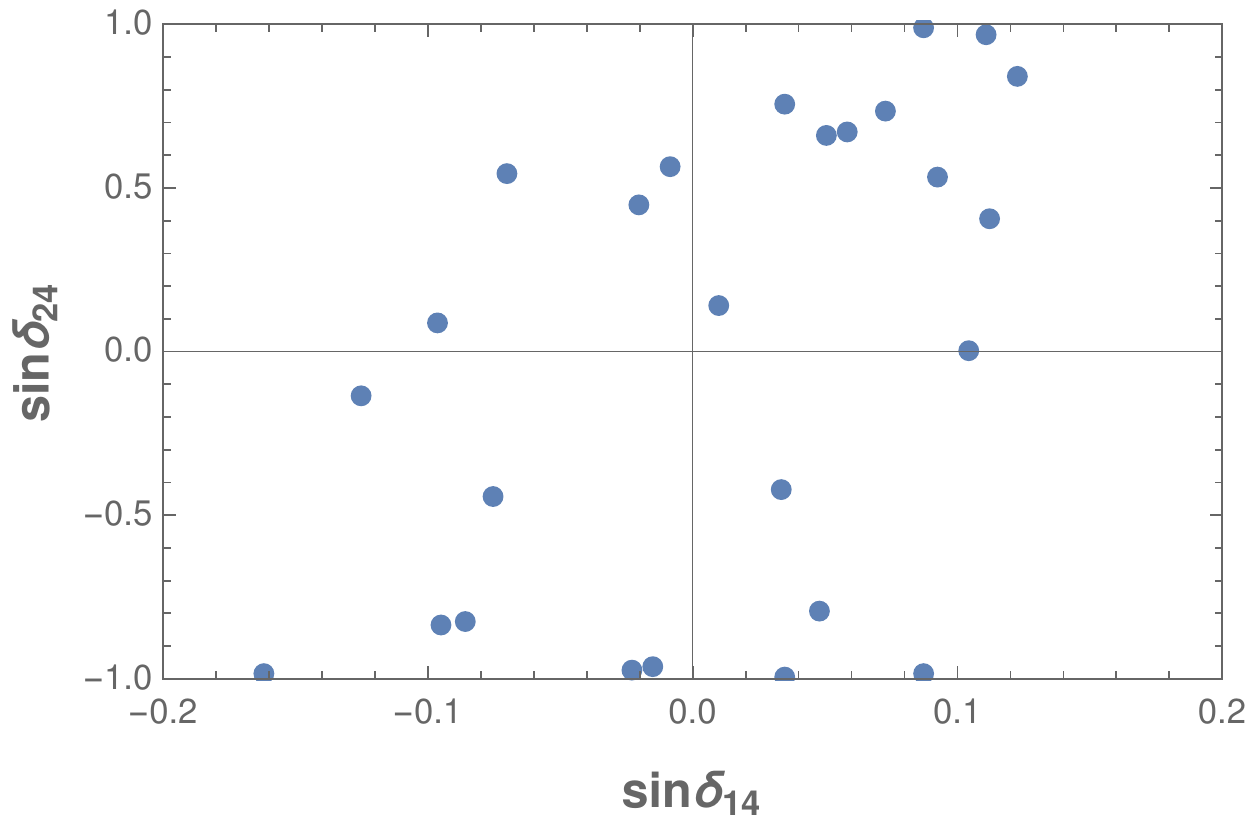}
   \caption{Neutrino oscillation parameters in active-sterile sector for case (ix) from texture 1 zero category for NH.}
    \label{fig4}
 \end{figure}

\begin{figure}
  \centering

    \includegraphics[width=0.45\textwidth]{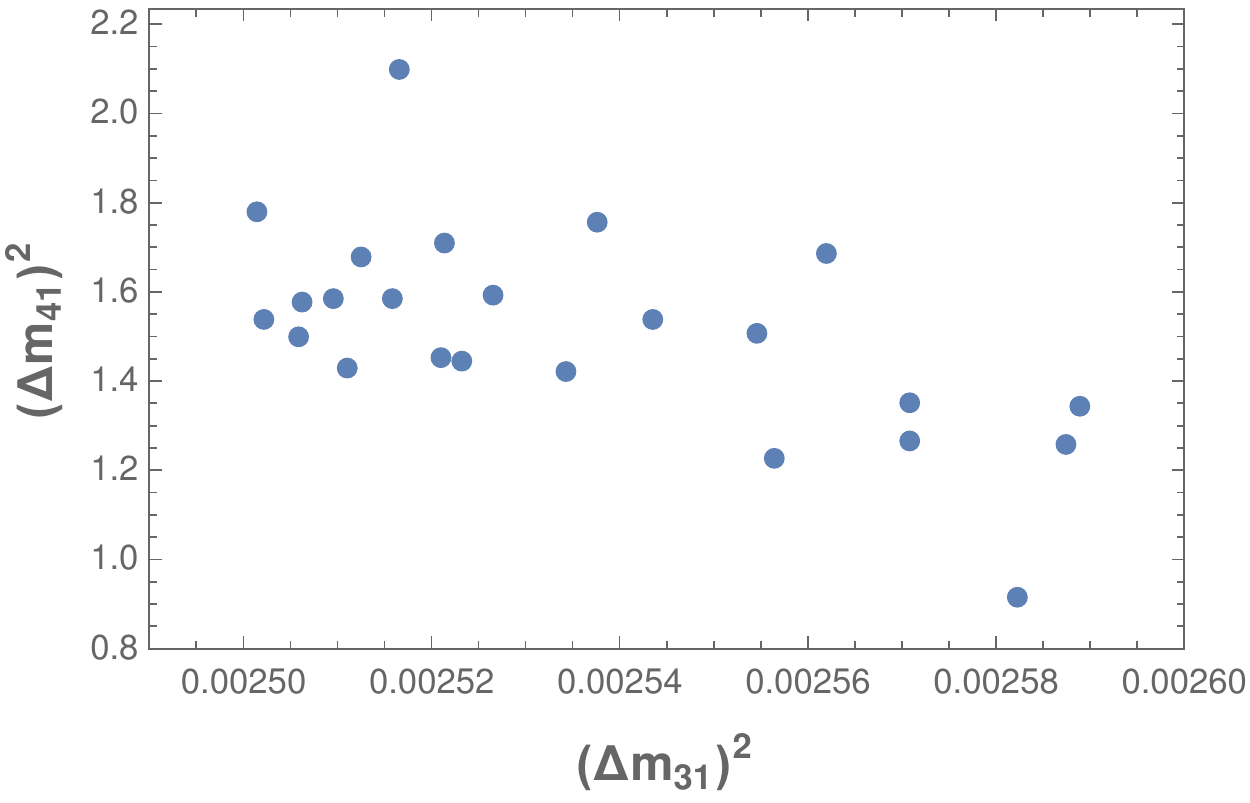}
    \includegraphics[width=0.45\textwidth]{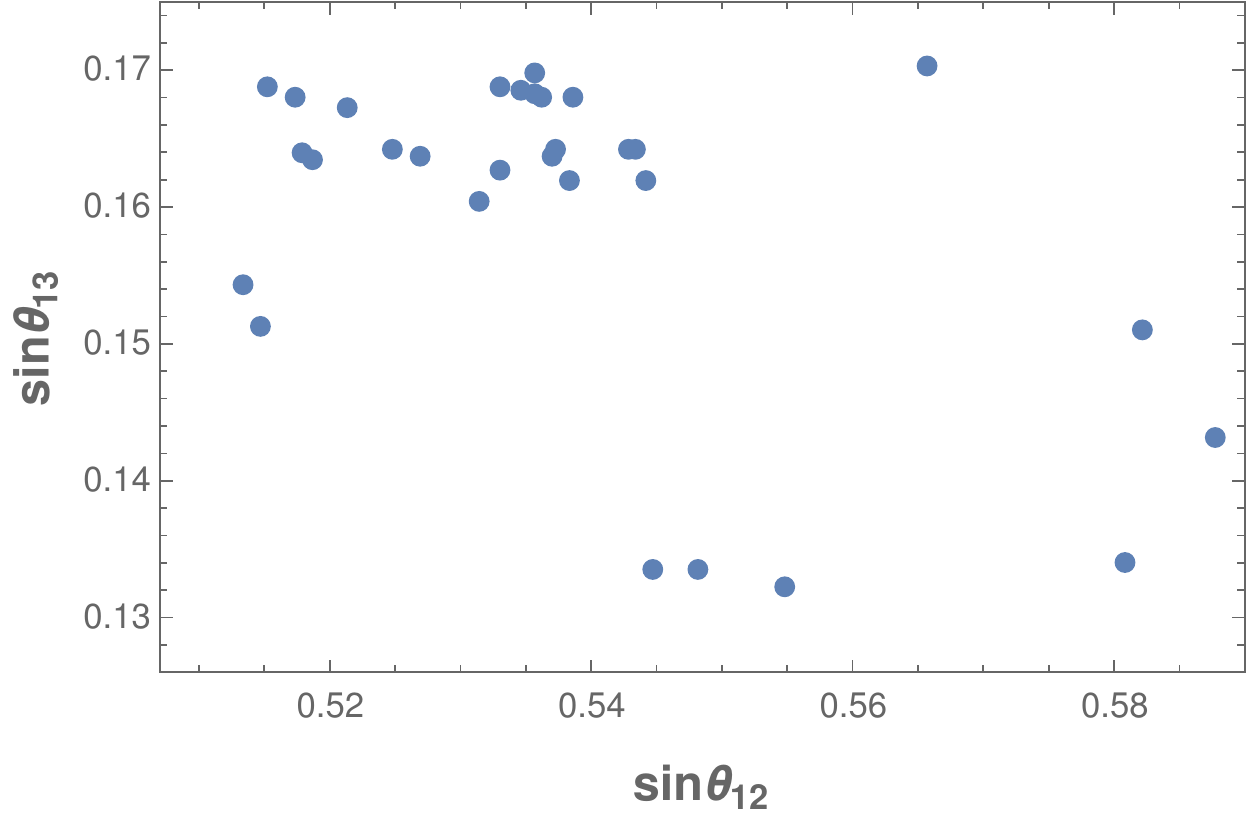} \\
    \includegraphics[width=0.45\textwidth]{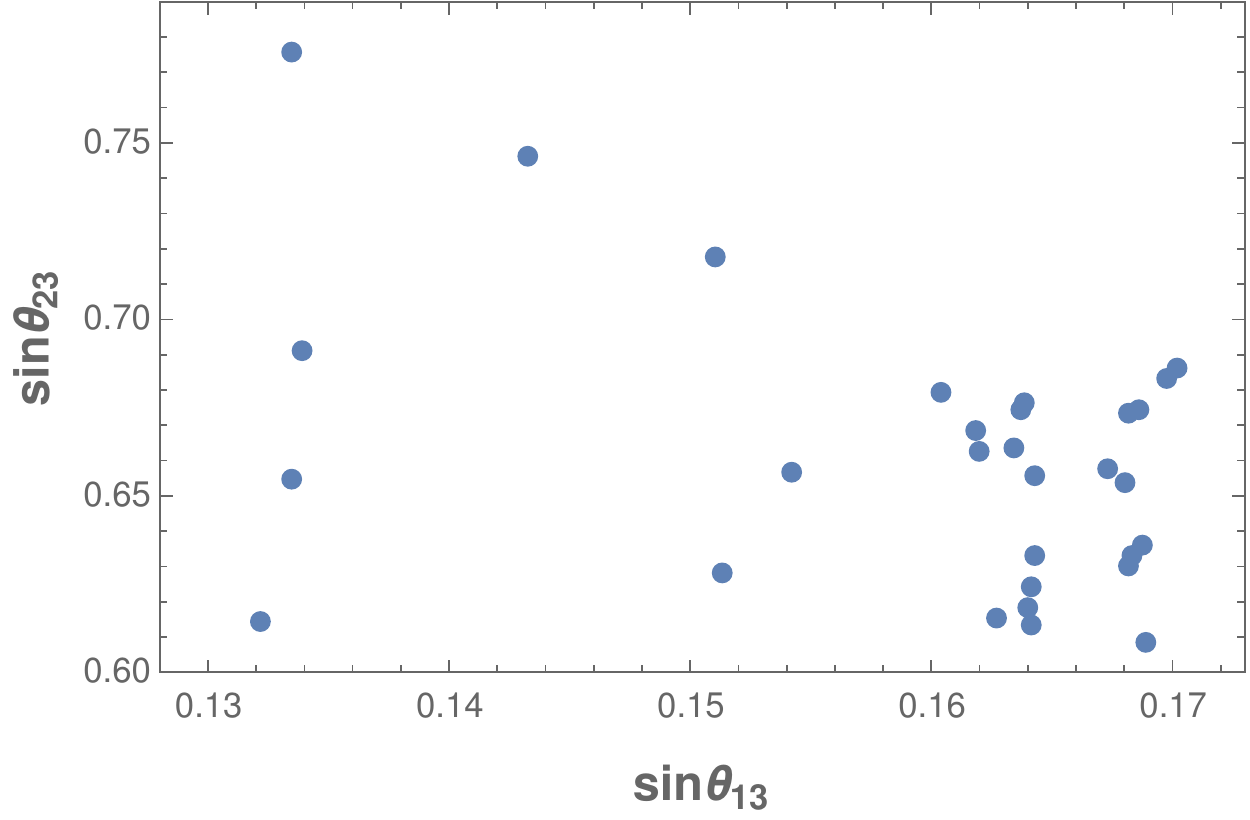}
    \includegraphics[width=0.45\textwidth]{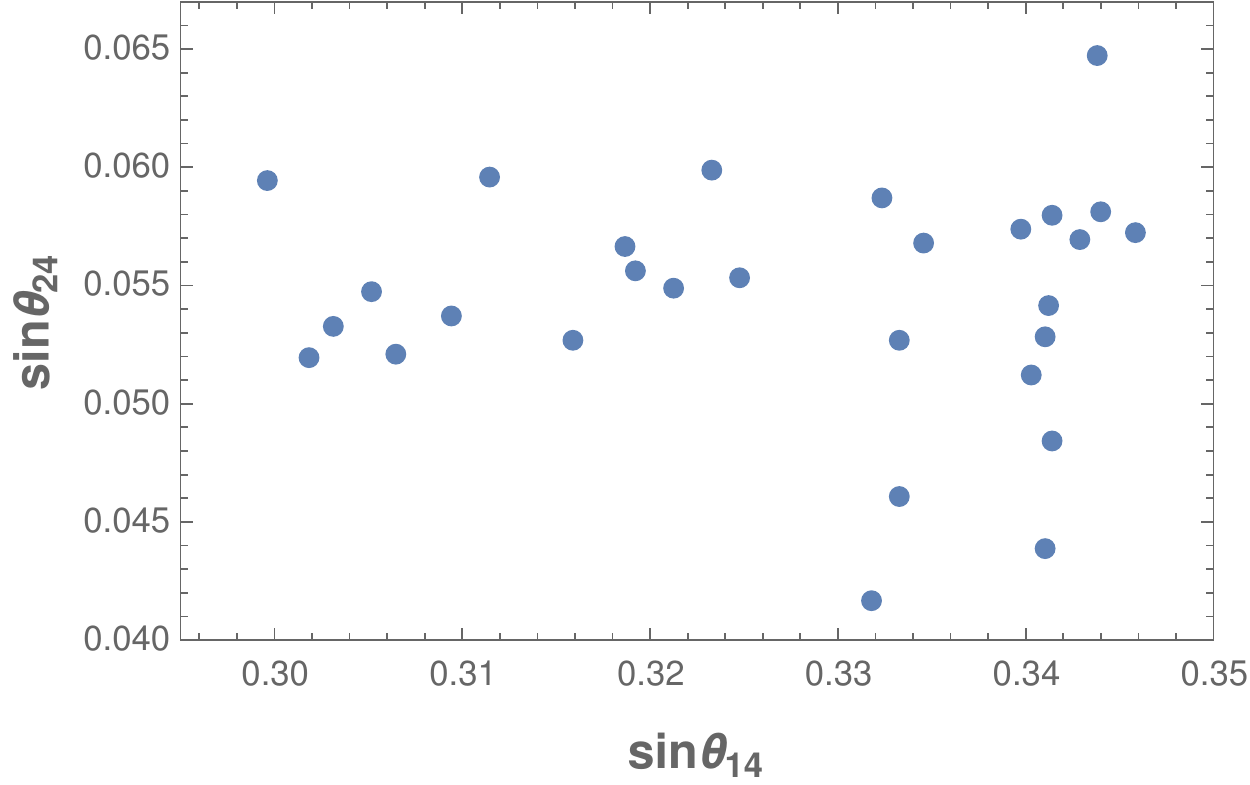} \\
    \includegraphics[width=0.45\textwidth]{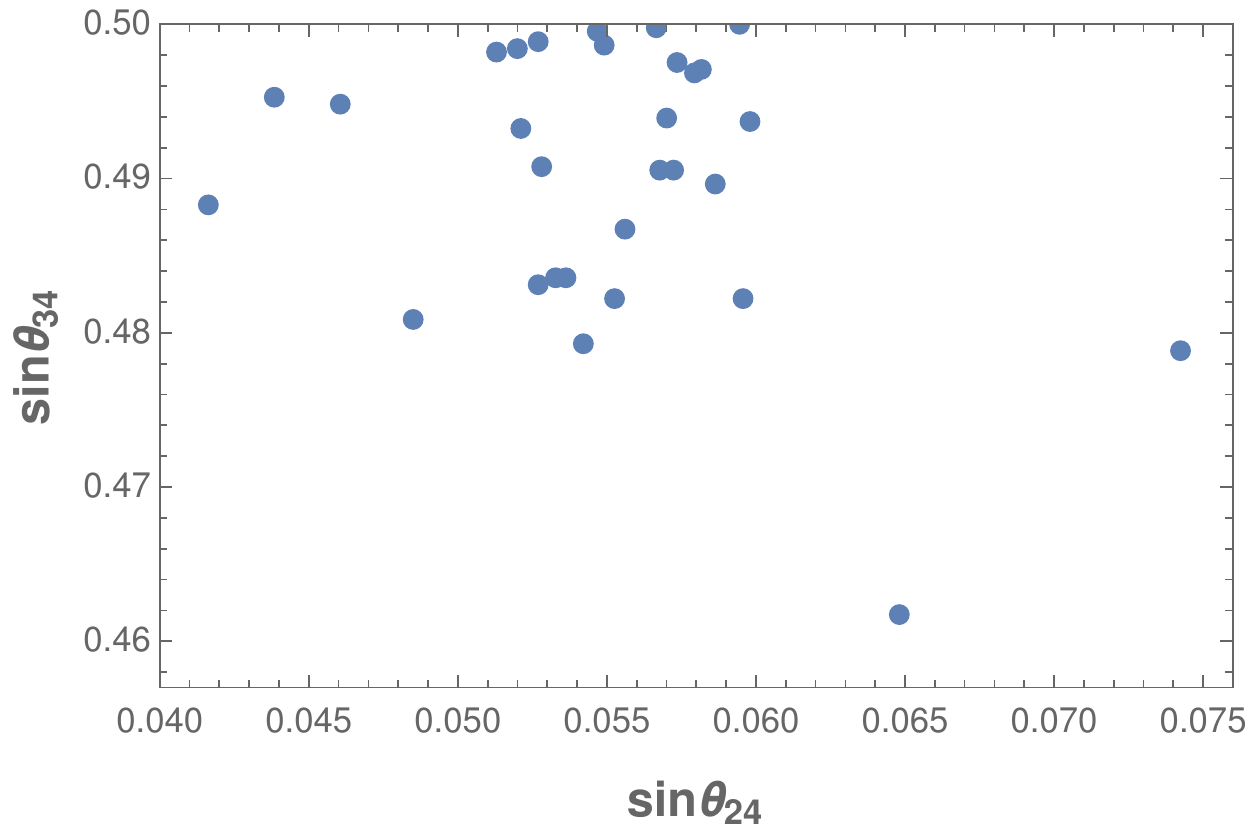}
    \includegraphics[width=0.45\textwidth]{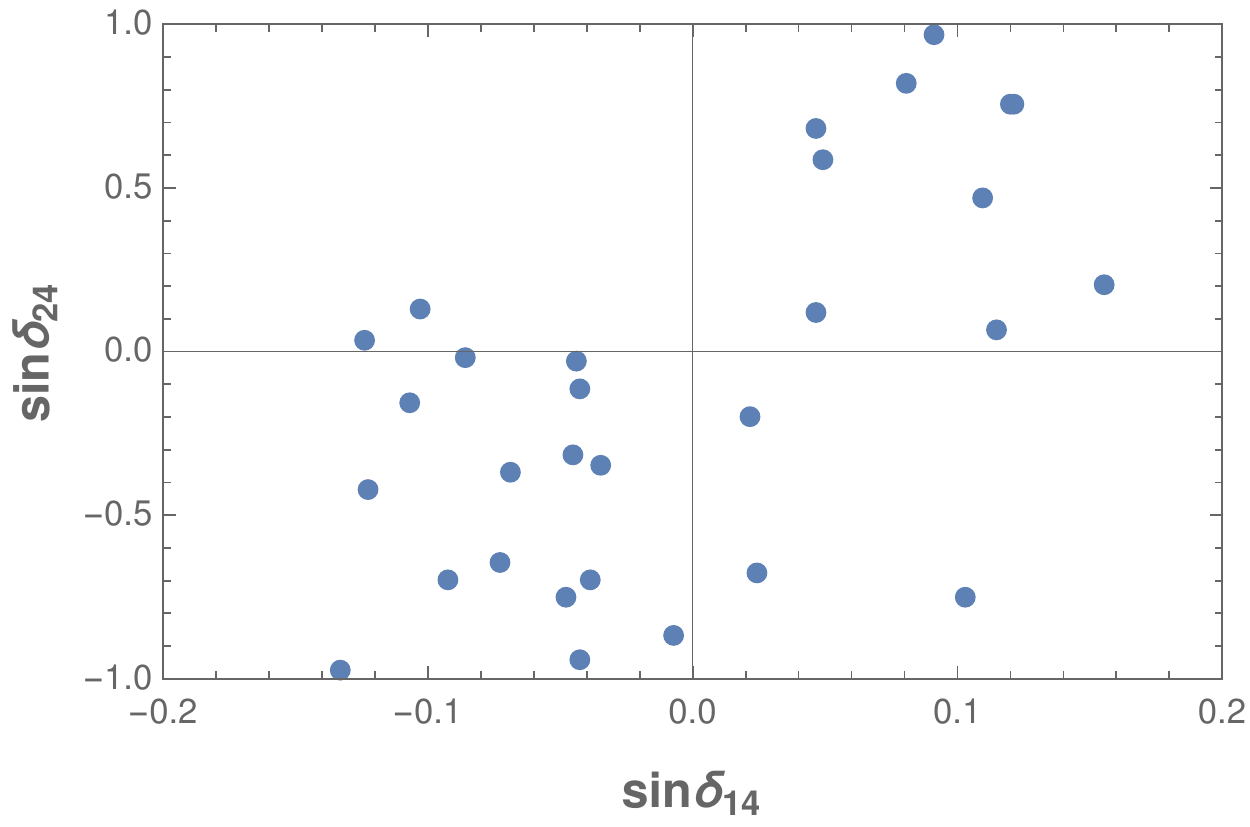} \\
    \includegraphics[width=0.45\textwidth]{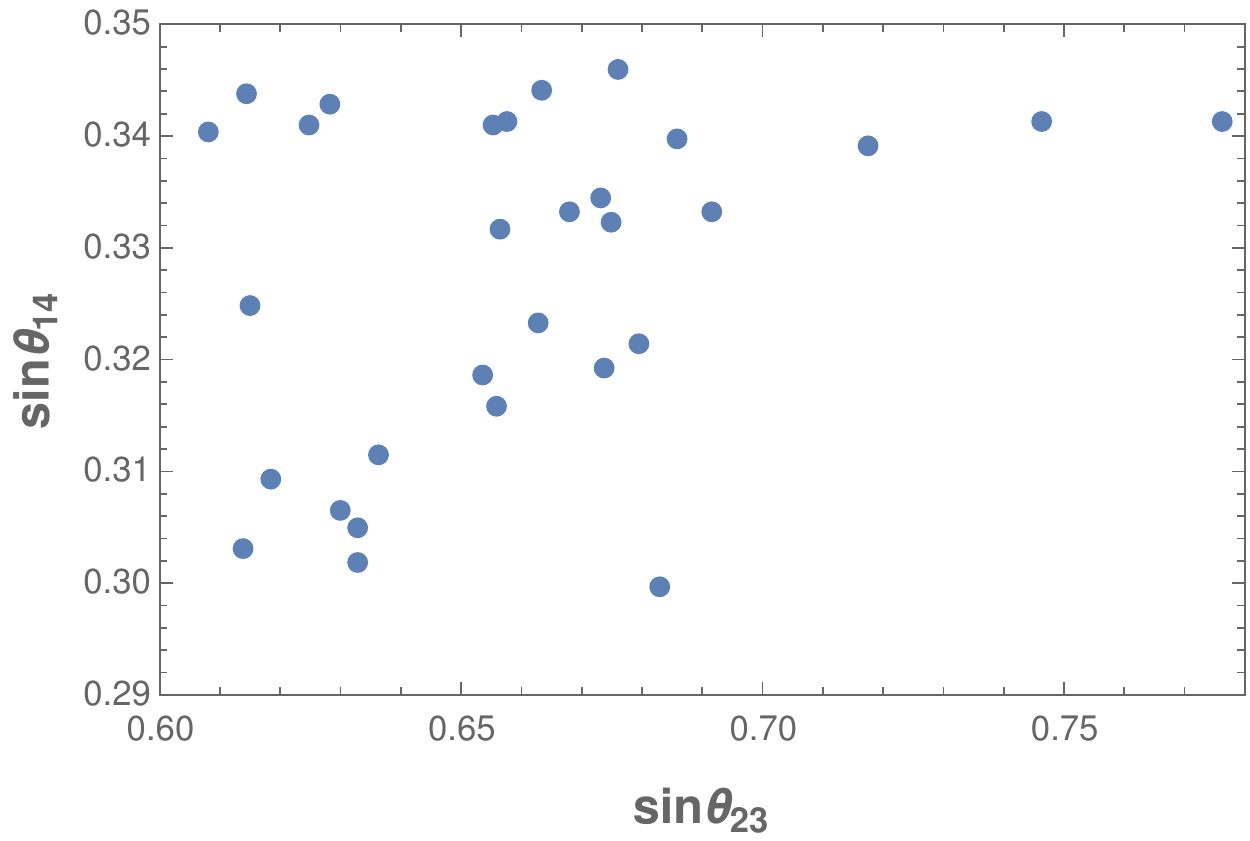}

   \caption{Neutrino oscillation parameters in active-sterile sector for case (x) from texture 1 zero category for NH.}
    \label{fig5}
 \end{figure}

Among the one zero texture category, only two subclasses namely (ix), (x) with NH are allowed. The subclass (ix) has three complex constraints $M_{\mu\tau} = 0, M_{e \mu} = -M_{\mu\mu}, M_{e \tau} = -M_{\tau \tau}$ which indeed give rise to six real equations. These six real equations are solved simultaneously for six unknown parameters: three active-sterile mixing angles $(\theta_{14}, \theta_{24}, \theta_{34})$ and three CP violating Majorana phases $(\alpha, \beta, \gamma)$. We took the lightest neutrino mass $m_{lightest}=m_{1}$ to be zero as before and varied the three active neutrino mixing angles $(\theta_{12}, \theta_{13}, \theta_{23})$, three mass squared differences $(\Delta m_{21}^{2}, \Delta m_{31}^{2}, \Delta m_{41}^{2})$ and three Dirac CP violating phases $(\delta_{13}, \delta_{14}, \delta_{24})$ in their respective $3 \sigma$ global fit ranges. The solution of the six real equations obtained like this give rise to a few correlation plots which are shown in figure \ref{fig4}. Now, for the subclass (x) under texture 1 zero category, we again have three complex constraints $M_{\mu\tau} = 0, M_{e \mu} = M_{\mu\mu}, M_{e \tau} = M_{\tau \tau}$ which give rise to six real equations. The corresponding correlation plots obtained from the solutions of these equations are shown in figure \ref{fig5}.

\begin{figure}
  \centering

    \includegraphics[width=0.45\textwidth]{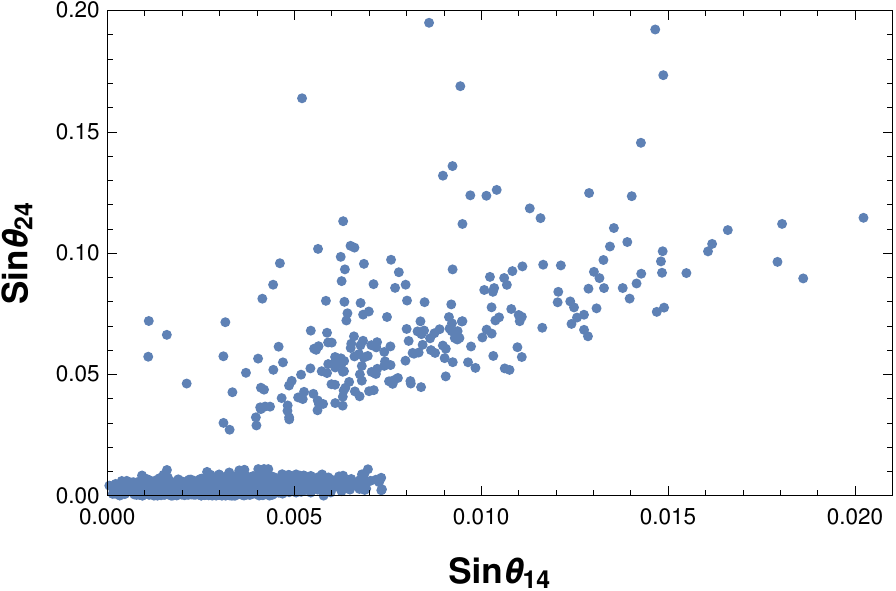}
    \includegraphics[width=0.45\textwidth]{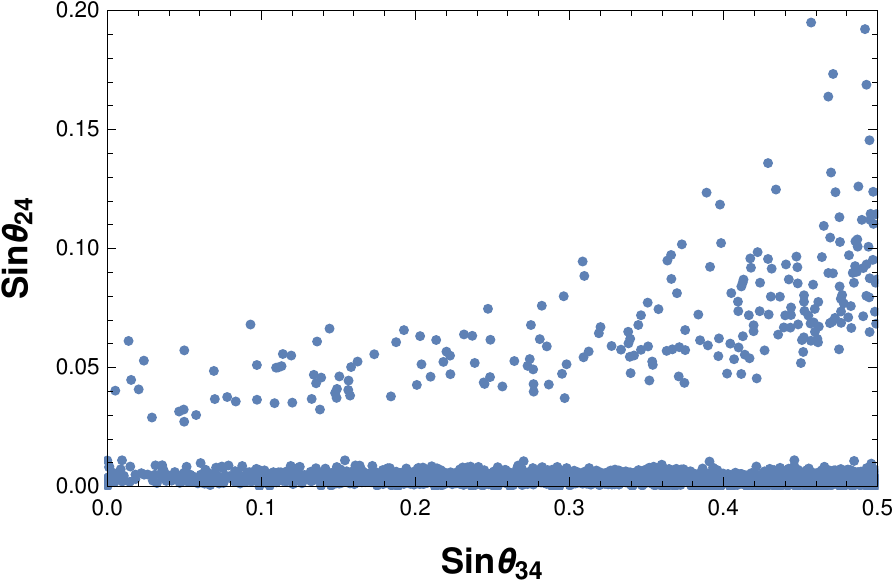} \\
    \includegraphics[width=0.45\textwidth]{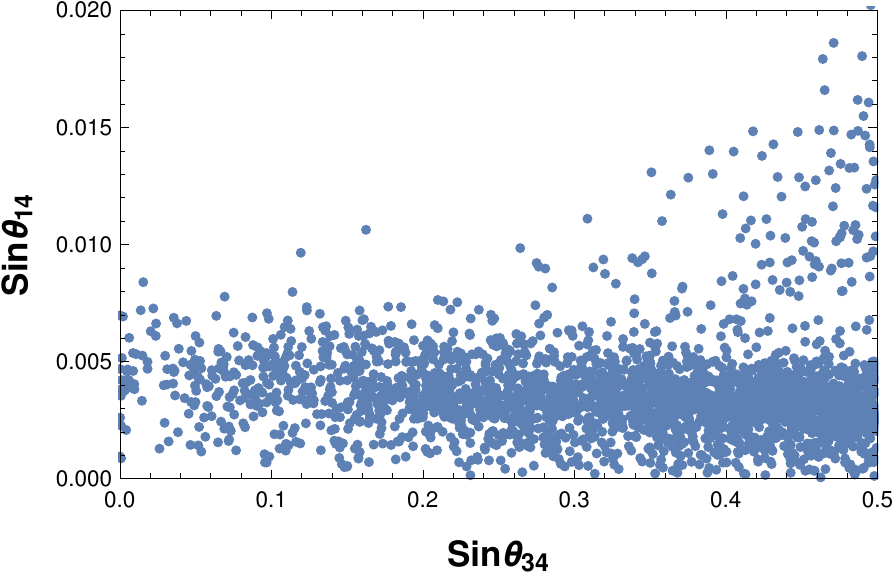}
    \includegraphics[width=0.45\textwidth]{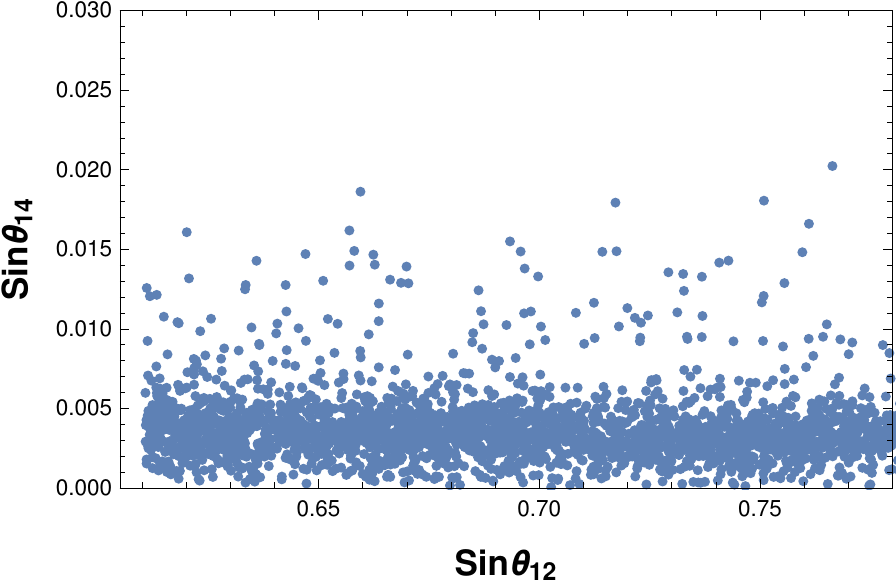} \\
    \includegraphics[width=0.45\textwidth]{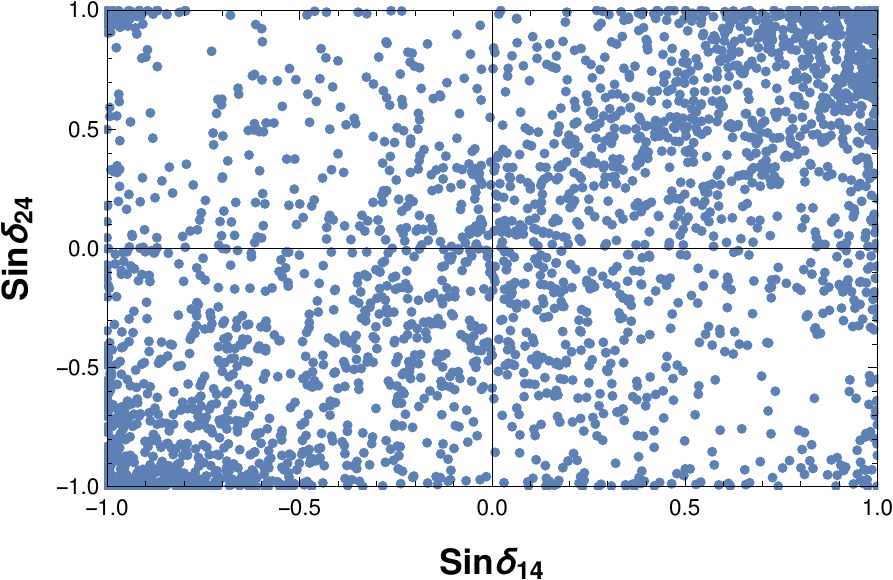}

   \caption{Neutrino oscillation parameters in active-sterile sector for case (i) from texture 2 zero category for NH.}
    \label{fig6}
 \end{figure}

\begin{figure}
  \centering

    \includegraphics[width=0.45\textwidth]{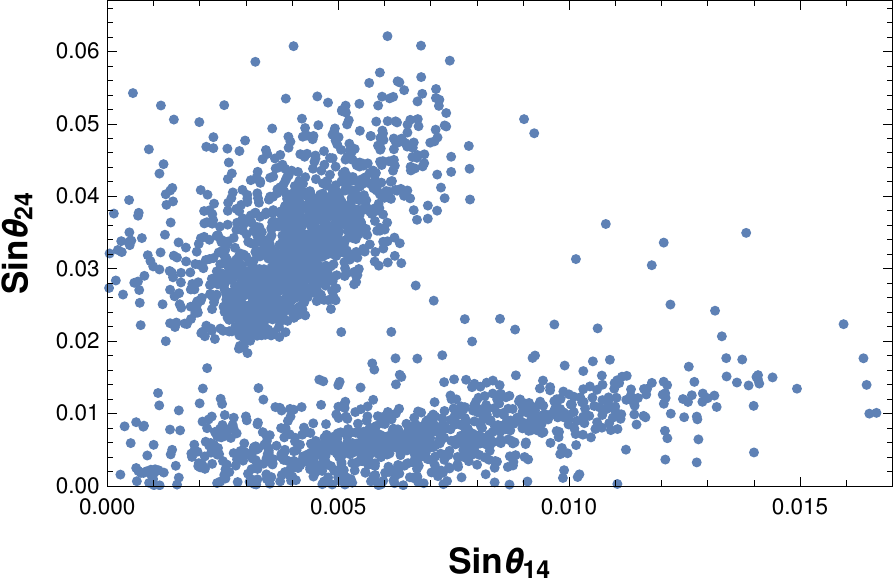}
    \includegraphics[width=0.45\textwidth]{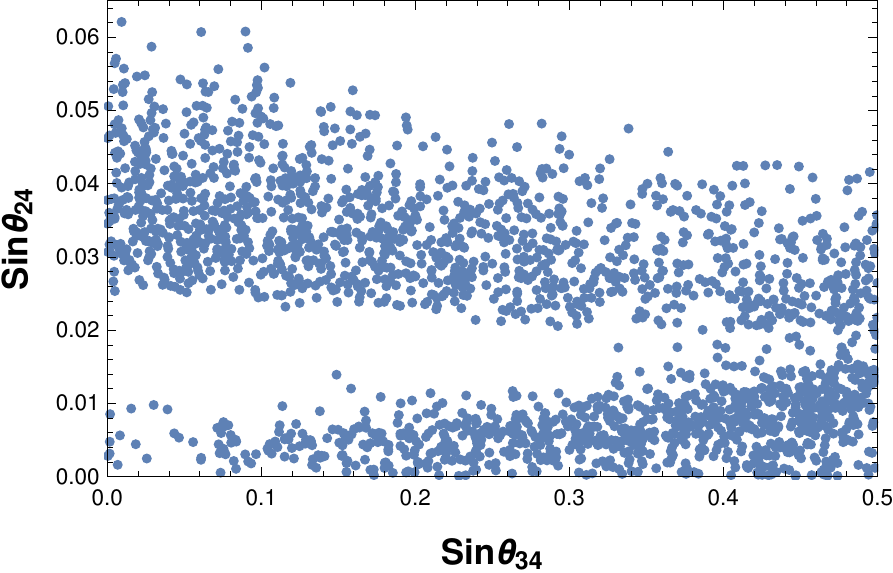} \\
    \includegraphics[width=0.45\textwidth]{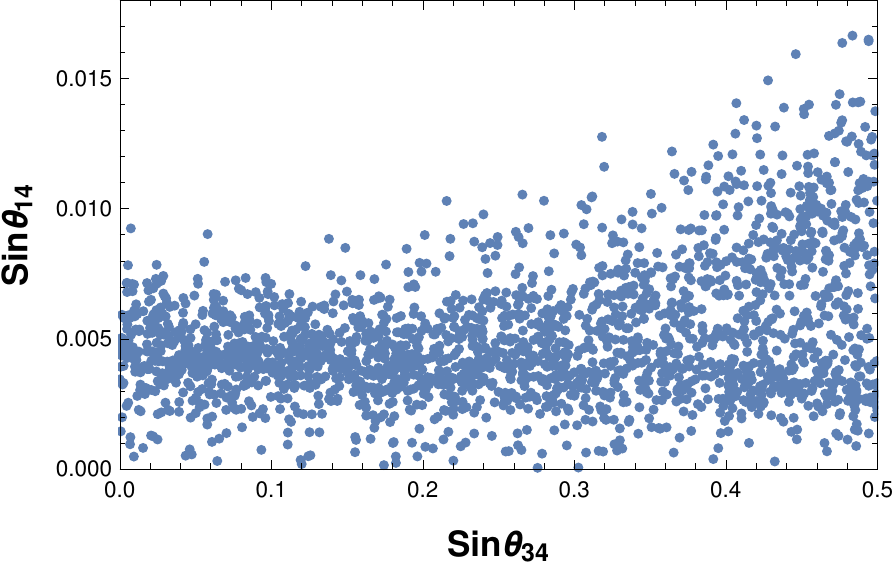}
    \includegraphics[width=0.45\textwidth]{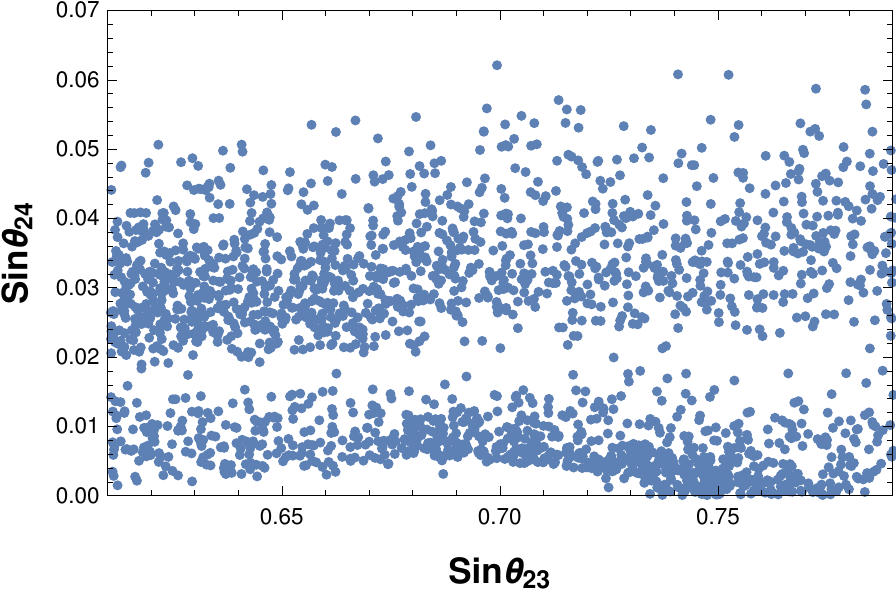} \\
    \includegraphics[width=0.45\textwidth]{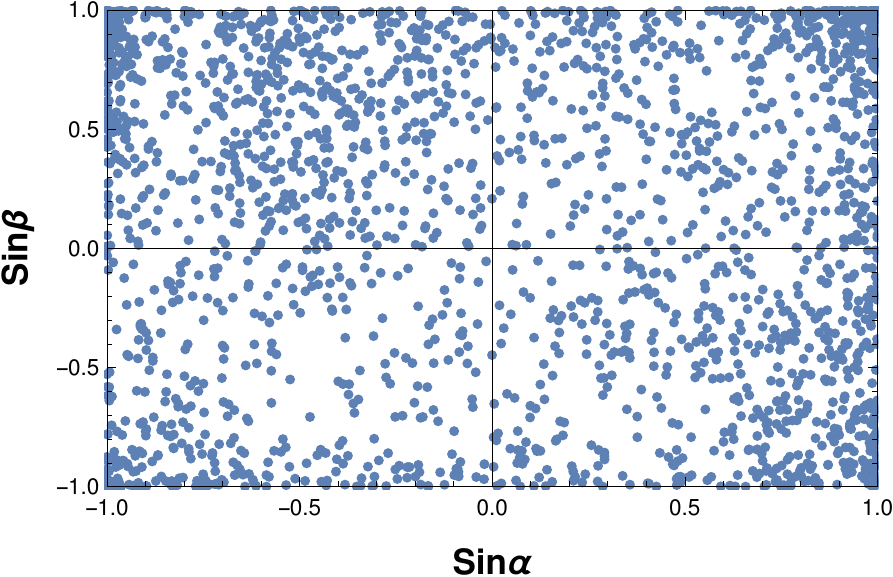}
    
     \caption{Neutrino oscillation parameters in active-sterile sector for case (ii) from texture 2 zero category for NH.}
      \label{fig7}
   \end{figure}

\begin{figure}
  \centering

    \includegraphics[width=0.45\textwidth]{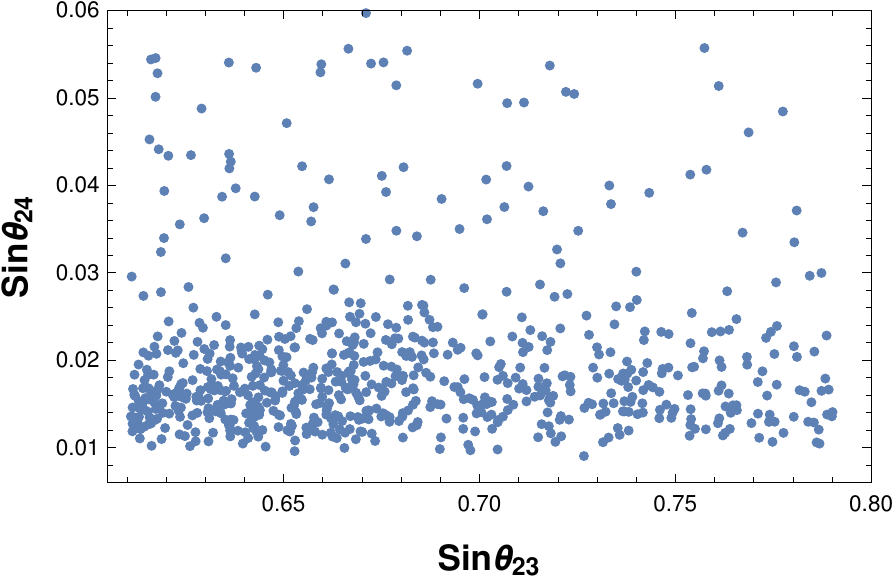}
    \includegraphics[width=0.45\textwidth]{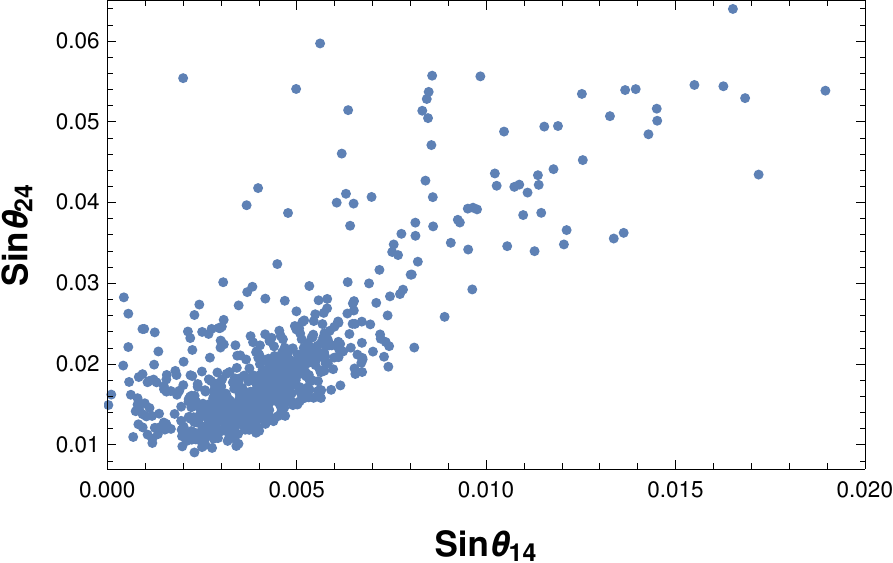} \\
    \includegraphics[width=0.45\textwidth]{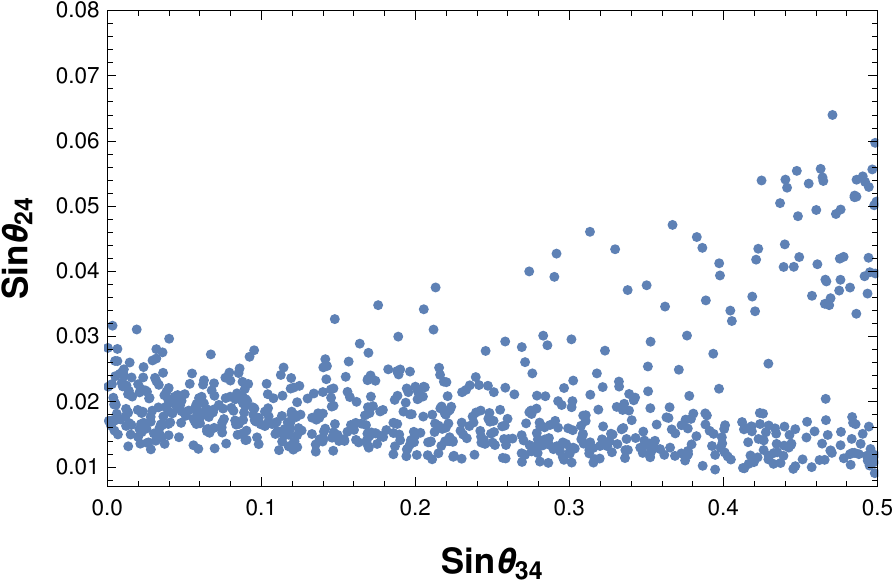}
    \includegraphics[width=0.45\textwidth]{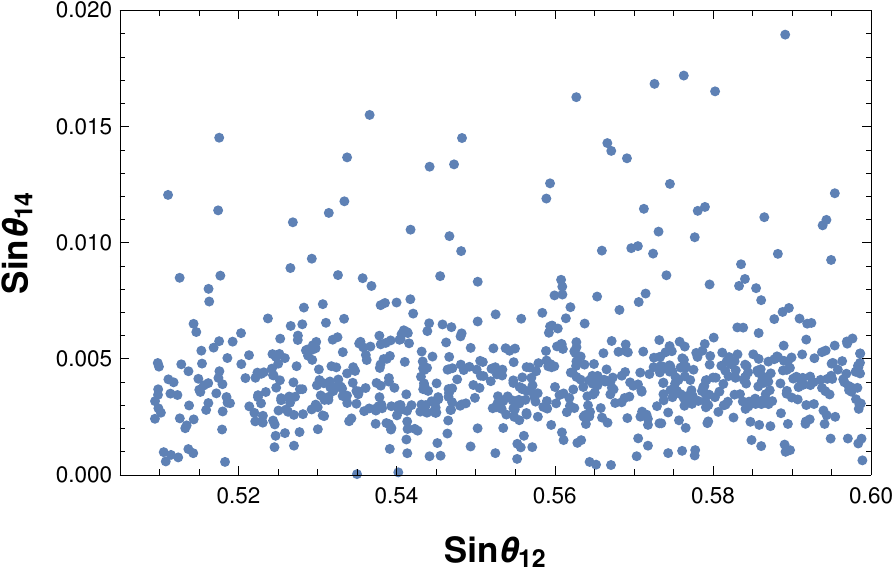} \\
    \includegraphics[width=0.45\textwidth]{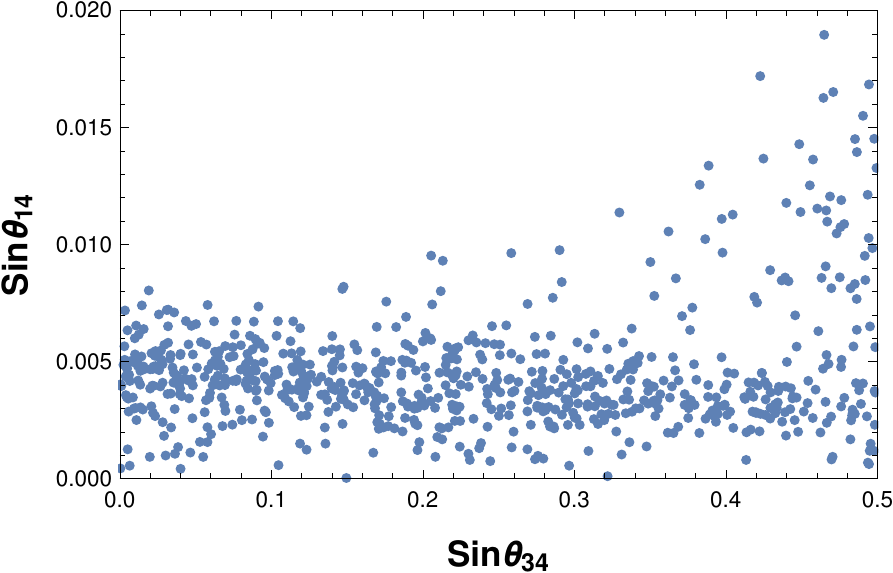}
    \includegraphics[width=0.45\textwidth]{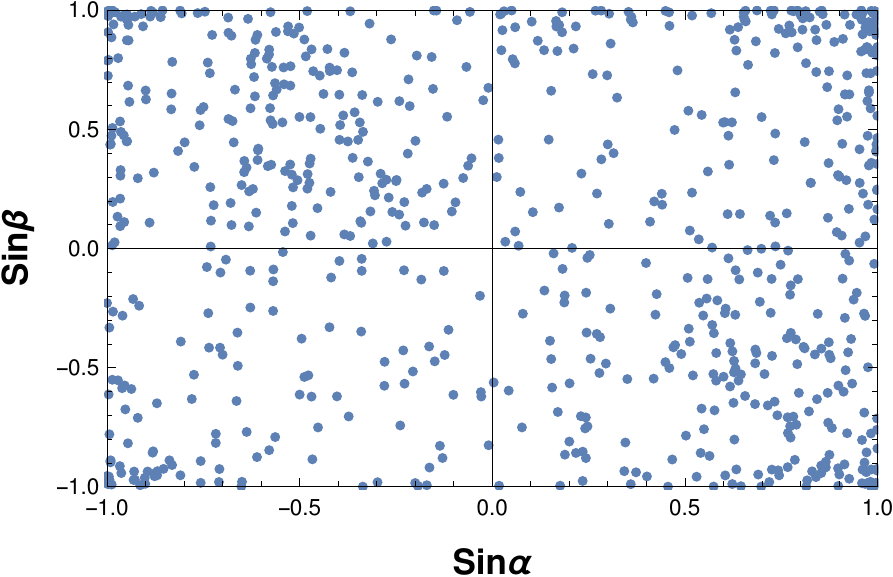} \\

     \caption{Neutrino oscillation parameters in active-sterile sector for texture 3 zero case for NH.}
     \label{fig8}
\end{figure}

Similar numerical analysis was done for two zero and three zero texture subclasses as well. Among the two zero texture subclasses only two namely, (i), (ii) with NH were found to be allowed. The subclass (i) has three complex constraints $M_{e\mu} = 0, M_{e \tau} = 0, M_{\mu \mu} = M_{\mu \tau}$ while subclass (ii) has $M_{e\mu} = 0 \quad M_{e \tau} = 0 \quad M_{\mu \mu} = -M_{\mu \tau}$. The correlations corresponding to the solutions for subclass (i), (ii) are shown in figure \ref{fig6}, \ref{fig7} respectively. Similarly, the three zero texture has the three complex constraint equations $M_{e\mu}= 0, M_{e\tau}= 0, M_{\mu\tau}=0$ and the corresponding correlations are shown in figure \ref{fig8}. As can be seen from these plots, we get more allowed solutions for two zero and three zero texture cases out of one million iterations compared to what we had obtained for $\mu-\tau$ symmetric and one zero texture cases.


 
 
\section{Conclusion}
 \label{sec:conclude}
 To summarise, we have studied the viability of different possible textures in light neutrino mass matrix within the framework of $3+1$ light neutrino scenario by considering a $A_4$ flavour symmetric minimal extended seesaw mechanism. While the minimal extended seesaw mechanism naturally explains $3+1$ light neutrino scenario in an economical way predicting the lightest neutrino to be massless, presence of the $A_4$ flavour symmetry dictates the flavour structure of the $4\times 4$ light neutrino mass matrix. In addition to that, an additional discrete symmetry $Z_4$ is also chosen in order to forbid certain unwanted terms from the Lagrangian. Considering generic $A_4$ flavon alignments where a triplet flavon acquires VEV like $\langle \phi \rangle = v (n_1, n_2, n_3), n_i \in (-1, 0, 1)$, we first consider all possible combinations of such alignments and find the analytical form of the light neutrino mass matrix for each such case. For two triplet flavons taking part in generating the light neutrino mass matrix, while the other triplet alignment is kept fixed for diagonal charged lepton mass matrix, we get $1 \times 27 \times 27 = 729$ possible cases. Based on previous studies on $3+1$ neutrino textures, we first point out the disallowed textures out of these 729 mass matrices and discarded these 225 mass matrices from our analysis. From the remaining cases, we classify 96 of them as one zero texture, 64 as two zero texture, 8 as three zero texture and 296 of them as hybrid textures (which do not contain any zeros). The remaining 40 mass matrices correspond to an interesting category where the $3\times3$ active neutrino block of the $3+1$ light neutrino mass matrix possess $\mu-\tau$ symmetry whereas the active-sterile block breaks it explicitly.

We then analyse the mass matrices with texture zeros and $\mu-\tau$ symmetry by numerically solving the constraint equations in each case and comparing the resulting solution with the $3+1$ neutrino data for consistency. Although there are large number of mass matrices for each such cases, they belong to a smaller number of subclasses where each subclass is a group of mass matrices giving rise to the same set of constraint equations. The number of such subclasses is 4, 12, 8 and 1 for $\mu-\tau$ symmetric, one zero, two zero and three zero texture mass matrices respectively. We therefore numerically solved the constraint equations for these 25 cases in total. We find that only 8 out of these 25 subclasses are allowed by the $3+1$ global fit data and all of them have normal hierarchical pattern of light neutrino masses. Out of these 8 allowed subclasses, 3 belong to the $\mu-\tau$ symmetric category, 2 belong to the one zero texture category, 2 belong to the two zero texture category and 1 belong to the remaining three zero texture category. We also find interesting correlation plots between different light neutrino parameters for each of these allowed subclasses. The final results are summarised in the table \ref{table2}. Compared to the usual texture zero scenarios discussed in previous works, the textures in the present scenario are more constrained due to additional constraints apart from texture zero conditions or $\mu-\tau$ symmetry alone and the requirement of vanishing lightest neutrino mass. This is reflected in our results of getting only 8 out of 25 subclasses studied numerically.

While the fate of an additional light neutrino having mass around the eV scale is yet to be confirmed by other neutrino experiments, our analysis show how difficult it is to realise such a scenario in the minimal extended seesaw if $A_4$ flavour symmetry with generic vacuum alignment is present. If the existence of such light sterile neutrino gets well established later, the predictions for unknown neutrino parameters obtained in our analysis can be tested for further scrutiny of the model, in a way similar to \cite{DUNEtexture} where the possibility of probing texture zeros in three neutrino scenarios at neutrino oscillation experiments was studied.
 
 \begin{table}
\begin{center}
\begin{tabular}{| c | c | c | c |} \hline
Texture & Subclass & Normal Hierarchy  & Inverted Hierarchy\\ \hline 
\multirow{5}{4cm}{ \begin{center} $(\mu-\tau)$ symmetric \end{center}}
    & (i)           &  $\times$   &  $\times$   \\
    & (ii)           & $\surd$    & $\times$  \\
    & (iii)            & $\surd$     & $\times$   \\
    & (iv)            &   $\surd$  &  $\times$ \\ \hline
\multirow{12}{4cm}{\begin{center}  One zero \end{center} }
    & (i)           & $\times$    &   $\times$  \\
    & (ii)           & $\times$    & $\times$ \\
    & (iii)            & $\times$     & $\times$    \\
    & (iv)            & $\times$   & $\times$     \\
     & (v)            & $\times$    & $\times$  \\
      & (vi)            & $\times$    & $\times$  \\
       & (vii)            & $\times$   & $\times$    \\
        & (viii)            & $\times$   & $\times$  \\
    & (ix)            & $\surd$      & $\times$ \\ 
      & (x)            & $\surd$       & $\times$  \\ 
        & (xi)            & $\times$    & $\times$   \\ 
         & (xii)            & $\times$   &  $\times$    \\  \hline
\multirow{8}{4cm}{\begin{center} Two zero \end{center} }
    & (i)           & $\surd$      &  $\times$ \\
    & (ii)           & $\surd$      & $\times$ \\
    & (iii)            & $\times$    &  $\times$   \\
    & (iv)            & $\times$      & $\times$ \\ 
    & (v)            & $\times$   &  $\times$   \\
    & (vi)            & $\times$   &  $\times$  \\
    & (vii)            & $\times$   & $\times$  \\
    & (viii)            & $\times$   & $\times$   \\\hline
 Three zero & $(i)$  & $\surd$  & $\times$ \\ \hline
\end{tabular}
\caption{Table showing allowed and disallowed texture subclasses which have been analysed numerically. Here $(\surd)$ indicates allowed cases and $(\times)$ indicates disallowed cases.}
\label{table2}
\end{center}
\end{table}

\begin{acknowledgments}
NS and KB would like to acknowledge the facilities provided by IIT Guwahati during academic visits when this work was initiated. NS is also thankful to the organisers of Nu HoRIzons VII at HRI Allahabad (February 21-23, 2018) for considering a preliminary version of this work to be presented as a poster. KB would like to thank Gauhati University, to support her visit to Physics Deptt, IIT Guwahati, during August 2 - September 30, 2017, as visiting scientist.

\end{acknowledgments}

\appendix
\section{$A_4$ product rules}
\label{appen1}
$A_4$, the symmetry group of a tetrahedron, is a discrete non-abelian group of even permutations of four objects. It has four irreducible representations: three one-dimensional and one three dimensional which are denoted by $\bf{1}, \bf{1'}, \bf{1''}$ and $\bf{3}$ respectively, being consistent with the sum of square of the dimensions $\sum_i n_i^2=12$. Their product rules are given as
$$ \bf{1} \otimes \bf{1} = \bf{1}$$
$$ \bf{1'}\otimes \bf{1'} = \bf{1''}$$
$$ \bf{1'} \otimes \bf{1''} = \bf{1} $$
$$ \bf{1''} \otimes \bf{1''} = \bf{1'}$$
$$ \bf{3} \otimes \bf{3} = \bf{1} \otimes \bf{1'} \otimes \bf{1''} \otimes \bf{3}_a \otimes \bf{3}_s $$
where $a$ and $s$ in the subscript corresponds to anti-symmetric and symmetric parts respectively. Denoting two triplets as $(a_1, b_1, c_1)$ and $(a_2, b_2, c_2)$ respectively, their direct product can be decomposed into the direct sum mentioned above as
$$ \bf{1} \backsim a_1a_2+b_1c_2+c_1b_2$$
$$ \bf{1'} \backsim c_1c_2+a_1b_2+b_1a_2$$
$$ \bf{1''} \backsim b_1b_2+c_1a_2+a_1c_2$$
$$\bf{3}_s \backsim (2a_1a_2-b_1c_2-c_1b_2, 2c_1c_2-a_1b_2-b_1a_2, 2b_1b_2-a_1c_2-c_1a_2)$$
$$ \bf{3}_a \backsim (b_1c_2-c_1b_2, a_1b_2-b_1a_2, c_1a_2-a_1c_2)$$

\newpage

\section{\small{Vacuum alignment of flavon fields $\phi'$, $\phi''$ of allowed cases. VEV of the flavon field $\phi$, $\langle \phi \rangle = (1,0,0)$ has been used.}}
\label{appen2}
\begin{table}[htbp]
\tiny
\begin{minipage}[b]{0.20\linewidth}\centering
\begin{tabular}{| c | c | c |}
\hline
Sl no. & $\phi^{'}$ & $\phi^{''}$ \\
\hline\hline
1.    &   $(0,1,-1)$    &  $(0,1,-1)$ \\
2.       &  $(0,-1,1)$  & $(0,-1,1)$ \\
3.     &  $(1,0,-1)$  & $(1,0,-1)$ \\
4.    &   $(1,-1,0)$   & $(1,-1,0)$ \\
5.   &   $(-1,0,1)$   & $(-1,0,1)$ \\
6.  & $(-1,1,0)$   & $(-1,1,0)$ \\
7.  & $(0,1,-1)$   & $(0,-1,1)$ \\
8.    & $(0,1,-1)$   & $(1,0,-1)$ \\
9.     & $(0,1,-1)$   & $(-1,0,1)$ \\
10.   & $(0,-1,1)$   & $(0,1,-1)$ \\
11.   &  $(0,-1,1)$   & $(1,0,-1)$ \\
12.     & $(0,-1,1)$   & $(-1,0,1)$ \\
13.   &  $(1,0,-1)$   & $(1,-1,0)$ \\
14.     &  $(1,0,-1)$   & $(-1,0,1)$ \\
15.    & $(1,0,-1)$   & $(-1,1,0)$ \\
16.    &   $(1,-1,0)$   & $(0,1,-1)$ \\
17.    &   $(1,-1,0)$   & $(0,-1,1)$ \\
18.   &  $(1,-1,0)$   & $(-1,1,0)$ \\
19.    & $(-1,0,1)$   & $(1,0,-1)$ \\
20.    & $(-1,0,1)$   & $(1,-1,0)$ \\
21.   & $(-1,0,1)$   & $(-1,1,0)$ \\
22.   & $(-1,1,0)$   & $(0,1,-1)$ \\
23.    &  $(-1,1,0)$   & $(0,-1,1)$ \\
24.     &  $(-1,1,0)$   & $(1,-1,0)$ \\
25.   &  $(0,1,1)$   & $(0,1,1)$ \\
26.    & $(0,1,1)$   & $(0,1,-1)$ \\
27.    & $(0,1,1)$   & $(0,-1,1)$ \\
28.  & $(0,1,1)$  &  $(0,-1,-1)$  \\
29.   & $(0,1,1)$  &  $(1,0,1)$   \\
30.   & $(0,1,1)$  &  $(1,0,-1)$   \\
31.   & $(0,1,1)$  &  $(-1,0,1)$  \\
32.   & $(0,1,1)$  &  $(-1,0,-1)$ \\
33.   & $(0,1,-1)$  &  $(0,1,1)$ \\
34.    & $(0,1,-1)$  &  $(0,-1,-1)$   \\
35.   & $(0,1,-1)$  &  $(1,0,1)$   \\
36.   & $(0,1,-1)$  &  $(-1,0,-1)$   \\
37.   & $(0,-1,1)$  &  $(0,1,1)$   \\
38.   & $(0,-1,1)$  &  $(0,-1,-1)$   \\
39.   & $(0,-1,1)$  &  $(1,0,1)$   \\
40.   & $(0,-1,-1)$  &  $(0,1,1)$   \\
41.   & $(0,-1,-1)$  &  $(0,1,-1)$   \\
42.   & $(0,-1,-1)$  &  $(0,-1,1)$   \\
43.   & $(0,-1,-1)$  &  $(0,-1,-1)$   \\
44.   & $(0,-1,-1)$  &  $(1,0,1)$   \\
45.   & $(0,-1,-1)$  &  $(1,0,-1)$   \\
46.   & $(0,-1,-1)$  &  $(-1,0,1)$   \\
47.   & $(0,-1,-1)$  &  $(-1,0,-1)$   \\
48.   & $(1,0,1)$  &  $(1,0,1)$   \\[1ex]\hline
\end{tabular}
\caption{Texture one zero.}
\end{minipage}
\hspace{0.4cm}
\begin{minipage}[b]{0.20\linewidth}\centering
\begin{tabular}{| c | c | c |}
\hline
Sl no. & $\phi^{'}$ & $\phi^{''}$ \\
\hline\hline
49.  & $(1,0,1)$  &  $(1,0,-1)$   \\
50.  & $(1,0,1)$  &  $(1,1,0)$   \\ 
51.    & $(1,0,1)$  &  $(1,-1,0)$ \\
52.   & $(1,0,1)$  &  $(-1,0,1)$   \\
53.   & $(1,0,1)$  &  $(-1,1,0)$   \\
54.   & $(1,0,1)$  &  $(-1,-1,0)$   \\
55.   & $(1,0,1)$  &  $(-1,0,-1)$   \\
56.   & $(1,0,-1)$  &  $(1,0,1)$   \\
57.   & $(1,0,-1)$  &  $(1,1,0)$   \\
58.    & $(1,0,-1)$  &  $(-1,0,-1)$   \\
59.    & $(1,0,-1)$  &  $(-1,-1,0)$   \\
60.   & $(1,1,0)$  &  $(0,1,1)$   \\
61.   & $(1,1,0)$  &  $(0,1,-1)$   \\
62.    & $(1,1,0)$  &  $(0,-1,1)$   \\
63.     & $(1,1,0)$  &  $(0,-1,-1)$   \\
64.    & $(1,1,0)$  &  $(1,1,0)$   \\
65.   & $(1,1,0)$  &  $(1,-1,0)$ \\
66.   & $(1,1,0)$  &  $(-1,1,0)$   \\
67.  & $(1,1,0)$  &  $(-1,-1,0)$   \\
68.   & $(1,-1,0)$  &  $(0,1,1)$   \\
69.   & $(1,-1,0)$  &  $(0,-1,-1)$   \\
70.   & $(1,-1,0)$  &  $(1,1,0)$   \\
71.   & $(-1,0,1)$  &  $(1,0,1)$   \\
72.   & $(-1,0,1)$  &  $(1,1,0)$   \\
73.   & $(-1,0,1)$  &  $(-1,0,-1)$   \\
74.   & $(-1,0,1)$  &  $(-1,-1,0)$   \\
75.   & $(-1,0,-1)$  &  $(1,0,1)$   \\
76.   & $(-1,0,-1)$  &  $(1,0,-1)$   \\
77.   & $(-1,0,-1)$  &  $(1,1,0)$   \\
78.   & $(-1,0,-1)$  &  $(1,-1,0)$   \\
79.   & $(-1,0,-1)$  &  $(-1,0,-1)$   \\
80.   & $(-1,0,-1)$  &  $(-1,1,0)$   \\
81.  & $(-1,0,-1)$  &  $(-1,-1,0)$   \\
82.  & $(-1,1,0)$  &  $(0,-1,-1)$   \\
83.  & $(-1,1,0)$  &  $(1,1,0)$   \\
84.   & $(-1,1,0)$  &  $(-1,-1,0)$   \\
85.  & $(-1,-1,0)$  &  $(0,1,1)$   \\
86.   & $(-1,-1,0)$  &  $(0,1,-1)$   \\
87.   & $(-1,-1,0)$  &  $(0,-1,1)$   \\
88.   & $(-1,-1,0)$  &  $(0,-1,-1)$   \\
89.  & $(-1,-1,0)$  &  $(-1,1,0)$   \\
90.   & $(-1,-1,0)$  &  $(-1,-1,0)$   \\
91.   & $(-1,-1,0)$  &  $(1,1,0)$   \\
92.   & $(-1,-1,0)$  &  $(1,-1,0)$   \\
93.   & $(-1,1,0)$  &  $(0,1,1)$   \\
94.  & $(0,-1,1)$  & $(-1,0,-1)$ \\
95.  & $(1,-1,0)$  & $(1,1,0)$ \\
96   & $(1,-1,0)$  & $(-1,-1,0)$ \\[1ex] \hline
\end{tabular}
\caption{Texture one zero.}
\end{minipage}
\hspace{0.4cm}
\begin{minipage}[b]{0.20\linewidth}\centering
\begin{tabular}{| c | c | c |}
\hline
Sl no.  & $\phi^{'}$ & $\phi^{''}$ \\
\hline\hline
1.  &   $(0,0,1)$    &  $(0,0,1)$ \\
2.   &   $(0,0,1)$    &  $(0,0,-1)$ \\
3. &   $(0,0,-1)$    &  $(0,0,1)$ \\
4.  &   $(0,0,-1)$    &  $(0,0,-1)$ \\
5.  &   $(1,0,0)$    &  $(0,1,0)$ \\
6.  &   $(1,0,0)$    &  $(0,-1,0)$ \\
7.  &   $(-1,0,0)$    &  $(0,1,0)$ \\
8.  &   $(-1,0,0)$    &  $(0,-1,0)$ \\[1ex]
\hline
\end{tabular}
\caption{Texture three zero.}
\end{minipage}
\hspace{0.4cm}
\begin{minipage}[b]{0.20\linewidth}\centering
\begin{tabular}{| c | c | c |}
\hline
Sl no. & $\phi^{'}$ & $\phi^{''}$ \\
\hline\hline
1.   &   $(0,0,1)$    &  $(0,1,1)$ \\
2.   &   $(0,0,1)$    &  $(0,1,-1)$ \\
3.  &   $(0,0,1)$    &  $(0,-1,1)$ \\
4.  &   $(0,0,1)$    &  $(0,-1,-1)$ \\
5.  &   $(0,0,1)$    &  $(1,0,1)$ \\
6.  &   $(0,0,1)$    &  $(1,0,-1)$ \\
7.  &   $(0,0,1)$    &  $(-1,0,1)$ \\
8.  &   $(0,0,1)$    &  $(-1,0,-1)$ \\
9.  &   $(0,0,-1)$    &  $(0,1,1)$ \\
10.  &   $(0,0,-1)$    &  $(0,1,-1)$ \\
11.  &   $(0,0,-1)$    &  $(0,-1,1)$ \\
12.  &   $(0,0,-1)$    &  $(0,-1,-1)$ \\
13.  &   $(0,0,-1)$    &  $(1,0,1)$ \\
14.  &   $(0,0,-1)$    &  $(1,0,-1)$ \\
15.  &   $(0,0,-1)$    &  $(-1,0,1)$ \\
16.  &   $(0,0,-1)$    &  $(-1,0,-1)$ \\
17.  &   $(0,1,1)$    &  $(0,0,1)$ \\
18.  &   $(0,1,1)$    &  $(0,0,-1)$ \\
19.  &   $(0,1,-1)$    &  $(0,0,1)$ \\
20.  &   $(0,1,-1)$    &  $(0,0,-1)$ \\
21.  &   $(0,-1,1)$    &  $(0,0,1)$ \\
22.  &   $(0,-1,1)$    &  $(0,0,-1)$ \\
23.   &   $(0,-1,-1)$    &  $(0,0,1)$ \\
24.   &   $(0,-1,-1)$    &  $(0,0,-1)$ \\
25.   &   $(1,0,0)$    &  $(0,1,1)$ \\
26.   &   $(1,0,0)$    &  $(0,1,-1)$ \\
27.   &   $(1,0,0)$    &  $(0,-1,1)$ \\
28.   &   $(1,0,0)$    &  $(0,-1,-1)$ \\
29.   &   $(1,0,0)$    &  $(1,1,0)$ \\
30.   &   $(1,0,0)$    &  $(1,-1,0)$ \\
31.   &   $(1,0,0)$    &  $(-1,1,0)$ \\
32.   &   $(1,0,0)$    &  $(-1,-1,0)$ \\[1ex]\hline
\end{tabular}
\caption{Texture two zero.}
\end{minipage}
\end{table}

\begin{table}[ht]
\tiny
\begin{minipage}[b]{0.20\linewidth}\centering
\begin{tabular}{| c | c | c |}
\hline
Sl no. & $\phi^{'}$ & $\phi^{''}$ \\
\hline\hline
33.    &   $(1,0,1)$    &  $(0,0,1)$ \\
34.   &   $(1,0,1)$    &  $(0,0,-1)$ \\
35.  &   $(1,0,1)$    &  $(0,1,0)$ \\
36.  &   $(1,0,1)$    &  $(0,-1,0)$ \\
37.  &   $(1,0,-1)$    &  $(0,0,1)$ \\
38.   &   $(1,0,-1)$    &  $(0,0,-1)$ \\
39.   &   $(1,0,-1)$    &  $(0,1,0)$ \\
40.    &   $(1,0,-1)$    &  $(0,-1,0)$ \\
41.   &   $(1,1,0)$    &  $(0,1,0)$ \\
42.   &   $(1,1,0)$    &  $(0,-1,0)$ \\
43.    &   $(1,-1,0)$    &  $(0,1,0)$ \\
44.   &   $(1,-1,0)$    &  $(0,-1,0)$ \\
45.  &   $(-1,0,0)$    &  $(0,1,1)$ \\
46.   &   $(-1,0,0)$    &  $(0,1,-1)$\\
47.   &   $(-1,0,0)$    &  $(0,-1,1)$\\
48.  &   $(-1,0,0)$    &  $(0,-1,-1)$\\
49.   &   $(-1,0,0)$    &  $(1,1,0)$\\
50.   &   $(-1,0,0)$    &  $(1,-1,0)$\\
51.   &   $(-1,0,0)$    &  $(-1,1,0)$\\
52.   &   $(-1,0,0)$    &  $(-1,-1,0)$\\
53.   &   $(-1,0,1)$    &  $(0,0,1)$\\
54.   &   $(-1,0,1)$    &  $(0,0,-1)$\\
55.   &   $(-1,0,1)$    &  $(0,1,0)$\\
56.   &   $(-1,0,1)$    &  $(0,-1,0)$\\
57.   &   $(-1,0,-1)$    &  $(0,0,1)$\\
58.   &   $(-1,0,-1)$    &  $(0,0,-1)$\\
59.   &   $(-1,0,-1)$    &  $(0,1,0)$\\
60.   &   $(-1,0,-1)$    &  $(0,-1,0)$\\
61.   &   $(-1,1,0)$    &  $(0,1,0)$\\
62.   &   $(-1,1,0)$    &  $(0,-1,0)$\\
63.   &   $(-1,-1,0)$    &  $(0,1,0)$\\
64.   &   $(-1,-1,0)$    &  $(0,-1,0)$ \\[1ex]\hline
\end{tabular}
\caption{Texture two zero (contd).}
\end{minipage}
\hspace{0.5cm}
\begin{minipage}[b]{0.20\linewidth}\centering
\begin{tabular}{| c | c | c |}
\hline
Sl no.  & $\phi^{'}$ & $\phi^{''}$ \\
\hline\hline
1.   & $(1,0,-1)$    &  $(0,1,-1)$ \\
2.   &   $(1,0,-1)$    &  $(0,-1,1)$ \\
3.   &   $(-1,0,1)$    &  $(0,1,-1)$ \\
4.  &   $(-1,0,1)$    &  $(0,-1,1)$ \\
5.  &   $(0,0,0)$    &  $(0,1,-1)$ \\
6.  &   $(0,0,0)$    &  $(0,-1,1)$ \\
7.  &   $(0,1,0)$    &  $(0,1,-1)$ \\
8.  &   $(0,1,0)$    &  $(0,-1,1)$ \\
9.  &   $(0,-1,0)$    &  $(0,1,-1)$ \\
10.   &   $(0,-1,0)$    &  $(0,-1,1)$ \\
11.    &   $(1,0,1)$    &  $(0,1,-1)$ \\
12.    &   $(1,0,1)$    &  $(0,-1,1)$ \\
13.   &   $(1,0,-1)$    &  $(0,0,0)$ \\
14.   &   $(1,0,-1)$    &  $(0,1,1)$ \\
15.   &   $(1,0,-1)$    &  $(0,-1,-1)$ \\
16.   &   $(1,0,-1)$    &  $(1,0,0)$ \\
17.   &   $(1,0,-1)$    &  $(1,1,1)$ \\
18.   &   $(1,0,-1)$    &  $(1,-1,-1)$ \\
19.   &   $(1,0,-1)$    &  $(-1,0,0)$ \\
20.   &   $(1,0,-1)$    &  $(-1,1,1)$ \\
21.  &   $(1,0,-1)$    &  $(-1,-1,-1)$ \\
22.  &   $(1,1,1)$    &  $(0,1,-1)$ \\
23.   &   $(1,1,1)$    &  $(0,-1,1)$ \\
24.   &   $(1,-1,1)$    &  $(0,1,-1)$ \\
25.   &   $(1,-1,1)$    &  $(0,-1,1)$ \\
26.   &   $(-1,0,1)$    &  $(0,0,0)$ \\
27.   &   $(-1,0,1)$    &  $(0,1,1)$ \\
28.   &   $(-1,0,1)$    &  $(0,-1,-1)$ \\
29.   &   $(-1,0,1)$    &  $(1,0,0)$ \\
30.   &   $(-1,0,1)$    &  $(1,1,1)$ \\
31.   &   $(-1,0,1)$    &  $(1,-1,-1)$ \\
32.   &   $(-1,0,1)$    &  $(-1,0,0)$ \\
33.   &   $(-1,0,1)$    &  $(-1,1,1)$ \\
34.   &   $(-1,0,1)$    &  $(-1,-1,-1)$ \\
35.   &   $(-1,0,-1)$    &  $(0,1,-1)$ \\
36.   &   $(-1,0,-1)$    &  $(0,-1,1)$ \\
37.   &   $(-1,-1,-1)$    &  $(0,1,-1)$ \\
38.   &   $(-1,1,-1)$    &  $(0,1,-1)$ \\
39.   &   $(-1,1,-1)$    &  $(0,-1,1)$ \\
40.   &   $(-1,-1,-1)$    &  $(0,-1,1)$ \\[1ex]\hline
\end{tabular}
\caption{$(\mu - \tau)$ symmetry (in $3 \times 3$ block).}
\end{minipage}
\hspace{0.8cm}
\begin{minipage}[b]{0.20\linewidth}\centering
\begin{tabular}{| c | c | c |}
\hline
Sl no. & $\phi^{'}$ & $\phi^{''}$ \\
\hline\hline
1. &  $(0,0,0)$  &  $(1,1,-1)$ \\
2.   &   $(0,0,0)$    &  $(1,-1,1)$  \\
3.   &   $(0,0,0)$    &  $(-1,1,-1)$  \\
4.   &   $(0,0,0)$    &  $(-1,-1,1)$  \\
5.   &   $(0,0,1)$    &  $(1,1,1)$  \\
6.   &   $(0,0,1)$    &  $(1,1,-1)$  \\
7.    &   $(0,0,1)$    &  $(1,-1,1)$  \\
\hline
\end{tabular}
\caption{Hybrid texture.}
\end{minipage}
\hspace{0.4cm}
\begin{minipage}[b]{0.20\linewidth}\centering
\begin{tabular}{| c | c | c |}
\hline
Sl no. & $\phi^{'}$ & $\phi^{''}$ \\
\hline\hline
49.  &   $(0,-1,1)$    &  $(-1,1,1)$  \\
50.    &   $(0,-1,1)$    &  $(-1,1,-1)$  \\
51.    &   $(0,-1,1)$    &  $(-1,-1,1)$  \\
52.    &   $(0,-1,1)$    &  $(-1,-1,-1)$  \\
53.  &   $(0,-1,-1)$    &  $(1,1,1)$  \\
54.     &   $(0,-1,-1)$    &  $(1,1,-1)$  \\
55.     &   $(0,-1,-1)$    &  $(1,-1,1)$  \\
\hline
\end{tabular}
\caption{Hybrid texture (contd).}
\end{minipage}
\end{table}
\begin{table}[ht]
\tiny
\begin{minipage}[b] {0.20\linewidth}\centering
\begin{tabular}{| c | c | c |}
\hline
Sl no.  & $\phi^{'}$ & $\phi^{''}$ \\
\hline\hline
97.  &   $(1,1,1)$    &  $(-1,1,0)$  \\
98.  &   $(1,1,1)$    &  $(-1,1,-1)$  \\
99.   &   $(1,1,1)$    &  $(-1,-1,0)$  \\
100.   &   $(1,1,1)$    &  $(-1,-1,1)$  \\
101.   &   $(1,1,-1)$    &  $(0,0,0)$  \\
102.   &   $(1,1,-1)$    &  $(0,0,1)$  \\
103.  &   $(1,1,-1)$    &  $(0,0,-1)$  \\
\hline
\end{tabular}
\caption{Hybrid texture (contd).}
\end{minipage}
\hspace{0.6cm}
\begin{minipage}[b]{0.20\linewidth}\centering
\begin{tabular}{ | c | c | c |}
\hline
Sl no.  & $\phi^{'}$ & $\phi^{''}$ \\
\hline\hline
148.   &   $(1,-1,1)$    &  $(-1,1,-1)$ \\ 
149.    &   $(1,-1,1)$   &  $(-1,-1,0)$ \\
150.   &   $(1,-1,1)$    &  $(-1,-1,1)$  \\
151.   &   $(1,-1,-1)$    &  $(0,0,0)$  \\
152.   &   $(1,-1,-1)$    &  $(0,0,1)$  \\
153.   &   $(1,-1,-1)$    &  $(0,0,-1)$  \\
154.   &   $(1,-1,-1)$    &  $(0,1,0)$  \\
\hline
\end{tabular}
\caption{Hybrid texture (contd).}
\end{minipage}
\hspace{0.6cm}
\begin{minipage}[b]{0.20\linewidth}\centering
\begin{tabular}{| c | c | c |}
\hline
Sl no.  & $\phi^{'}$ & $\phi^{''}$ \\
\hline\hline
199.   &   $(-1,1,0)$    &  $(-1,1,-1)$  \\  
200.    &   $(-1,1,0)$    &  $(-1,-1,1)$  \\  
201.    &   $(-1,1,0)$    &  $(-1,-1,-1)$  \\  
202.    &   $(-1,1,1)$    &  $(0,0,0)$  \\  
203.    &   $(-1,1,1)$    &  $(0,0,1)$  \\  
204.    &   $(-1,1,1)$    &  $(0,0,-1)$  \\  
205.    &   $(-1,1,1)$    &  $(0,1,0)$  \\
\hline
\end{tabular}
\caption{Hybrid texture (contd).}
\end{minipage}
\hspace{0.6cm}
\begin{minipage}[b]{0.20\linewidth}\centering
\begin{tabular}{| c | c | c |}
\hline
Sl no. & $\phi^{'}$ & $\phi^{''}$ \\
\hline\hline
248.    &   $(-1,-1,0)$    &  $(1,-1,-1)$  \\  
249.    &   $(-1,-1,0)$    &  $(-1,1,1)$  \\  
250.    &   $(-1,-1,0)$    &  $(-1,1,-1)$  \\
251.    &   $(-1,-1,0)$    &  $(-1,-1,1)$  \\  
252.    &   $(-1,-1,0)$    &  $(-1,-1,-1)$  \\  
253.    &   $(-1,-1,1)$    &  $(0,0,0)$  \\  
254.   &   $(-1,-1,1)$    &  $(0,0,1)$  \\
\hline
\end{tabular}
\caption{Hybrid texture (contd)}
\end{minipage}
\end{table}

\section{\small{Vacuum alignment of $\phi'$, $\phi''$ of disallowed cases. VEV of the flavon field $\phi$, $\langle \phi \rangle = (1,0,0)$ has been used.}}
\label{appen3}
\begin{table}[htbp]
\tiny
\begin{minipage}[b]{0.20\linewidth}\centering
\begin{tabular}{| c | c | c |}
\hline
Sl no. & $\phi^{'}$ & $\phi^{''}$ \\
\hline\hline
1.    &   $(0,0,0)$    &  $(0,0,0)$ \\
2.    &  $(0,0,0)$  & $(1,0,0)$ \\
3.    &  $(0,0,0)$  & $(-1,0,0)$ \\
4.   &   $(0,1,0)$   & $(0,0,0)$ \\
5.   &   $(0,1,0)$   & $(1,0,0)$ \\
6.   &   $(0,1,0)$   & $(-1,0,0)$ \\
7.   &   $(0,-1,0)$   & $(0,0,0)$ \\
8.   &   $(0,-1,0)$   & $(1,0,0)$ \\
9.  &   $(0,-1,0)$   & $(-1,0,0)$ \\[1ex]\hline
\end{tabular}
\caption{Texture zero in the entire 2nd and 3rd row and column of $4 \times 4$ matrix.}
\end{minipage}
\hspace{0.6cm}
\begin{minipage}[b]{0.20\linewidth}\centering
\begin{tabular}{| c | c | c |}
\hline
Sl no.  & $\phi^{'}$ & $\phi^{''}$ \\
\hline\hline
1.   &   $(0,1,-1)$   & $(1,-1,0)$ \\
2.    &   $(0,1,-1)$   & $(-1,1,0)$ \\
3.    &   $(0,-1,1)$   & $(1,-1,0)$ \\
4.   &   $(0,-1,1)$   & $(-1,1,0)$ \\
5.   &   $(0,0,0)$   & $(0,1,0)$ \\
6.    &   $(0,0,0)$   & $(0,-1,0)$ \\
7.    &   $(0,0,0)$   & $(1,-1,0)$ \\
8.    &   $(0,0,0)$   & $(-1,1,0)$ \\
9.    &   $(0,0,0)$   & $(-1,-1,0)$ \\
10.    &   $(0,0,1)$   & $(0,0,0)$ \\
11.    &   $(0,0,1)$   & $(0,1,0)$ \\
12.    &   $(0,0,1)$   & $(0,-1,0)$ \\
13.    &   $(0,0,1)$   & $(1,0,0)$ \\
14.    &   $(0,0,1)$   & $(1,1,0)$ \\
15.    &   $(0,0,1)$   & $(1,-1,0)$ \\
16.    &   $(0,0,1)$   & $(-1,0,0)$ \\
17.    &   $(0,0,1)$   & $(-1,1,0)$ \\
18.   &   $(0,0,1)$   & $(-1,-1,0)$ \\
19.    &   $(0,0,-1)$   & $(0,0,0)$ \\
20.   &   $(0,0,-1)$   & $(0,1,0)$ \\
21.   &   $(0,0,-1)$   & $(1,0,0)$ \\
22.   &   $(0,0,-1)$   & $(1,1,0)$ \\
23.   &   $(0,0,-1)$   & $(1,-1,0)$ \\
24.   &   $(0,0,-1)$   & $(-1,0,0)$ \\
25.   &   $(0,0,-1)$   & $(-1,1,0)$ \\
26.    &   $(0,0,-1)$   & $(-1,-1,0)$ \\
27.   &   $(0,1,0)$   & $(0,1,0)$ \\
28.   &   $(0,1,0)$   & $(0,-1,0)$ \\
29.   &   $(0,1,0)$   & $(1,1,0)$ \\
30.    &   $(0,1,0)$   & $(1,-1,0)$ \\
31.    &   $(0,1,0)$   & $(-1,1,0)$ \\
32.    &   $(0,1,0)$   & $(-1,-1,0)$ \\
33.    &   $(0,1,1)$   & $(0,0,0)$ \\
34.    &   $(0,1,1)$   & $(0,1,0)$ \\
35.    &   $(0,1,1)$   & $(0,-1,0)$ \\
36.    &   $(0,1,1)$   & $(1,0,0)$ \\
37.    &   $(0,1,1)$   & $(1,1,0)$ \\
38.     &   $(0,1,1)$   & $(1,-1,0)$ \\
39.   &   $(0,1,1)$   & $(-1,0,0)$ \\
40.    &   $(0,1,1)$   & $(-1,1,0)$ \\[1ex]\hline
\end{tabular}
\caption{Texture zero in the entire 2nd row and column of $4 \times 4$ matrix.}
\end{minipage}
\hspace{0.6cm}
\begin{minipage}[b]{0.20\linewidth}\centering
\begin{tabular}{| c | c | c |}
\hline
Sl no.  & $\phi^{'}$ & $\phi^{''}$ \\
\hline\hline
41.    &   $(0,1,1)$   & $(-1,-1,0)$ \\
42.    &   $(0,1,-1)$   & $(0,0,0)$ \\
43.  &   $(0,1,-1)$   & $(0,1,0)$ \\
44.   &   $(0,1,-1)$   & $(0,-1,0)$ \\
45.   &   $(0,1,-1)$   & $(1,0,0)$ \\
46.   &   $(0,1,-1)$   & $(1,1,0)$ \\
47.   &   $(0,1,-1)$   & $(-1,0,0)$ \\
48.   &   $(0,1,-1)$   & $(-1,-1,0)$ \\
49.   &   $(0,-1,0)$   & $(0,1,0)$ \\
50.   &   $(0,-1,0)$   & $(0,-1,0)$ \\
51.   &   $(0,-1,0)$   & $(1,1,0)$ \\
52.   &   $(0,-1,0)$   & $(1,-1,0)$ \\
53.   &   $(0,-1,0)$   & $(-1,1,0)$ \\
54.   &   $(0,-1,0)$   & $(-1,-1,0)$ \\
55.   &   $(0,-1,1)$   & $(0,0,0)$ \\
56.   &   $(0,-1,1)$   & $(0,1,0)$ \\
57.   &   $(0,-1,1)$   & $(0,-1,0)$ \\
58.   &   $(0,-1,1)$   & $(1,0,0)$ \\
59.   &   $(0,-1,1)$   & $(1,1,0)$ \\
60.   &   $(0,-1,1)$   & $(-1,0,0)$ \\
61.    &   $(0,-1,1)$   & $(-1,-1,0)$ \\
62.   &   $(0,-1,-1)$   & $(0,0,0)$ \\
63.   &   $(0,-1,-1)$   & $(0,1,0)$ \\
64.  &   $(0,-1,-1)$   & $(0,-1,0)$ \\
65.  &   $(0,-1,-1)$   & $(1,0,0)$ \\
66.  &   $(0,-1,-1)$   & $(1,1,0)$ \\
67.  &   $(0,-1,-1)$   & $(1,-1,0)$ \\
68.  &   $(0,-1,-1)$   & $(-1,0,0)$ \\
69.  &   $(0,-1,-1)$   & $(-1,1,0)$ \\
70.  &   $(0,-1,-1)$   & $(-1,-1,0)$ \\
71.  &   $(0,0,-1)$   & $(0,-1,0)$ \\[1ex]\hline
\end{tabular}
\caption{Texture zero in the entire 2nd row and column of $4 \times 4$ matrix (contd).}
\end{minipage}
\hspace{0.6cm}
\begin{minipage}[b]{0.20\linewidth}
\centering
\begin{tabular}{| c | c | c |}
\hline
Sl no. & $\phi^{'}$ & $\phi^{''}$ \\
\hline\hline
1.  &   $(1,-1,0)$   & $(1,0,-1)$ \\
2.    &   $(1,-1,0)$   & $(-1,0,1)$ \\
3.    &   $(-1,1,0)$   & $(1,0,-1)$ \\
4.   &   $(-1,1,0)$   & $(-1,0,1)$ \\
5.   &   $(0,0,0)$   & $(0,0,1)$ \\
6.   &   $(0,0,0)$   & $(0,0,-1)$ \\
7.  &   $(0,0,0)$   & $(1,0,1)$ \\
8.   &   $(0,0,0)$   & $(1,0,-1)$ \\
9.   &   $(0,0,0)$   & $(1,1,0)$ \\
10.   &   $(0,0,0)$   & $(-1,0,1)$ \\
11.   &   $(0,0,0)$   & $(-1,0,-1)$ \\
12.   &   $(0,1,0)$   & $(0,0,1)$ \\
13.   &   $(0,1,0)$   & $(0,0,-1)$ \\
14.   &   $(0,1,0)$   & $(1,0,1)$ \\
15.   &   $(0,1,0)$   & $(1,0,-1)$ \\
16.  &   $(0,1,0)$   & $(-1,0,1)$ \\
17.   &   $(0,1,0)$   & $(-1,0,-1)$ \\
18.  &   $(0,-1,0)$   & $(0,0,1)$ \\
19.   &   $(0,-1,0)$   & $(0,0,-1)$ \\
20.   &   $(0,-1,0)$   & $(1,0,1)$ \\
21.   &   $(0,-1,0)$   & $(1,0,-1)$ \\
22.  &   $(0,-1,0)$   & $(-1,0,1)$ \\
23.   &   $(0,-1,0)$   & $(-1,0,-1)$ \\
24.  &   $(1,0,0)$   & $(0,0,0)$ \\
25.   &   $(1,0,0)$   & $(0,0,1)$ \\
26.    &   $(1,0,0)$   & $(0,0,-1)$ \\
27.   &   $(1,0,0)$   & $(1,0,0)$ \\
28.   &   $(1,0,0)$   & $(1,0,1)$ \\
29.   &   $(1,0,0)$   & $(1,0,-1)$ \\
30.   &   $(1,0,0)$   & $(-1,0,0)$ \\
31.   &   $(1,0,0)$   & $(-1,0,1)$ \\
32.   &   $(1,0,0)$   & $(-1,0,-1)$ \\
33.   &   $(1,1,0)$   & $(0,0,0)$ \\
34.   &   $(1,1,0)$   & $(0,0,1)$ \\
35.   &   $(1,1,0)$   & $(0,0,-1)$ \\
36.   &   $(1,1,0)$   & $(1,0,0)$ \\
37.   &   $(1,1,0)$   & $(1,0,1)$ \\
38.   &   $(1,1,0)$   & $(1,0,-1)$ \\
39.   &   $(1,1,0)$   & $(-1,0,0)$ \\
40.   &   $(1,1,0)$   & $(-1,0,1)$ \\[1ex]\hline
\end{tabular}
\caption{Texture zero in the entire 3rd row and column of $4 \times 4$ matrix. }
\end{minipage}
\end{table}

\begin{table}[ht]
\tiny
\begin{minipage}[b]{0.20\linewidth}
\centering
%
\caption{Texture zero in the entire 3rd row and column of $4 \times 4$ matrix (contd). }
\end{minipage}
\hspace{0.6cm}
\begin{minipage}[b]{0.20\linewidth}
\centering
%
\caption{$(\mu-\tau)$ symmetry in the entire $4 \times 4$ matrix.}
\end{minipage}
\hspace{0.5cm}
\begin{minipage}[b]{0.30\linewidth}
\centering
%
\caption{$(\mu-\tau)$ symmetry in the entire $4 \times 4$ matrix (contd).}
\end{minipage}
\end{table}

\vspace{-2.2cm}
\section{Light neutrino mass matrix elements}
\label{appen4}
\vspace{-1.2cm}{{\small \begin{widetext}
\begin{equation}
M_{ee} = c_{12}^2 c_{13}^2 c_{14}^2 m_1+e^{- i \alpha } c_{13}^2 c_{14}^2 m_2 s_{12}^2+e^{- i \beta } c_{14}^2 m_3 s_{13}^2+e^{-i \gamma } m_4 s_{14}^2 \nonumber
\end{equation}
\begin{eqnarray}
M_{e\mu} &=&-e^{-i \delta _{24}} c_{14} \big(e^{i \delta _{24}} c_{12} c_{13} c_{23} c_{24} \big(m_1-e^{- i \alpha } m_2\big) s_{12}-e^{i \big(\delta
_{13}+\delta _{24}\big)} c_{13} c_{24} \big(e^{- i \beta } m_3-e^{- i \alpha } m_2 s_{12}^2\big) s_{13} s_{23} \nonumber \\
 && +e^{i \big(2 \alpha +\delta _{14}\big)}M
c_{13}^2 m_2 s_{12}^2 s_{14} s_{24}-e^{i \delta _{14}} \big(e^{- i \gamma } m_4-e^{- i \beta } m_3 s_{13}^2\big) s_{14} s_{24}+c_{12}^2 c_{13}
m_1 \big(e^{i \big(\delta _{13}+\delta _{24}\big)} c_{24} s_{13} s_{23} \nonumber  \\
&& +e^{i \delta _{14}} c_{13} s_{14} s_{24}\big)\big) \nonumber
\end{eqnarray}
\begin{eqnarray}
M_{e\tau}&=&c_{14} \big(-e^{i \big(- \alpha +\delta _{14}\big)} c_{13}^2 c_{24} m_2 s_{12}^2 s_{14} s_{34}+e^{i \delta _{14}} c_{24} \big(e^{- i \gamma
} m_4-e^{- i \beta } m_3 s_{13}^2\big) s_{14} s_{34} \nonumber \\
&& +c_{12} c_{13} \big(m_1-e^{- i \alpha } m_2\big) s_{12} \big(c_{34} s_{23}+e^{i \delta
_{24}} c_{23} s_{24} s_{34}\big)+e^{i \delta _{13}} c_{13} \big(e^{- i \beta } m_3-e^{- i \alpha } m_2 s_{12}^2\big) s_{13} \big(c_{23} c_{34} \nonumber \\
&& -e^{i
\delta _{24}} s_{23} s_{24} s_{34}\big)-c_{12}^2 c_{13} m_1 \big(e^{i \delta _{13}} c_{23} c_{34} s_{13}+\big(e^{i \delta _{14}} c_{13} c_{24}
s_{14}-e^{i \big(\delta _{13} +\delta _{24}\big)} s_{13} s_{23} s_{24}\big) s_{34}\big)\big) \nonumber
\end{eqnarray}
\begin{eqnarray}
M_{\mu\mu} &=&e^{ i \big(-\gamma + 2\delta_{14}- 2 \delta_{24}\big)} c_{14}^2 m_4 s_{24}^2+e^{- i \beta } m_3 \big(e^{i \delta _{13}} c_{13} c_{24} s_{23}-e^{i
\big(\delta _{14}-\delta _{24}\big)} s_{13} s_{14} s_{24}\big){}^2+e^{- i \alpha } m_2 \big(c_{12} c_{23} c_{24}\nonumber \\
&& +s_{12} \big(-e^{i \delta
_{13}} c_{24} s_{13} s_{23}-e^{i \big(\delta _{14}-\delta _{24}\big)} c_{13} s_{14} s_{24}\big)\big){}^2+m_1 \big(c_{23} c_{24} s_{12}+c_{12}
\big(e^{i \delta _{13}} c_{24} s_{13} s_{23} \nonumber \\
&& +e^{i \big(\delta _{14}-\delta _{24}\big)} c_{13} s_{14} s_{24}\big)\big){}^2 \nonumber
\end{eqnarray}
\begin{eqnarray}
M_{\mu\tau} &=& e^{i \big(- \gamma +2 \delta _{14}-\delta _{24}\big)} c_{14}^2 c_{24} m_4 s_{24} s_{34}+e^{i \big(2 \beta +\delta _{13}\big)} m_3 \big(e^{i
\delta _{13}} c_{13} c_{24} s_{23}-e^{i \big(\delta _{14}-\delta _{24}\big)} s_{13} s_{14} s_{24}\big)  \nonumber \\
&&\big(-e^{-i \big(\delta _{13}-\delta
_{14}\big)} c_{24} s_{13} s_{14} s_{34}+c_{13} \big(c_{23} c_{34}-e^{i \delta _{24}} s_{23} s_{24} s_{34}\big)\big)+m_1 \big(-c_{23} c_{24}
s_{12}+c_{12} \big(-e^{i \delta _{13}} c_{24} s_{13} s_{23} \nonumber \\
&& -e^{i \big(\delta _{14}-\delta _{24}\big)} c_{13} s_{14} s_{24}\big)\big) \big(s_{12}
\big(c_{34} s_{23}+e^{i \delta _{24}} c_{23} s_{24} s_{34}\big)+c_{12} \big(-e^{i \delta _{14}} c_{13} c_{24} s_{14} s_{34}-e^{i \delta _{13}}
s_{13} \big(c_{23} c_{34} \nonumber \\
&&-e^{i \delta _{24}} s_{23} s_{24} s_{34}\big)\big)\big)+e^{- i \alpha } m_2 \big(c_{12} c_{23} c_{24}+s_{12} \big(-e^{i
\delta _{13}} c_{24} s_{13} s_{23}-e^{i \big(\delta _{14}-\delta _{24}\big)} c_{13} s_{14} s_{24}\big)\big) \big(-c_{12} \big(c_{34} s_{23} \nonumber \\
&&+e^{i
\delta _{24}} c_{23} s_{24} s_{34}\big)+s_{12} \big(-e^{i \delta _{14}} c_{13} c_{24} s_{14} s_{34}-e^{i \delta _{13}} s_{13} \big(c_{23} c_{34}-e^{i
\delta _{24}} s_{23} s_{24} s_{34}\big)\big)\big) \nonumber
\end{eqnarray}
\begin{eqnarray}
M_{\tau \tau} &=& e^{ i \big(-\gamma + 2 \delta_{14}\big)} c_{14}^2 c_{24}^2 m_4 s_{34}^2+e^{ i \big(-\beta +2\delta_{13}\big)} m_3 \big(e^{-i \big(\delta
_{13}-\delta _{14}\big)} c_{24} s_{13} s_{14} s_{34}+c_{13} \big(-c_{23} c_{34}+e^{i \delta _{24}} s_{23} s_{24} s_{34}\big)\big){}^2 \nonumber \\
&& +m_1\big(s_{12} \big(c_{34} s_{23}+e^{i \delta _{24}} c_{23} s_{24} s_{34}\big)+c_{12} \big(-e^{i \delta _{14}} c_{13} c_{24} s_{14} s_{34}-e^{i
\delta _{13}} s_{13} \big(c_{23} c_{34}-e^{i \delta _{24}} s_{23} s_{24} s_{34}\big)\big)\big){}^2 \nonumber \\
&& +e^{- i \alpha } m_2 \big(c_{12} \big(c_{34}
s_{23}+e^{i \delta _{24}} c_{23} s_{24} s_{34}\big)-s_{12} \big(-e^{i \delta _{14}} c_{13} c_{24} s_{14} s_{34}-e^{i \delta _{13}} s_{13} \big(c_{23}
c_{34} -e^{i \delta _{24}} s_{23} s_{24} s_{34}\big)\big)\big){}^2 \nonumber
\end{eqnarray}
\begin{eqnarray}
M_{es} &=& c_{14} \big(e^{i \delta _{14}} c_{24} c_{34} \big(e^{- i \gamma } m_4-e^{- i \alpha } c_{13}^2 m_2 s_{12}^2-e^{- i \beta } m_3 s_{13}^2\big)
s_{14}-e^{i \delta _{13}} c_{13} \big(e^{- i \beta } m_3-e^{- i \alpha } m_2 s_{12}^2\big) s_{13}  \nonumber \\
&& \big(e^{i \delta _{24}} c_{34} s_{23} s_{24}+c_{23}
s_{34}\big)+c_{12} c_{13} \big(m_1-e^{- i \alpha } m_2\big) s_{12} \big(e^{i \delta _{24}} c_{23} c_{34} s_{24}-s_{23} s_{34}\big) \nonumber \\
&& -c_{12}^2
c_{13} m_1 \big(e^{i \delta _{14}} c_{13} c_{24} c_{34} s_{14}-e^{i \delta _{13}} s_{13} \big(e^{i \delta _{24}} c_{34} s_{23} s_{24}+c_{23} s_{34}\big)\big)\big) \nonumber
\end{eqnarray}
\begin{eqnarray}
M_{\mu s} &=& e^{i \big(2 \gamma +2 \delta _{14}-\delta _{24}\big)} c_{14}^2 c_{24} c_{34} m_4 s_{24}+e^{i \big(2 \beta +\delta _{13}\big)} m_3 \big(e^{i
\delta _{13}} c_{13} c_{24} s_{23}-e^{i \big(\delta _{14}-\delta _{24}\big)} s_{13} s_{14} s_{24}\big) \nonumber \\
&& \big(-e^{-i \big(\delta _{13}-\delta
_{14}\big)} c_{24} c_{34} s_{13} s_{14}-c_{13} \big(e^{i \delta _{24}} c_{34} s_{23} s_{24}+c_{23} s_{34}\big)\big)+m_1 \big(-c_{23} c_{24}
s_{12}+c_{12} \big(-e^{i \delta _{13}} c_{24} s_{13} s_{23} \nonumber \\
&&  -e^{i \big(\delta _{14}-\delta _{24}\big)} c_{13} s_{14} s_{24}\big)\big)\big(s_{12}
\big(e^{i \delta _{24}} c_{23} c_{34} s_{24}-s_{23} s_{34}\big)+c_{12} \big(-e^{i \delta _{14}} c_{13} c_{24} c_{34} s_{14}+e^{i \delta _{13}}
s_{13} \big(e^{i \delta _{24}} c_{34} s_{23} s_{24} \nonumber \\
&& +c_{23} s_{34}\big)\big)\big)+e^{- i \alpha } m_2 \big(c_{12} c_{23} c_{24}+s_{12}\big(-e^{i
\delta _{13}} c_{24} s_{13} s_{23}-e^{i \big(\delta _{14}-\delta _{24}\big)} c_{13} s_{14} s_{24}\big)\big) \big(c_{12} \big(-e^{i \delta
_{24}} c_{23} c_{34} s_{24} \nonumber \\
&&+s_{23} s_{34}\big)+s_{12} \big(-e^{i \delta _{14}} c_{13} c_{24} c_{34} s_{14}+e^{i \delta _{13}} s_{13} \big(e^{i
\delta _{24}} c_{34} s_{23} s_{24}+c_{23} s_{34}\big)\big)\big) \nonumber
\end{eqnarray}
\begin{eqnarray}
M_{\tau s} &=& e^{ i \big(-\gamma +2\delta_{14}\big)} c_{14}^2 c_{24}^2 c_{34} m_4 s_{34}+e^{ i \big(-\beta +2 \delta_{13}\big)} m_3 \big(-e^{-i \big(\delta
_{13}-\delta _{14}\big)} c_{24} c_{34} s_{13} s_{14}-c_{13} \big(e^{i \delta _{24}} c_{34} s_{23} s_{24}+c_{23} s_{34}\big)\big)  \nonumber \\
&& \big(-e^{-i
\big(\delta _{13}-\delta _{14}\big)} c_{24} s_{13} s_{14} s_{34}+c_{13} \big(c_{23} c_{34}-e^{i \delta _{24}} s_{23} s_{24} s_{34}\big)\big)+m_1
\big(s_{12} \big(e^{i \delta _{24}} c_{23} c_{34} s_{24}-s_{23} s_{34}\big) \nonumber \\
&& +c_{12} \big(-e^{i \delta _{14}} c_{13} c_{24} c_{34} s_{14}+e^{i
\delta _{13}} s_{13} \big(e^{i \delta _{24}} c_{34} s_{23} s_{24}+c_{23} s_{34}\big)\big)\big) \big(s_{12} \big(c_{34} s_{23}+e^{i \delta
_{24}} c_{23} s_{24} s_{34}\big) \nonumber \\
&& +c_{12} \big(-e^{i \delta _{14}} c_{13} c_{24} s_{14} s_{34}-e^{i \delta _{13}} s_{13} \big(c_{23} c_{34}-e^{i
\delta _{24}} s_{23} s_{24} s_{34}\big)\big)\big)+e^{- i \alpha } m_2 \big(c_{12} \big(-e^{i \delta _{24}} c_{23} c_{34} s_{24}+s_{23} s_{34}\big) \nonumber \\
&& +s_{12}
\big(-e^{i \delta _{14}} c_{13} c_{24} c_{34} s_{14}+e^{i \delta _{13}} s_{13} \big(e^{i \delta _{24}} c_{34} s_{23} s_{24}+c_{23} s_{34}\big)\big)\big)
\big(-c_{12} \big(c_{34} s_{23}+e^{i \delta _{24}} c_{23} s_{24} s_{34}\big) \nonumber \\
&& +s_{12} \big(-e^{i \delta _{14}} c_{13} c_{24} s_{14} s_{34} -e^{i
\delta _{13}} s_{13} \big(c_{23} c_{34}-e^{i \delta _{24}} s_{23} s_{24} s_{34}\big)\big)\big) \nonumber
\end{eqnarray}
\begin{eqnarray}
M_{ss} &=& e^{- i \big(\gamma +\delta _{14}\big)} c_{14}^2 c_{24}^2 c_{34}^2 m_4+e^{ i \big(-\beta + 2 \delta_{13}\big)} m_3 \big(e^{-i \big(\delta
_{13}-\delta _{14}\big)} c_{24} c_{34} s_{13} s_{14}+c_{13} \big(e^{i \delta _{24}} c_{34} s_{23} s_{24}+c_{23} s_{34}\big)\big){}^2 \nonumber \\
&& +m_1 \big(s_{12}
\big(e^{i \delta _{24}} c_{23} c_{34} s_{24}-s_{23} s_{34}\big)+c_{12} \big(-e^{i \delta _{14}} c_{13} c_{24} c_{34} s_{14}+e^{i \delta _{13}}
s_{13} \big(e^{i \delta _{24}} c_{34} s_{23} s_{24} +c_{23} s_{34}\big)\big)\big){}^2 \nonumber \\
&&+e^{- i \alpha } m_2 \big(c_{12} \big(-e^{i \delta _{24}}
c_{23} c_{34} s_{24}+s_{23} s_{34}\big)  \nonumber \\
&& +s_{12} \big(-e^{i \delta _{14}} c_{13} c_{24} c_{34} s_{14}+e^{i \delta _{13}} s_{13} \big(e^{i \delta
_{24}} c_{34} s_{23} s_{24}+c_{23} s_{34}\big)\big)\big){}^2 \nonumber
\end{eqnarray}

\end{widetext}}

\bibliography{apssamp}

\end{document}